\documentclass[aps,prc,showkeys,nofootinbib]{revtex4}

\usepackage{color} 

\usepackage[dvips]{graphicx}
\usepackage{dcolumn}
\usepackage{amsthm}
\usepackage{bm}

\usepackage{isotope}
\usepackage{physics}

\newcommand{\sigmabf}{\mbox{\boldmath $\sigma$}}
\newcommand{\xibf}{\mbox{\boldmath $\xi$}}
\newcommand{\rhobf}{\mbox{\boldmath $\rho$}}

\def\sunyatsenlp{$^{(1)}$Sino-French Institute of Nuclear Engineering and Technology, Sun Yat-Sen University, Zhuhai 519082, China}
\def\wigner{$^{(2)}$Wigner Research Centre for Physics, Budapest, 1121, Hungary}
\def\sunyatsen{$^{(3)}$School of Physics and Astronomy, Sun Yat-Sen University, Zhuhai, 519082, China}
\def\kyiv{$^{(4)}$Institute for Nuclear Research, National Academy of Sciences of Ukraine, Kyiv, 03680, Ukraine}

\begin{document}

\title{Nucleon microscopy in proton-nucleus scattering via analysis of bremsstrahlung emission: role of incoherent emission}


\author{Sergei~P.~Maydanyuk$^{1,2,4}$}\email{sergei.maydanyuk@wigner.hu}
\author{Li-Ping~Zou$^{1}$}\email{zoulp5@mail.sysu.edu.cn}
\author{Peng-Ming~Zhang$^{3}$}\email{zhangpm5@mail.sysu.edu.cn}

\affiliation{\sunyatsenlp}
\affiliation{\wigner}
\affiliation{\sunyatsen}
\affiliation{\kyiv}

\date{\small\today}

\begin{abstract}
We study electromagnetic form factors of protons in proton-nucleus scattering via analysing of experimental cross-sections of accompanying bremsstrahlung photons.
A new bremsstrahlung model for proton-nucleus scattering is developed, where a main focus is given on incoherent bremsstrahlung that has not been considered previously.
%
In analysis we choose experimental bremsstrahlung data of $p + \isotope[197]{Au}$ scattering at proton beam energy of 190~MeV obtained by TAPS collaboration.
We find the following.
(1) 
Inclusion of incoherent emission to calculations improves agreements with experimental data essentially,
contribution of incoherent bremsstrahlung is essentially larger than coherent one.
%
(2) Inclusion of form factors of the scattered proton improves agreement with experimental data in comparison with calculations with coherent and incoherent contributions without form factors.
%
%
%
(3) Sensitivity of model in study of form factors of the scattered proton is high.
This demonstrates a new opportunity to study internal structure of protons under influence of nuclear forces in nuclear scattering.
%


\end{abstract}




\maketitle

\section{Introduction
\label{sec.introduction}}


Form factors of nucleons provide unique information about internal structure, distribution of electrical charge, magnetization, other important physical properties of these objects~\cite{A1Collaboration.2014.PRCv90p015206,Friedrich.2003.EPJA}.
%
High energy lepton-nucleon scattering plays a key role in determination of these characteristics.
Relevant information is summarized in reports [see Review PDG~\cite{Review_particle_phys.2018},
reviews~\cite{DeRoeck.2011.PPNP,Forte.2013.ARNPS,Blumlein.2013.PPNP,Perez.2013.RPP,Ball.2013.JHEP}].
Investigations of nucleon-nucleon collisions at TEVATRON (Fermilab), RHIC (Brookhaven), LHC (CERN) provide important information~\cite{Review_particle_phys.2018}.
Three-nucleon (and even two-nucleon) interactions, clusterization in nuclei playing important roles in understanding of nuclear processes, cannot be obtained just from analysis of lepton-nucleon processes.
It is known that complete information about nuclear interactions (which are under a basis of mechanisms and dynamics of reactions)
cannot be obtained from analysis of reactions between two nucleons
(fusion is a good demonstration, for example, see Ref.~\cite{Maydanyuk_Zhang_Zou.2017.PRC} with many demonstrations, also
reviews~\cite{Back.2014.RMP,Birkelund.1979.PRep,Vaz.1981.PRep,Birkelund.1983.ARNPS,Beckerman.1985.PRep,Steadman.1986.ARNPS,
Beckerman.1988.RPP,Rowley.1991.PLB,Vandenbosch.1992.ARNPS,Reisdorf.1994.JPG,Dasgupta.1998.ARNPS,Balantekin.1998.RMP,Liang.2005.IJMPE,
Canto.2006.PRep,Keeley.2007.PPNP,Hagino.2012.PTP}).
This situation leads naturally to investigations of interactions between nucleons and nuclei. 
In particular, bremsstrahlung photon emitted in nuclear process 
provide additional useful information about those reactions and related questions above.

A lot of properties of nuclei, nuclear mechanisms and reactions have been studied by many people on the basis of analysis of the bremsstrahlung emission.
In our investigations dynamics of the nuclear processes, nuclear deformations, interactions between nucleons, nuclei, hypernuclei, nuclear reactions with pions,
types of nuclear interactions,
quantum effects,
different scenarios of nuclear decays and fissions, etc.
were studied via bremsstrahlung analysis~\cite{Maydanyuk.2003.PTP,Maydanyuk.2006.EPJA,Maydanyuk.2008.EPJA,Maydanyuk.2008.MPLA,Maydanyuk.2009.NPA,%
Maydanyuk.2009.JPS,Maydanyuk.2010.PRC,Maydanyuk.2011.JPCS}.
Measurements of such photons provide necessary information about these aspects, models suitability can be verified.
So, analysis of bremsstrahlung photons is an useful tool to investigate questions above.



Investigations of internal structure of nucleons are not restricted by high energies.
As an example, note investigations of pion-nucleus interactions.
In QCD, momentum transfer below 1~GeV/c is low to distinguish structures of quarks and gluons.
So, a question could be asked how QCD can be relevant for pion-nucleon interactions at energies below few 100~MeV.
An answer is that structure of Lagrangian and its symmetries characterize the pion-nucleon and consequently pion-nucleus interactions at energy scale below 100 MeV
(see review~\cite{Kluge.1991.RepProgPhys}, reference therein).
%
A lot of experimental data of pion-nucleus interactions below 100 MeV (and higher) have been obtained.
%
Measurements have been performed
at CERN Synchro-cyclotron~\cite{Binon.1970.NPB},
in $M 13$ pion channel using the quadrupole-quadrupole-dipole pion spectrometer at TRIUMF (Vancouver)~\cite{Sobie.1984.PRC,Oram.1981.NIM,Sobie.1984.NIM},
at 7 GeV proton synchrotron NIMROD at Rutherford Laboratory~\cite{Clough.1974.NPB},
in pion channel and spectrometer (EPICS) at Clinton P. Anderson Meson Physics Facility (LAMPF, Los Alamos)~\cite{Clough.1974.NPB,Morris.1981.PRC,Boyer.1981.PRC,Preedom.1981.PRC},
at pion spectrometer facility at Swiss Institute for Nuclear Research (SIN, Switzerland)~\cite{Preedom.1979.NPA,Albanese.1980.NPA}.
%
%
%
Pion-nucleus cross-sections for
$^{6,7}{\rm Li}$ and $^{9}{\rm Be}$ at the incident pion energy $T_{\pi}$ of 90--860~MeV~\cite{Clough.1974.NPB},
\isotope[12]{C} at $T_{\pi} = 30$~MeV~\cite{Preedom.1981.PRC},
50~MeV~\cite{Sobie.1984.PRC,Preedom.1981.PRC},
180~MeV~\cite{Morris.1981.PRC},
120--280 ~MeV~\cite{Binon.1970.NPB},
90--860~MeV~\cite{Clough.1974.NPB},
$^{16}{\rm O}$ at $T_{\pi}=30, 50$~MeV~\cite{Preedom.1981.PRC},
79, 114, 163, 240, 343~MeV \cite{Albanese.1980.NPA} and
88--228~MeV~\cite{Clough.1974.NPB},
$^{28}{\rm Si}$ at $T_{\pi} = 130, 180, 226$~MeV~\cite{Preedom.1979.NPA},
\isotope[32,34]{S} at $T_{\pi}=50$~MeV~\cite{Sobie.1984.PRC},
$^{40}{\rm Ca}$ at $T_{\pi}=30, 50$~MeV~\cite{Preedom.1981.PRC} and 180~MeV\cite{Morris.1981.PRC},
$^{42,44,48}{\rm Ca}$ at $T_{\pi}=180$~MeV~\cite{Boyer.1981.PRC},
\isotope[40,42,44,48]{Ca} and $^{54}{\rm Fe}$ at 116, 180, 292.5~MeV~\cite{Boyer.1984.PRC},
$^{90}{\rm Zr}$ and $^{208}{\rm Pb}$ at $T_{\pi}= 30, 50$~MeV~\cite{Preedom.1981.PRC}, etc.
have been measured.
Note that theoretical description of these experimental data on the non-relativistic basis
(optical model \cite{Preedom.1979.NPA,Preedom.1981.PRC,Sobie.1984.PRC,Khallaf.2000.PRC,Akhter.2001.JPG,Khallaf.2002.PRC},
$\alpha$-cluster model~\cite{Ebrahim.2011.BrJP},
DWIA collective model~\cite{Preedom.1979.NPA,Morris.1981.PRC,Boyer.1981.PRC,Boyer.1984.PRC},
microscopical form factor model~\cite{Boyer.1981.PRC}, etc.)
and the relativistic basis (here, there are many papers related with formalism of Satchler and Johnson \cite{Satchler.1992.NPA,Johnson.1996.AP}, etc.)
have been obtained with similar accuracy at such a energy range.
However, the first investigations on the basis of bremsstrahlung analysis of these reactions have shown with good sensitivity how to distinguish role of relativistic and non-relativistic formalisms~\cite{Maydanyuk_Zhang_Zou.2018.PRC}.
Bremsstrahlung real photons and even virtual photons in electron scattering at low energies was studied for a long time ago by Schwinger and Suura~\cite{Schwinger.1949.PhysRev,Suura.1955.PhysRev}.
In particular, deformation of low energy part of peak (related for elastic electron scattering) in the spectrum was explained by such emission~\cite{Hofstadter.1956.RevModPhys}.

In Ref.~\cite{Maydanyuk_Zhang_Zou.2019.PRC.microscopy} a new approach to study internal structure of proton in its scattering off nucleus-target via analysis of emission of the bremsstrahlung photons was proposed.
We found visible sensitivity of the calculated spectra on parameters of internal structure of the scattered proton (see Fig.~1-3 in that paper).
But, in Ref.~\cite{Maydanyuk_Zhang_Zou.2019.PRC.microscopy} incoherent emission was not included to the model and calculations.
Although additional calculations (related with many-nucleon generalization) for incoherent processes were not done, idea connecting the bremsstrahlung spectrum with form factors of the scattered proton without such a type of emission was realized.
However, results of experimental study of bremsstrahlung in proton-nucleus scattering by TAPS collaboration~\cite{Goethem.2002.PRL} indicate that incoherent processes play an important role on the bremsstrahlung emission.
As we shown in Refs.~\cite{Maydanyuk_Zhang.2015.PRC,Maydanyuk_Zhang_Zou.2016.PRC,Maydanyuk.2012.PRC},
inclusion of relations between spin of the scattered proton and momenta of nucleons of nucleus-target to the model improves essentially agreement between theory and experimental data.
Without such an inclusion it is impossible to explain platheou in experimental data~\cite{Goethem.2002.PRL}.
As it was estimated, after inclusion of incoherent bremsstrahlung emission to the previous formalism~\cite{Maydanyuk_Zhang_Zou.2019.PRC.microscopy},
level of agreement between calculations and experimental data should be higher, analysis of internal structure of the scattered proton should be proper.
So, this is a main aim of current research.


Photons emitted from incoherent processes have been studied in many topics of nuclear and particle physics.
Physical picture of the studied reaction becomes more complete, after inclusion of such a type of photons.
%
For example, authors of Ref.~\cite{Zhu.2015.PRC} found not negligible role of incoherent photons in photoproduction of heavy quarks (charm, bottom) and heavy quarkonia [$J/\psi$, $\Upsilon(1S)$] in ultra-peripheral collisions of heavy nuclei Pb-Pb and proton-proton collisions at high energies
(see also Refs.~\cite{Zhu.2016.NPB,Ma_Zhu.2018.PRD}, review~\cite{Baur.2002.PhysRep}).
Incoherent photons have also an important place in study of interactions between dark matter and nuclear matter
\cite{Bell_Dent.2020.PRD}
(see also Refs.~\cite{Dent.2015.PRD,Bell_Cai_Dent.2015.PRD,Anand.2014.PRC,Ibe.2018.JHEP}, reference therein).
%
Generally it is difficult to realize calculations of the spectra including the incoherent contribution.
Researchers usually apply different approximations to calculate the coherent processes or incoherent ones
(see research~\cite{Remington.1987.PRC}).
So, in new research we focus on construction of unified formalism (with corresponding calculations and analysis) with description of the coherent and incoherent bremsstrahlung emissions.
We find that the calculated spectra are changed essentially after taking incoherent emission into account.

Note perspectives on studying electromagnetic observables of light nuclei ($A \le 8$) based on chiral effective field theory \cite{Pastore.2008.PRC}
(see research~\cite{Eden.1996.PRC} for $pp$ bremsstrahlung).
But, detailed description on quantum fluxes with good accuracy provided by our approach in space representation can be applied for different nuclear processes (with nuclei from light up to heavy).
Such a advance of formalism allows to study not small quantum effects in different nuclear processes
(for example, see demonstrations in Refs.~\cite{Maydanyuk_Zhang_Zou.2017.PRC,Maydanyuk.2015.NPA} for fusion, reference therein).
Such quantum effects have not been described by approaches based on chiral effective field theory.

The paper is organized in the following way.
In Sec.~\ref{sec.1}--\ref{sec.18} we present a new model of the bremsstrahlung photons emitted during proton nucleus scattering.
In Sec.~\ref{sec.results} we analyze experimental bremsstrahlung data in the scattering of $p + \isotope[197]{Au}$ at proton beam energy of 190~MeV.
We summarize conclusions in Sec.~\ref{sec.conclusions}.
Details of new calculations are presented in Appendixes~\ref{sec.app.1}--\ref{sec.app.4}.
\section{Many-nucleon generalization of Pauli equation with form factors
\label{sec.1}}

\subsection{Generalized Pauli equation for fermion with mass $m$ in field $V(\vb{r})$ with electromagnetic form factors
\label{sec.1.1}}

In Ref.~\cite{Maydanyuk_Zhang_Zou.2019.PRC.microscopy} [see Eqs.~(28)--(30) in that paper]
a generalization of Pauli equation for one nucleon with mass $m$ inside field $V(\vb{r})$ was obtained
[along logics in Ref.~\cite{Ahiezer.1981}, see Eqs.~(1.3.5)--(1.3.7) in p.~33, 48--60],
where electromagnetic form factors $F_{1}$ and $F_{2}$ of this nucleon are included:
\begin{equation}
  i\hbar\, \Bigl\{ F_{1}^{2} - F_{2}^{2}\, \vb{q}^{2} \Bigr\}\, f(|\vb{q}|) \displaystyle\frac{\partial \varphi}{\partial t} =
  A \cdot f(|\vb{q}|) \cdot \varphi + B \cdot \varphi.
\label{eq.4.1.1}
\end{equation}
Here, functions $A$, $B$ are defined by Eqs.~(29), (30) in Ref.~\cite{Maydanyuk_Zhang_Zou.2019.PRC.microscopy},
function $f(|\vb{q}|)$ is given by Eq.~(26) in Ref.~\cite{Maydanyuk_Zhang_Zou.2019.PRC.microscopy}.

In this paper we use approximation of component $A_{0} = 0$ for the four-potential of the electromagnetic field.
So, Eq.~(\ref{eq.4.1.1}) is transformed to
[see Eqs.~(B1), (B4), (B5), (B6) in Ref.~\cite{Maydanyuk_Zhang_Zou.2019.PRC.microscopy}, for details]
\begin{equation}
  i\hbar\, \Bigl\{ F_{1}^{3} (1 + F_{1}) - F_{1} F_{2}^{2}\, \vb{q}^{2} - F_{2}^{4}\, \mathbf{q}^{4} \Bigr\}\, \displaystyle\frac{\partial \varphi}{\partial t} =
  \bigl( h_{0} + h_{\gamma 0} + h_{\gamma 1} \bigr) \cdot \varphi,
\label{eq.4.1.8}
\end{equation}
where
\begin{equation}
\begin{array}{lll}
\vspace{0.1mm}
  h_{0} & = &
  \displaystyle\frac{a_{1}\, \mathbf{p}_{i}^{2}}{m} +
  \Bigl(
    F_{1}^{3}\, (1 + F_{1}) +
    F_{1} F_{2}^{2}\, (3 - 2 F_{1})\, \mathbf{q}^{2} +
    F_{2}^{4}\, \mathbf{q}^{4}
  \Bigr)\, V(\mathbf{r})\; + \\
\vspace{1.5mm}
  & + &
  mc^{2}
  \Bigl[
    F_{1}^{2}\, (1 - F_{1}^{2}) -
    (1 - 2\,F_{1}^{2})\, F_{2}^{2}\, \mathbf{q}^{2} -
    F_{2}^{4}\, \mathbf{q}^{4}
  \Bigr]\; - \\

\vspace{0.1mm}
  & - &
  i\, \displaystyle\frac{2F_{1}F_{2}}{mc}\,
  \Bigl\{
    F_{1} \bigl( F_{1} + F_{2} q^{4} \bigr)\, q^{m} +
    i\, \bigl( F_{1}^{2} + F_{1} F_{2} q^{4} - F_{2}^{2}\, \mathbf{q}^{2} \bigr) \varepsilon_{mjl}\, q^{j}\, \sigma_{l}
  \Bigr\}\, V(\mathbf{r})\, \mathbf{p}_{m}\; + \\
\vspace{0.1mm}
  & + &
  \displaystyle\frac{1}{m}\,
    \Bigl\{
      \bigl(F_{2}^{4} \mathbf{q}^{2} + a_{2} + a_{3}\bigr)\, \bigl( \mathbf{qp} \bigr)^{2} +
      i\, \varepsilon_{lm^{\prime}k}\, \sigma_{k}\: q^{l} q^{m}\, (a_{2}-a_{3}) p_{m} p_{m'}
    \Bigr\}\; + \\
\vspace{0.1mm}
  & + &
  i\, cF_{2}\,
  \Bigl\{
  \Bigl[
    F_{1}^{2} (1 - F_{1}) + F_{1} F_{2} (3F_{1} - 1)\, q^{4} -
    F_{2}^{2} \bigl( F_{1} + F_{2} q^{4} \bigr)\, \mathbf{q}^{2}
  \Bigr]\, q^{m}\; + \\
  & + &
  i\, \Bigl[
    2\, F_{1}^{2} +
    F_{1} F_{2} \bigl( 3 F_{1} - 1 \bigr)\, q^{4} -
    F_{2}^{2} \bigl( 2 + F_{2} q^{4} \bigr)\, \mathbf{q}^{2}
  \Bigr]\, \varepsilon_{mjl}\, q^{j}\, \sigma_{l} \Bigr\}\, \mathbf{p}_{m},
\end{array}
\label{eq.4.2.2}
\end{equation}
\begin{equation}
\begin{array}{lll}
\vspace{1.1mm}
  h_{\gamma 0} & = &
  \displaystyle\frac{a_{1}}{m}\,
    \Bigl[
      - \displaystyle\frac{z_{i}e}{c} (-i\hbar\, \mathbf{div A} + 2\,\mathbf{Ap}) +
      \displaystyle\frac{z_{i}^{2}e^{2}}{c^{2}}\, \mathbf{A}^{2} -
      \displaystyle\frac{z_{i}e}{c}\, \sigmabf \mathbf{H}
    \Bigr], \\

\vspace{0.5mm}
  h_{\gamma 1} & = &
  -\, i\, cF_{2}\,
    \Bigl\{ b_{1}\, q^{m} + b_{2}\, \varepsilon_{mjl}\, q^{j}\, \sigma_{l} \Bigr\}\,
    \displaystyle\frac{ze}{c} \mathbf{A}_{m} +
  \displaystyle\frac{1}{m}\,
  \Bigl\{
    \bigl(F_{2}^{4} \mathbf{q}^{2} + a_{2} + a_{3}\bigr)\,
      \Bigl[ -2 \mathbf{(qp)} \displaystyle\frac{ze}{c}\, \mathbf{(qA)} + \displaystyle\frac{z^{2}e^{2}}{c^{2}}\, \mathbf{(qA)}^{2} \Bigr]\; + \\
  & + &
  i\, \varepsilon_{lm^{\prime}k}\, \sigma_{k}\: q^{l} q^{m}\,
  \Bigl[
    (a_{2}-a_{3}) \Bigl(- \displaystyle\frac{ze}{c} ( A_{m'} p_{m} + A_{m} p_{m'} ) + \displaystyle\frac{z^{2}e^{2}}{c^{2}}\, A_{m} A_{m'} \Bigr) +
    \displaystyle\frac{i\hbar ze}{c} \Bigl( a_{2}\, \displaystyle\frac{dA_{m'}}{dx_{m}} - a_{3}\, \displaystyle\frac{dA_{m}}{dx_{m'}} \Bigr) \Bigr]
  \Bigr\}.
\end{array}
\label{eq.4.3.2}
\end{equation}
Here, the first term $h_{\gamma 0}$ describes electric and magnetic emissions of the bremsstrahlung photons without form factors of the scattered proton.
Such types of the coherent and incoherent bremsstrahlung emission in the proton-nucleus scattering were studied in details in Refs.~\cite{Maydanyuk.2012.PRC,Maydanyuk_Zhang.2015.PRC}.
The second term $h_{\gamma 1}$ is emission operator, describing contribution to the full bremsstrahlung spectrum taking the form factors of the scattered proton into account.
Coefficients $a_{1}$, $a_{2}$ and $a_{3}$ are defined in Eqs.~(A48) and
$b_{1}$, $b_{2}$ and $b_{3}$ are defined in Eqs.~(A27) in Ref.~\cite{Maydanyuk_Zhang_Zou.2019.PRC.microscopy}.


\subsection{Elastic scattering of virtual photon on proton 
\label{sec.6}}

As in Ref.~\cite{Maydanyuk_Zhang_Zou.2019.PRC.microscopy}, in this paper we will study virtual photons in approximation of elastic scattering with nucleons.
So, energy of proton in beam in not changed by virtual photon and we have
[for example, see Ref.~\cite{Greiner.Chromodynamics.2002}, p.~79--83]
\begin{equation}
\begin{array}{lll}
  q_{4} = 0, &
  \vb{q}^{2} = - q^{2} = Q^{2}.
\end{array}
\label{eq.6.1.2}
\end{equation}
Coefficient $a_{1}$, $a_{2}$, $a_{3}$ and $b_{1}$, $b_{2}$, $b_{3}$ can be simplified [see solutions (39)--(40) in Ref.~\cite{Maydanyuk_Zhang_Zou.2019.PRC.microscopy}].
%
%
%
%
%
We define real photons of bremsstrahlung emission in QED representation as Eq.~(42) in Ref.~\cite{Maydanyuk_Zhang_Zou.2019.PRC.microscopy},
with Coulomb gauge and two independent polarizations $\vb{e}^{(1)}$ and $\vb{e}^{(2)}$.
%
%
After calculations we obtain the following terms for hamiltonian
[see Eqs.~(45), (46), also Eqs.~(C1)--(C3), (C8) in Ref.~\cite{Maydanyuk_Zhang_Zou.2019.PRC.microscopy}, for details]:
\begin{equation}
\begin{array}{lll}
\vspace{0.1mm}
  h_{0} & = &
  \displaystyle\frac{a_{1}\, \mathbf{p}^{2}}{m} +
  \Bigl( F_{1}^{3}\, (1 + F_{1}) + F_{1} F_{2}^{2}\, (3 - 2 F_{1})\, Q^{2} + F_{2}^{4}\, Q^{4} + i\, \displaystyle\frac{F_{1}^{3}F_{2}\, Q^{2}}{mc} \Bigr) \bigl[ ze\, A_{0} + V(\mathbf{r}) \bigr]\; + \\
\vspace{1.1mm}
  & + &
  mc^{2} \Bigl[  F_{1}^{2}\, (1 - F_{1}^{2}) - (1 - 2\,F_{1}^{2})\, F_{2}^{2}\, Q^{2} - F_{2}^{4}\, Q^{4} \Bigr] +
  \displaystyle\frac{Q^{4}}{4m}\, \bigl(F_{2}^{4} Q^{2} + 2a_{2}\bigr) -
  \displaystyle\frac{i\, cF_{2}\, Q^{2}}{2}\, \Bigl[ F_{1}^{2} (1 - F_{1}) - F_{2}^{2} F_{1} \, Q^{2} \Bigr]\; + \\
  & + &
  \Bigl\{ \displaystyle\frac{2F_{1}F_{2}}{mc}\, \bigl[ ze\, A_{0} + V(\mathbf{r}) \bigr]\, - 2cF_{2} \Bigr\}
    \bigl( F_{1}^{2} - F_{2}^{2}\, Q^{2} \bigr) \varepsilon_{mjl}\, q^{j}\, \sigma_{l}\, \mathbf{p}_{m},
\end{array}
\label{eq.6.3.4}
\end{equation}
\begin{equation}
\begin{array}{lll}
  h_{\gamma 1} & = &
  ze\, F_{2}\, \sqrt{\displaystyle\frac{\pi\hbar c^{2}}{w_{\rm ph}}}\; e^{-i\, \mathbf{k_{\rm ph}r}}\;
  \Bigl\{
    2 Q\, \sin \varphi_{ph}\,
      \Bigl[ -\, i\, b_{1} + \displaystyle\frac{F_{2}\, Q^{2}}{mc}\, \bigl( 2F_{1}^{2} - F_{2}^{2}\, Q^{2} \bigr) \Bigr]\; - \\
  & - &
    i\, \sqrt{2}\, b_{2}\, \varepsilon_{mjl}\, q^{j}\, \sigma_{l} \sum\limits_{\alpha=1,2} \mathbf{e}_{m}^{(\alpha),\,*} +
    \displaystyle\frac{4ze}{mc}\, \sqrt{\displaystyle\frac{\pi\hbar}{w_{\rm ph}}}\;
      F_{2}Q^{2}\, \bigl( 2F_{1}^{2} - F_{2}^{2}\, Q^{2} \bigr)\;
      e^{-i\, \mathbf{k_{\rm ph}r}}\, \sin^{2} \varphi_{ph}
  \Bigr\},
\end{array}
\label{eq.6.3.5}
\end{equation}
$h_{\gamma 0}$ is not changed,
$\varphi_{ph}$ is angle between vectors $\vb{q}$ and $\vb{A}$
[see Eq.~(44) in Ref.~\cite{Maydanyuk_Zhang_Zou.2019.PRC.microscopy}].
%
%

\subsection{Many-nucleon generalization
\label{sec.7}}

We write many-nucleon generalization of Pauli equation on the scattering of proton off nucleus-target with $A$ nucleons in the laboratory frame as
(formalism is along Refs.~\cite{Maydanyuk.2012.PRC,Maydanyuk_Zhang.2015.PRC,Maydanyuk_Zhang_Zou.2016.PRC}, reference therein)
\begin{equation}
  i\hbar\, \displaystyle\frac{\partial \varphi}{\partial t} =
  \Bigl\{
    \displaystyle\sum\limits_{i=1}^{A+1}
      \bigl[ \bar{h}_{0} (\vb{r}_{i}) + \bar{h}_{\gamma 0} (\vb{r}_{i}) + \bar{h}_{\gamma 1} (\vb{r}_{i}) \bigr] +
    V (\vb{r}_{1} \ldots \vb{r}_{A}, \vb{r}_{\rm p})
  \Bigr\} \cdot \varphi,
\label{eq.7.1.5}
\end{equation}
where
\begin{equation}
\begin{array}{llllll}
  \bar{h}_{0} = \displaystyle\frac{h_{0}}{f(F_{1}, F_{2}, \vb{q}^{2})}, &
  \bar{h}_{\gamma 0} = \displaystyle\frac{h_{\gamma 0}}{f(F_{1}, F_{2}, \vb{q}^{2})}, &
  \bar{h}_{\gamma 1} = \displaystyle\frac{h_{\gamma 1}}{f(F_{1}, F_{2}, \vb{q}^{2})}.
\end{array}
\label{eq.7.1.2}
\end{equation}
Here,
$V(\vb{r}_{1} \ldots \vb{r}_{A}, \vb{r}_{p})$ is a general form of the potential of interactions between nucleons of nucleus and the scattered proton
[potential of relative motion (with possible tunneling after emission of photon at high energies) of the scattered proton concerning to nucleus-target
is defined according to formalism in
Refs.~\cite{Maydanyuk_Zhang_Zou.2019.PRC.microscopy,Maydanyuk.2012.PRC,Maydanyuk_Zhang.2015.PRC,Maydanyuk_Zhang_Zou.2016.PRC,Maydanyuk_Zhang_Zou.2018.PRC,Maydanyuk.2011.JPG}.
From Eq.~(\ref{eq.4.1.8}) and (\ref{eq.6.1.2}) we find:
\begin{equation}
\begin{array}{llllll}
  f(F_{1}, F_{2}, Q^{2}) = F_{1}^{3} (1 + F_{1}) - F_{1} F_{2}^{2}\, Q^{2} - F_{2}^{4}\, Q^{4}.
\end{array}
\label{eq.7.1.4}
\end{equation}
We substitute formulas (\ref{eq.7.1.2}), (\ref{eq.6.3.4}), (\ref{eq.4.3.2}) and (\ref{eq.6.3.5}) for components of operators of emission to this equation and calculate each term as
\begin{equation}
\begin{array}{llllll}
\vspace{0.1mm}
  & \displaystyle\sum\limits_{i=1}^{A+1} \bar{h}_{0} (\vb{r}_{i}) =
  \displaystyle\sum\limits_{i=1}^{A+1}
    \displaystyle\frac{a_{1i}\, \mathbf{p}_{i}^{2}}{f_{i}\,m_{i}} + V(\vb{r}_{1} \ldots \vb{r}_{A}, \vb{r}_{\rm p}) +
    \Delta\, h_{0}, \\

  & \Delta\, h_{0} =
  \displaystyle\sum\limits_{i=1}^{A+1}
  \displaystyle\frac{1}{f_{i}}
  \Bigl\{

  \Bigl( F_{1i}^{3}\, (1 + F_{1i}) + F_{1i} F_{2i}^{2}\, (3 - 2 F_{1i})\, Q^{2} + F_{2i}^{4}\, Q^{4} +
  i\, \displaystyle\frac{F_{1i}^{3}F_{2i}\, Q^{2}}{m_{i}c} \Bigr) \bigl[ z_{i}e\, A_{0i} + V(\vb{r}_{i}) \bigr]\; + \\
\vspace{1.1mm}
  + &
  m_{i}c^{2} \Bigl[  F_{1i}^{2}\, (1 - F_{1i}^{2}) - (1 - 2\,F_{1i}^{2})\, F_{2i}^{2}\, Q^{2} - F_{2i}^{4}\, Q^{4} \Bigr] +
  \displaystyle\frac{Q^{4}}{4m_{i}}\, \bigl(F_{2i}^{4} Q^{2} + 2a_{2i}\bigr) -
  \displaystyle\frac{i\, cF_{2i}\, Q^{2}}{2}\, \Bigl[ F_{1i}^{2} (1 - F_{1i}) - F_{2i}^{2} F_{1i} \, Q^{2} \Bigr]\; + \\
  + &
  \Bigl\{ \displaystyle\frac{2F_{1i}F_{2i}}{m_{i}c}\, \bigl[ z_{i}e\, A_{0i} + V(\vb{r}_{i}) \bigr]\, - 2cF_{2i} \Bigr\}
    \bigl( F_{1i}^{2} - F_{2i}^{2}\, Q^{2} \bigr) \varepsilon_{mjl}\, q^{j}\, \sigma_{l}\, \vb{p}_{mi}
  \Bigr\},
\end{array}
\label{eq.8.1.1}
\end{equation}
\begin{equation}
\begin{array}{lll}
  & \displaystyle\sum\limits_{i=1}^{A+1} \bar{h}_{\gamma 0} (\vb{r}_{i}) =
  \displaystyle\sum\limits_{i=1}^{A+1}
    \displaystyle\frac{a_{1i}}{f_{i}\,m_{i}}\,
    \Bigl[
      - \displaystyle\frac{z_{i}e}{c} (-i\hbar\, \vb{div A}_{i} + 2\,\vb{A_{i}p_{i}}) +
      \displaystyle\frac{z_{i}^{2}e^{2}}{c^{2}}\, \vb{A}_{i}^{2} -
      \displaystyle\frac{z_{i}e}{c}\, \sigmabf \vb{H}_{i}
    \Bigr], 
\end{array}
\label{eq.7.1.6b}
\end{equation}
\begin{equation}
\begin{array}{llllll}
\vspace{1.0mm}
  & \displaystyle\sum\limits_{i=1}^{A+1} \bar{h}_{\gamma 1} (\vb{r}_{i}) =
  \displaystyle\sum\limits_{i=1}^{A+1}
    \displaystyle\frac{1}{f_{i}}\, \cdot

  z_{i}e\, F_{2i}\, \sqrt{\displaystyle\frac{\pi\hbar c^{2}}{w_{\rm ph}}}\; e^{-i\, \mathbf{k_{\rm ph}r_{i}}}\;
  \Bigl\{
    2 Q\, \sin \varphi_{ph,i}\,
      \Bigl[ -\, i\, b_{1i} + \displaystyle\frac{F_{2i}\, Q^{2}}{m_{i}c}\, \bigl( 2F_{1i}^{2} - F_{2i}^{2}\, Q^{2} \bigr) \Bigr]\; - \\

  - &
    i\, \sqrt{2}\, b_{2i}\, \varepsilon_{mjl}\, q^{j}\, \sigma_{l} \sum\limits_{\alpha=1,2} \mathbf{e}_{m}^{(\alpha),\,*} +
    \displaystyle\frac{4z_{i}e}{m_{i}c}\, \sqrt{\displaystyle\frac{\pi\hbar}{w_{\rm ph}}}\;
      F_{2i}Q^{2}\, \bigl( 2F_{1i}^{2} - F_{2i}^{2}\, Q^{2} \bigr)\;
      e^{-i\, \mathbf{k_{\rm ph}r_{i}}}\, \sin^{2} \varphi_{ph,i}
  \Bigr\}.
\end{array}
\label{eq.7.1.6c}
\end{equation}

In Ref.~\cite{Maydanyuk_Zhang_Zou.2019.PRC.microscopy} we studied internal structure of proton in the proton-nucleus scattering on the basis of coherent bremsstrahlung analysis only.
But, inclusion of incoherent contribution to the full bremsstrahlung spectrum (without internal structure of proton) improves agreement with experimental data \cite{Maydanyuk_Zhang.2015.PRC}.
So, an important question is to understand role of incoherent emission in study of internal structure of the scattered proton.
By such a motivation, now we introduce the following approximation
\begin{equation}
\begin{array}{llllll}
  \Delta\, h_{0} = 0, &
  \displaystyle\frac{a_{1i}}{f_{i}} = \displaystyle\frac{1}{2}.
\end{array}
\label{eq.8.1.2}
\end{equation}
In frameworks of this approximation, without internal structure of nucleons (i.e., at $F_{1} = 1$, $F_{2} = 0$) we obtain standard formalism with coherent and incoherent bremsstrahlung
(for example, see Ref.~\cite{Maydanyuk_Zhang.2015.PRC}):
\begin{equation}
\begin{array}{llllll}
\vspace{0.1mm}
  \displaystyle\sum\limits_{i=1}^{A+1} \bar{h}_{0} (\vb{r}_{i}) =
  \displaystyle\sum\limits_{i=1}^{A+1}
    \displaystyle\frac{\mathbf{p}_{i}^{2}}{2\,m_{i}} + V(\vb{r}_{1} \ldots \vb{r}_{A+1}), \\

  \displaystyle\sum\limits_{i=1}^{A+1} \bar{h}_{\gamma 0} (\vb{r}_{i}) =
  \displaystyle\sum\limits_{i=1}^{A+1}
  \Bigl[
    - \displaystyle\frac{z_{i}e}{2\,m_{i}c} (-i\hbar\, \vb{div A}_{i} + 2\,\vb{A_{i}p_{i}}) +
    \displaystyle\frac{z_{i}^{2}e^{2}}{2\,m_{i}c^{2}}\, \vb{A}_{i}^{2} -
    \displaystyle\frac{z_{i}e}{2\,m_{i}c}\, \sigmabf \vb{H}_{i}
  \Bigr].
\end{array}
\label{eq.8.1.4}
\end{equation}
But, additional operator (\ref{eq.7.1.6c}) adds internal structure of nucleons to analysis.

\section{Operator of emission with relative coordinates
\label{sec.2}}

\subsection{Coordinates of relative distances
\label{sec.2.4}}

Let us rewrite formalism above via coordinates of relative distances and momenta (in lab frame).
Following to Ref.~\cite{Liu_Maydanyuk_Zhang_Liu.2019.PRC.hypernuclei}
[adapting formalism in Eqs.~(11)--(14) and Appendix~A in that paper for proton-nucleus scattering],
we define coordinates of center-of-mass for the nucleus as $\vb{R}_{A}$, and for the complete system as $\vb{R}$:
\begin{equation}
\begin{array}{lll}
  \vb{R}_{A}  = \displaystyle\frac{1}{m_{A}} \displaystyle\sum_{k=1}^{A} m_{Ak}\, \vb{r}_{A k}, &
  \vb{R}      = \displaystyle\frac{m_{A}\vb{R}_{A} + m_{\rm p}\vb{r}_{\rm p}}{m_{A}+m_{\rm p}} = c_{A}\, \vb{R}_{A} + c_{\rm p}\, \vb{r}_{\rm p},
\end{array}
\label{eq.2.4.1}
\end{equation}
where $m_{\rm p}$ and $m_{A}$ are masses of proton scattered and nucleus-target,
$m_{Ak}$ is mass of nucleon of nucleus with number $k$, and
we introduced new coefficients $c_{A} = \frac{m_{A}}{m_{A}+m_{\rm p}}$ and $c_{\rm p} = \frac{m_{\rm p}}{m_{A} + m_{\rm p}}$.
Introducing new relative coordinate $\vb{r}$,
new relative coordinates $\rhobf_{A j}$ for nucleons for the nucleus
as
\begin{equation}
\begin{array}{lllll}
  \vb{r} = \vb{r}_{\rm p} - \vb{R}_{A}, &
  \rhobf_{Aj} = \vb{r}_{Aj} - \vb{R}_{A}, &
  \rhobf_{AA} = -\, \displaystyle\frac{1}{m_{AA}} \displaystyle\sum_{k=1}^{A-1} m_{Ak}\, \rhobf_{A k},
\end{array}
\label{eq.2.4.2}
\end{equation}
we obtain new independent variables $\vb{R}$, $\vb{r}$ and
$\rhobf_{Aj}$ ($j=1 \ldots A-1$).
%
We rewrite old coordinates $\vb{r}_{Ak}$ of nucleons via new coordinates as
\begin{equation}
\begin{array}{llllll}
  \vb{r}_{\rm p} = \vb{R} + c_{A}\, \vb{r}, &
  \vb{r}_{Aj} = \rhobf_{A j} + \vb{R} - c_{\rm p}\, \vb{r}, &
  \vb{r}_{AA} = \vb{R} - c_{\rm p} \vb{r} - \displaystyle\frac{1}{m_{AA}} \displaystyle\sum_{k=1}^{A-1} m_{Ak}\, \rhobf_{A k}.
\end{array}
\label{eq.2.4.6}
\end{equation}
We calculate momenta $\vu{p}_{\rm p}$, $\vu{p}_{Aj}$, $\vu{p}_{AA}$ corresponding to independent variables $\vb{R}$, $\vb{r}$, $\rhobf_{A j}$.

\subsection{Coherent and incoherent terms without form factors
\label{sec.10.1}}

Using coordinates of relative distances in Eqs.~(\ref{eq.2.4.1})--(\ref{eq.2.4.6}) and corresponding momenta,
we calculate operator of emission of bremsstrahlung photons in scattering of proton off nucleus in the laboratory frame
(see solutions for operators $\hat{H}_{P}$, $\hat{H}_{p}$, $\hat{H}_{k}$ in Eqs.~(16)--(18) in Appendix~B in Ref.~\cite{Liu_Maydanyuk_Zhang_Liu.2019.PRC.hypernuclei},
solutions for $\Delta \hat{H}_{\gamma E}$, $\Delta \hat{H}_{\gamma M}$ in Appendix A in Ref.~\cite{Maydanyuk_Zhang_Zou.2019.brem_alpha_nucleus.arxiv},
we improve new calculations for last matrix element $\hat{H}_{k}$):
%
\begin{equation}
  \hat{H}_{\gamma 0} =
  \displaystyle\sum\limits_{i=1}^{A+1} \bar{h}_{\gamma 0} (\vb{r}_{i}) =
%
  \hat{H}_{P} + \hat{H}_{p} + \Delta \hat{H}_{\gamma E} + \Delta \hat{H}_{\gamma M} + \hat{H}_{k},
\label{eq.10.1.1}
\end{equation}
where
\begin{equation}
\begin{array}{lllll}
\vspace{-0.1mm}
  & \hat{H}_{P} =
  -\, \sqrt{\displaystyle\frac{2\pi c^{2}}{\hbar w_{\rm ph}}}\;
  \mu_{N}\, \displaystyle\frac{2 m_{\rm p}}{m_{A} + m_{p}}\;
  e^{-i\, \vb{k_{\rm ph}} \vb{R}}
  \displaystyle\sum\limits_{\alpha=1,2}
  \biggl\{
    z_{p}\, e^{-i\, c_{A}\, \vb{k_{\rm ph}} \vb{r}} +
    e^{i\, c_{p}\, \vb{k_{\rm ph}} \vb{r} } \displaystyle\sum_{j=1}^{A} z_{j}\, e^{-i\, \vb{k_{\rm ph}} \rhobf_{Aj}}
  \biggr\}\, \vb{e}^{(\alpha)} \cdot \vu{P}\; - \\
  - &
  \sqrt{\displaystyle\frac{2\pi c^{2}}{\hbar w_{\rm ph}}}\;
    \displaystyle\frac{i\, \mu_{N}}{m_{A} + m_{p}}\;
    e^{-i\, \vb{k_{\rm ph}} \vb{R}}\,
    \displaystyle\sum\limits_{\alpha=1,2}
    \biggl\{
      e^{-i\, c_{A}\, \vb{k_{\rm ph}} \vb{r}}\, \mu_{p}^{\rm (an)}\, m_{p}\, \sigmabf +
      e^{i\, c_{p}\, \vb{k_{\rm ph}} \vb{r}}\,
        \displaystyle\sum_{j=1}^{A} \mu_{j}^{\rm (an)}\, m_{Aj}\, e^{-i\, \vb{k_{\rm ph}} \rhobf_{Aj}}\, \sigmabf
  \biggr\}\, \cdot \bigl[ \vu{P} \times \vb{e}^{(\alpha)} \bigr], \\
\end{array}
\label{eq.10.1.2}
\end{equation}
\begin{equation}
\begin{array}{lll}
\vspace{-0.1mm}
  & \hat{H}_{p} =
  -\, \sqrt{\displaystyle\frac{2\pi c^{2}}{\hbar w_{\rm ph}}}\;
  2\, \mu_{N}\,  m_{\rm p}\,
  e^{-i\, \vb{k_{\rm ph}} \vb{R}}
  \displaystyle\sum\limits_{\alpha=1,2}
  \biggl\{
    e^{-i\, c_{A} \vb{k_{\rm ph}} \vb{r}}\, \displaystyle\frac{z_{p}}{m_{p}} -
    e^{i\, c_{p} \vb{k_{\rm ph}} \vb{r}}\,  \displaystyle\frac{1}{m_{A}}\, \displaystyle\sum_{j=1}^{A} z_{j}\, e^{-i\, \vb{k_{\rm ph}} \rhobf_{Aj}}
  \biggr\}\; \vb{e}^{(\alpha)} \cdot \vu{p}\; - \\
  - &
  i\, \sqrt{\displaystyle\frac{2\pi c^{2}}{\hbar w_{\rm ph}}}\: \mu_{N}\,
  e^{-i\, \vb{k_{\rm ph}} \vb{R}}
  \displaystyle\sum\limits_{\alpha=1,2}
  \biggl\{
    e^{-i\, c_{A} \vb{k_{\rm ph}} \vb{r}} \displaystyle\frac{1}{m_{p}} \cdot \mu_{p}^{\rm (an)}\, m_{p}\; \sigmabf -
    e^{i\, c_{p} \vb{k_{\rm ph}} \vb{r}} \displaystyle\frac{1}{m_{A}}
      \displaystyle\sum_{j=1}^{A} \mu_{j}^{\rm (an)}\, m_{Aj}\; e^{-i\, \vb{k_{\rm ph}} \rhobf_{Aj}}\, \sigmabf
  \biggr\} \cdot \bigl[ \vu{p} \times \vb{e}^{(\alpha)} \bigr].
\end{array}
\label{eq.10.1.3}
\end{equation}
\begin{equation}
\begin{array}{lll}
  & \hat{H}_{k} =
  i\, \hbar\,
  \sqrt{\displaystyle\frac{2\pi c^{2}}{\hbar w_{\rm ph}}}\:  \mu_{N}\,
  e^{-i\, \vb{k_{\rm ph}} \vb{R}}\,
  \displaystyle\sum\limits_{\alpha=1,2}
  \biggl\{
    e^{-i\, c_{A}\, \vb{k_{\rm ph}} \vb{r}}\, \mu_{p}^{\rm (an)}\, \sigmabf +
    e^{i\, c_{p}\, \vb{k_{\rm ph}} \vb{r}}\, \displaystyle\sum_{j=1}^{A} \mu_{j}^{\rm (an)}\, e^{-i\, \vb{k_{\rm ph}} \rhobf_{Aj}}\, \sigmabf
  \biggr\}
  \cdot \bigl[ \vb{k_{\rm ph}} \times \vb{e}^{(\alpha)} \bigr], \\
\end{array}
\label{eq.10.1.4}
\end{equation}
\begin{equation}
\begin{array}{lcl}
\vspace{0.4mm}
  \Delta \hat{H}_{\gamma E} & = &
  -\, \sqrt{\displaystyle\frac{2\pi c^{2}}{\hbar w_{\rm ph}}}\:
    2\, \mu_{N}\, e^{-i\, \vb{k_{\rm ph}}\vb{R}}\,
    \displaystyle\sum\limits_{\alpha=1,2} \vb{e}^{(\alpha)}\; \times \\
  & \times &
  \biggl\{
    e^{i\, c_{\rm p}\, \vb{k_{\rm ph}} \vb{r}}\,
    \displaystyle\sum_{j=1}^{A-1}
      \displaystyle\frac{z_{j}\,m_{\rm p}}{m_{Aj}}\, e^{-i\, \vb{k_{\rm ph}} \rhobf_{Aj}}\, \vb{\tilde{p}}_{Aj} -
    \displaystyle\frac{m_{\rm p}}{m_{A}}\, e^{i\, c_{\rm p}\, \vb{k_{\rm ph}} \vb{r}}\,
      \displaystyle\sum_{j=1}^{A} z_{j}\, e^{-i\, \vb{k_{\rm ph}} \rhobf_{Aj}}\, \displaystyle\sum_{k=1}^{A-1} \vb{\tilde{p}}_{Ak}
  \biggr\},
\end{array}
\label{eq.10.1.5}
\end{equation}
\begin{equation}
\begin{array}{lll}
\vspace{0.4mm}
  & \Delta \hat{H}_{\gamma M} =
  -\, i\, \sqrt{\displaystyle\frac{2\pi c^{2}}{\hbar w_{\rm ph}}}\:
   \mu_{N}\:
    e^{-i\, \vb{k_{\rm ph}} \vb{R}}\,
  \displaystyle\sum\limits_{\alpha=1,2} \; \times \\

  \times &
  \biggl\{
    e^{i\, \vb{k_{\rm ph}} c_{\rm p}\, \vb{r}}\,
    \displaystyle\sum_{j=1}^{A-1}
      \mu_{j}^{\rm (an)}\, e^{-i\, \vb{k_{\rm ph}} \rhobf_{Aj}}\, \sigmabf \cdot \bigl[ \vb{\tilde{p}}_{Aj} \times \vb{e}^{(\alpha)} \bigr] -

    e^{i\, \vb{k_{\rm ph}} c_{\rm p}\, \vb{r}}\,
    \displaystyle\sum_{j=1}^{A}
      \mu_{j}^{\rm (an)}\, \displaystyle\frac{m_{Aj}}{m_{A}}\,
      e^{-i\, \vb{k_{\rm ph}} \rhobf_{Aj}}\,
      \displaystyle\sum_{k=1}^{A-1} \sigmabf \cdot \bigl[ \vb{\tilde{p}}_{Ak} \times \vb{e}^{(\alpha)} \bigr]
  \biggr\}.
\end{array}
\label{eq.10.1.6}
\end{equation}
Here, $\mu_{N} = e\hbar / (2m_{\rm p}c)$ is nuclear magneton, 
$m_{i}$ and $z_{i}$ are mass and electric charge of nucleon with number $i$,
$\mu_{j}^{\rm (an)}$ are anomalous magnetic momenta of protons or neutrons of nucleus
[measured in units of nuclear magneton $\mu_{N}$].
$\vb{r}_{p}$, $\vb{R}_{A}$ and $\vb{R}$ are coordinates of the scattered proton, center of nucleus and center of masses of complete system in laboratory frame,
$\vb{\hat{P}}$, $\vb{\hat{p}}$ and $\vb{\tilde{p}}_{Aj}$ are momenta corresponding to $\vb{R}$, $\vb{r}$, $\rhobf_{Aj}$
(defined as $\vu{p}_{i} = -i\hbar\, \vb{d}/\vb{dr}_{i}$).
$\vb{e}^{(1)}$ and $\vb{e}^{(2)}$ are unit vectors of polarizations for photon with momentum $\vb{k}_{\rm ph}$
(
$\vb{e}^{(\alpha), *} = \vb{e}^{(\alpha)}$),
$\vb{k}_{\rm ph}$ is wave vector of the photon and $w_{\rm ph} = k_{\rm ph} c = \bigl| \vb{k}_{\rm ph}\bigr|c$.
$\vb{e}^{(\alpha)}$ are perpendicular to $\vb{k}_{\rm ph}$ in Coulomb gauge, satisfy Eq.~(8) in Ref.~\cite{Liu_Maydanyuk_Zhang_Liu.2019.PRC.hypernuclei}.
$\sigmabf$ are Pauli matrixes,
$A$ in summation is mass number of the nucleus-target.

We define the wave function of the full nuclear system as
$\Psi = \Phi (\vb{R}) \cdot \Phi_{\rm p - nucl} (\vb{r}) \cdot \psi_{\rm nucl} (\beta_{A}) \cdot \psi_{\rm p}$
following the formalism in Ref.~\cite{Maydanyuk_Zhang.2015.PRC}
(see Sect.~II.B, Eqs.~(10)--(13), we add new wave function $\psi_{\rm p}$ of the scattered proton).
Here,
$\beta_{A}$ is the set of numbers $1 \cdots A$ of nucleons of the nucleus,
$\Phi (\vb{R})$ is the function describing motion of center-of-mass of the full nuclear system in laboratory frame,
$\Phi_{\rm p - nucl} (\vb{r})$ is the function describing relative motion of the scattered proton concerning to nucleus (without description of internal relative motions of nucleons in the nucleus),
$\psi_{\rm p} (\beta_{p})$ is the wave function of the scattered proton,
$\psi_{\rm nucl} (\beta_{A})$ is the many-nucleon function of the nucleus,
defined in Eq.~(12) Ref.~\cite{Maydanyuk_Zhang.2015.PRC} on the basis of one-nucleon functions $\psi_{\lambda_{s}}(s)$.
One-nucleon functions $\psi_{\lambda_{s}}(s)$ represent the multiplication of space and spin-isospin
functions as $\psi_{\lambda_{s}} (s) = \varphi_{n_{s}} (\vb{r}_{s})\, \bigl|\, \sigma^{(s)} \tau^{(s)} \bigr\rangle$,
where
$\varphi_{n_{s}}$ is the space function of the nucleon with number $s$,
$n_{s}$ is the number of state of the space function of the nucleon with number $s$,
$\bigl|\, \sigma^{(s)} \tau^{(s)} \bigr\rangle$ is the spin-isospin function of the nucleon with number $s$.
We neglect corrections by full anti-symmetrization of all nucleons in the proton-nucleus wave function,
as this requires more complicated calculations.

\subsection{Terms with form factors
\label{sec.11}}

Now we consider operators of emission with included form factors of nucleons.
From Eq.~(\ref{eq.7.1.6c}) we write
\vspace{0.5mm}
\begin{equation}
\begin{array}{llllll}
\vspace{1.0mm}
  \displaystyle\sum\limits_{i=1}^{A+1} \bar{h}_{\gamma 1} (\vb{r}_{i}) =
  \displaystyle\sum\limits_{i=1}^{A+1}
    S_{1i}\, e^{-i\, \mathbf{k_{\rm ph}r_{i}}} +
  \displaystyle\sum\limits_{i=1}^{A+1}
    S_{2i}\, e^{-2i\, \mathbf{k_{\rm ph}r_{i}}},
\end{array}
\label{eq.11.1.2}
\end{equation}
where
\vspace{0.5mm}
\begin{equation}
\begin{array}{llllll}
\vspace{1.0mm}
  S_{1i} =
    \displaystyle\frac{z_{i}e\, F_{2i}}{f_{i}} \cdot
  \sqrt{\displaystyle\frac{\pi\hbar c^{2}}{w_{\rm ph}}}\;
  \Bigl\{
    2 Q\, \sin \varphi_{ph,i}\, \Bigl[ -\, i\, b_{1i} + \displaystyle\frac{F_{2i}\, Q^{2}}{m_{i}c}\, \bigl( 2F_{1i}^{2} - F_{2i}^{2}\, Q^{2} \bigr) \Bigr] -
    i\, \sqrt{2}\, b_{2i}\, \varepsilon_{mjl}\, q^{j}\, \sigma_{l} \sum\limits_{\alpha=1,2} \mathbf{e}_{m}^{(\alpha),\,*}
  \Bigr\}, \\

  S_{2i} =
    \displaystyle\frac{4z_{i}^{2}e^{2}\, F_{2i}}{f_{i}\,m_{i}} \cdot
    \displaystyle\frac{\pi\hbar}{w_{\rm ph}} \cdot
    F_{2i}Q^{2}\, \bigl( 2F_{1i}^{2} - F_{2i}^{2}\, Q^{2} \bigr)\;
    \sin^{2} \varphi_{ph,i}.
\end{array}
\label{eq.11.1.3}
\end{equation}
%
%
%
We transform formalism into relative distances, and using Eqs.~(\ref{eq.2.4.6}), we obtain:
\begin{equation}
\begin{array}{llllll}
  \hat{H}_{\gamma 1} =
  \displaystyle\sum\limits_{i=1}^{A+1} \bar{h}_{\gamma 1} (\vb{r}_{i}) =
  \bar{h}_{\gamma 1}^{\rm (proton)} (\vb{r}_{i}) + \bar{h}_{\gamma 1}^{\rm (nucleus)} (\vb{r}_{i}),
\end{array}
\label{eq.11.1.6}
\end{equation}
where
\vspace{0.5mm}
\begin{equation}
\begin{array}{llllll}
\vspace{1.5mm}
  \bar{h}_{\gamma 1}^{\rm (proton)} (\vb{r}_{i}) =
  e^{-i\, \vb{k}_{\rm ph}\, \vb{R}}\, e^{-i\, \vb{k}_{\rm ph}\, c_{A}\, \vb{r}}\, S_{1p} +
  e^{-2i\, \vb{k}_{\rm ph}\, \vb{R}}\, e^{-2i\, \vb{k}_{\rm ph}\, c_{A}\, \vb{r}}\, S_{2p}, \\

  \bar{h}_{\gamma 1}^{\rm (nucleus)} (\vb{r}_{i}) =
  e^{-i\, \vb{k}_{\rm ph}\, \vb{R}}
    e^{i\, \vb{k}_{\rm ph} c_{\rm p}\, \vb{r}} \displaystyle\sum\limits_{j=1}^{A} S_{1j}\, e^{-i\, \vb{k}_{\rm ph} \rhobf_{A j}} +
  e^{-2i\, \vb{k}_{\rm ph}\, \vb{R}}
    e^{2i\, \vb{k}_{\rm ph} c_{\rm p}\, \vb{r}} \displaystyle\sum\limits_{j=1}^{A} S_{2j}\, e^{-2i\, \vb{k}_{\rm ph} \rhobf_{A j}}.
\end{array}
\label{eq.11.1.7}
\end{equation}
In particular, the first term was partially studied in previous publication~\cite{Maydanyuk_Zhang_Zou.2019.PRC.microscopy} (without incoherent emission of other matrix elements).
The second term is new one and it describes influence of internal structure of nucleons inside nucleus-target on the full emission of bremsstrahlung photons.

\section{Matrix elements of emission of bremsstrahlung photons
\label{sec.13}}

We define the matrix element of emission, using the wave functions $\Psi_{i}$ and $\Psi_{f}$ of the full nuclear system in states before emission of photons ($i$-state) and after such emission ($f$-state),
as
\begin{equation}
  F = \langle \Psi_{f} |\, \hat{H}_{\gamma} |\, \Psi_{i} \rangle.
\label{eq.13.1.1}
\end{equation}
In this matrix element we integrate over all independent variables i.e.
space variables $\vb{R}$, $\vb{r}$, $\rhobf_{Am}$.
We should take into account space representation of momenta $\vu{P}$, $\vu{p}$, $\vb{\tilde{p}}_{A m}$
(as
$\vu{P} = -i\hbar\, \vb{d/dR}$,
$\vu{p} = -i\hbar\, \vb{d/dr}$,
$\vb{\tilde{p}}_{A m} = -i\hbar\, \vb{d/d} \rhobf_{Am}$).
Using formulas (\ref{eq.10.1.1})--(\ref{eq.10.1.6}) 
and (\ref{eq.11.1.6}) 
for operator of emission, we calculate
(see Eqs.~(23), (34) and Appendixes B--D in Ref.~\cite{Maydanyuk_Zhang_Zou.2019.brem_alpha_nucleus.arxiv} adapting calculations for proton-nucleus scattering):
%
\begin{equation}
  \langle \Psi_{f} |\, \hat{H}_{\gamma} |\, \Psi_{i} \rangle \;\; = \;\;
  \sqrt{\displaystyle\frac{2\pi\, c^{2}}{\hbar w_{\rm ph}}}\,
  \Bigl\{ M_{P} + M_{p}^{(E)} + M_{p}^{(M)} + M_{k} + M_{\Delta E} + M_{\Delta M} + M_{\rm form factor} \Bigr\},
\label{eq.13.1.2}
\end{equation}
where
\begin{equation}
\begin{array}{lll}
\vspace{-0.2mm}
  M_{p}^{(E)} & = &
  i \hbar\, (2\pi)^{3} \displaystyle\frac{\mu_{N}}{\mu}\;
  \displaystyle\sum\limits_{\alpha=1,2}
  \displaystyle\int\limits_{}^{}
    \Phi_{\rm p - nucl, f}^{*} (\vb{r})\;
    e^{-i\, \vb{k}_{\rm ph} \vb{r}} \cdot
    2\, m_{\rm p} \cdot Z_{\rm eff} (\vb{k}_{\rm ph}, \vb{r}) \cdot \vb{e}^{(\alpha)}\, \vb{\displaystyle\frac{d}{dr}} \cdot
    \Phi_{\rm p - nucl, i} (\vb{r})\; \vb{dr}, \\

  M_{p}^{(M)} & = &
  i \hbar\, (2\pi)^{3} \displaystyle\frac{\mu_{N}}{\mu}\;
  \displaystyle\sum\limits_{\alpha=1,2}
  \displaystyle\int\limits_{}^{}
    \Phi_{\rm p - nucl, f}^{*} (\vb{r})\;
    e^{-i\, \vb{k}_{\rm ph} \vb{r}} \cdot
    i\, \vb{M}_{\rm eff} (\vb{k}_{\rm ph}, \vb{r}) \cdot \Bigl[ \vb{\displaystyle\frac{d}{dr}} \times \vb{e}^{(\alpha)} \Bigr] \cdot
    \Phi_{\rm p - nucl, i} (\vb{r})\; \vb{dr},
\end{array}
\label{eq.13.1.3}
\end{equation}
\begin{equation}
\begin{array}{lllll}
\vspace{-0.1mm}
  M_{P} & = &
  \displaystyle\frac{\hbar\, (2\pi)^{3}}{m_{A} + m_{p}}\, \mu_{N}\,
  \displaystyle\sum\limits_{\alpha=1,2}
  \displaystyle\int\limits_{}^{}
    \Phi_{\rm p - nucl, f}^{*} (\vb{r})\;
  \biggl\{
    2\, m_{\rm p}\;
    \Bigl[
      e^{-i\, c_{A}\, \vb{k_{\rm ph}} \vb{r}} F_{p,\, {\rm el}} + e^{i\, c_{p}\, \vb{k_{\rm ph}} \vb{r}} F_{A,\, {\rm el}}
    \Bigr]\, \vb{e}^{(\alpha)} \cdot \vb{K}_{i}\; + \\

  & + &
    i\: \Bigl[
      e^{-i\, c_{A}\, \vb{k_{\rm ph}} \vb{r}}\, \vb{F}_{p,\, {\rm mag}} + e^{i\, c_{p}\, \vb{k_{\rm ph}} \vb{r}}\, \vb{F}_{A,\, {\rm mag}}
    \Bigr] \cdot
    \bigl[ \vb{K}_{i} \cp \vb{e}^{(\alpha)} \bigr]
  \biggr\} \cdot
  \Phi_{\rm p - nucl, i} (\vb{r})\; \vb{dr},
\end{array}
\label{eq.13.1.4}
\end{equation}
\begin{equation}
\begin{array}{lcl}
  M_{k} & = &
  i\, \hbar\, (2\pi)^{3}  \mu_{N}\,
  \displaystyle\sum\limits_{\alpha=1,2}
    \bigl[ \vb{k_{\rm ph}} \cp \vb{e}^{(\alpha)} \bigr]
  \displaystyle\int\limits_{}^{}
    \Phi_{\rm p - nucl, f}^{*} (\vb{r}) \cdot
    \Bigl\{ e^{-i\, c_{A}\, \vb{k_{\rm ph}} \vb{r}}\, \vb{D}_{p,\, {\rm k}} + e^{i\, c_{p}\, \vb{k_{\rm ph}} \vb{r}}\, \vb{D}_{A,\, {\rm k}} \Bigr\} \cdot
    \Phi_{\rm p - nucl, i} (\vb{r})\; \vb{dr},
\end{array}
\label{eq.13.1.5}
\end{equation}
\begin{equation}
\begin{array}{lll}
\vspace{-0.1mm}
  M_{\Delta E} & = &
  -\, (2\pi)^{3}\, 2\, \mu_{N}
  \displaystyle\sum\limits_{\alpha=1,2} \vb{e}^{(\alpha)}
  \displaystyle\int\limits_{}^{}
    \Phi_{\rm p - nucl, f}^{*} (\vb{r})\;
  \biggl\{
    \Bigl[ e^{-i\, c_{A}\, \vb{k_{\rm ph}} \vb{r}}\, \vb{D}_{p 1,\, {\rm el}} + e^{i\, c_{p}\, \vb{k_{\rm ph}} \vb{r}}\, \vb{D}_{A 1,\, {\rm el}} \Bigr]\; - \\

  &- &
    \Bigl[ e^{-i\, c_{A}\, \vb{k_{\rm ph}} \vb{r}}\, \vb{D}_{p 2,\, {\rm el}} +
    \displaystyle\frac{m_{\rm p}}{m_{A}}\, e^{i\, c_{p}\, \vb{k_{\rm ph}} \vb{r}}\, \vb{D}_{A 2,\, {\rm el}} \Bigr]
  \biggr\} \cdot
  \Phi_{\rm p - nucl, i} (\vb{r})\; \vb{dr},
\end{array}
\label{eq.13.1.6}
\end{equation}
\begin{equation}
\begin{array}{lll}
\vspace{-0.1mm}
  M_{\Delta M} & = &
  -\, i\, (2\pi)^{3}\,  \mu_{N}\,
  \displaystyle\sum\limits_{\alpha=1,2}
  \displaystyle\int\limits_{}^{}
    \Phi_{\rm p - nucl, f}^{*} (\vb{r})\;
  \biggl\{
    \Bigl[ e^{-i\, c_{A}\, \vb{k_{\rm ph}} \vb{r}}\; D_{p 1,\, {\rm mag}} (\vb{e}^{(\alpha)}) + e^{i\, c_{p}\, \vb{k_{\rm ph}} \vb{r}}\; D_{A 1,\, {\rm mag}} (\vb{e}^{(\alpha)}) \Bigr]\; - \\

  & - &
    \Bigl[ e^{-i\, c_{A}\, \vb{k_{\rm ph}} \vb{r}}\; D_{p 2,\, {\rm mag}} (\vb{e}^{(\alpha)}) + e^{i\, c_{p}\, \vb{k_{\rm ph}} \vb{r}}\; D_{A 2,\, {\rm mag}} (\vb{e}^{(\alpha)}) \Bigr]
  \biggr\} \cdot
  \Phi_{\rm p - nucl, i} (\vb{r})\; \vb{dr},
\end{array}
\label{eq.13.1.7}
\end{equation}
and $\vb{K}_{i} = \vb{K}_{f} + \vb{k}$.
Here, $\mu = m_{\rm p} m_{A} / (m_{\rm p} + m_{A})$ is reduced mass and
the effective electric charge and magnetic momentum are
\begin{equation}
\begin{array}{lll}
\vspace{1.0mm}
  Z_{\rm eff} (\vb{k}_{\rm ph}, \vb{r}) =
  e^{i\, \vb{k_{\rm ph}} \vb{r}}\,
  \Bigl[
    e^{-i\, c_{A} \vb{k_{\rm ph}} \vb{r}}\, \displaystyle\frac{m_{A}}{m_{p} + m_{A}}\, F_{p,\, {\rm el}} -
    e^{i\, c_{p} \vb{k_{\rm ph}} \vb{r}}\, \displaystyle\frac{m_{p}}{m_{p} + m_{A}}\, F_{A,\, {\rm el}}
  \Bigr], \\

  \vb{M}_{\rm eff} (\vb{k}_{\rm ph}, \vb{r}) =
  e^{i\, \vb{k_{\rm ph}} \vb{r}}\,
  \Bigl[
    e^{-i\, c_{A} \vb{k_{\rm ph}} \vb{r}}\,  \displaystyle\frac{m_{A}}{m_{p} + m_{A}}\, \vb{F}_{p,\, {\rm mag}} -
    e^{i\, c_{p} \vb{k_{\rm ph}} \vb{r}}\,  \displaystyle\frac{m_{p}}{m_{p} + m_{A}}\, \vb{F}_{A,\, {\rm mag}}
  \Bigr],
\end{array}
\label{eq.13.1.8}
\end{equation}
Here,
$\vb{D}_{A,\, {\rm k}}$,
$\vb{D}_{A 1,\, {\rm el}}$, $\vb{D}_{A 2,\, {\rm el}}$,
$D_{A 1,\, {\rm mag}}$, $D_{A 2,\, {\rm mag}}$,
$F_{K,\, {\rm el}}$,
$F_{A,\, {\rm el}}$,
$\vb{F}_{A,\, {\rm mag}}$
are electric and magnetic form factors defined in
Appendix~\ref{sec.app.form_factors} [see Eqs.~(\ref{eq.app.13.1.9.a})--(\ref{eq.app.13.1.9.d})].
Taking zero values of $\vb{D}_{p 1,\, {\rm el}}$, $\vb{D}_{p 2,\, {\rm el}}$, $D_{p 1,\, {\rm mag}} (\vb{e}^{(\alpha)})$, $D_{p 2,\, {\rm mag}} (\vb{e}^{(\alpha)})$ into account
in Eqs.~(\ref{eq.app.13.1.9.b}) and (\ref{eq.app.13.1.9.c}),
we simplify terms (\ref{eq.13.1.6})--(\ref{eq.13.1.7}) for the incoherent emission as
\begin{equation}
\begin{array}{lll}
  M_{\Delta E} & = &
  -\, (2\pi)^{3}\, 2\, \mu_{N}
  \displaystyle\sum\limits_{\alpha=1,2} \vb{e}^{(\alpha)}
  \displaystyle\int\limits_{}^{}
    \Phi_{\rm p - nucl, f}^{*} (\vb{r})\;
  \biggl\{
    e^{i\, c_{p}\, \vb{k_{\rm ph}} \vb{r}}\, \vb{D}_{A 1,\, {\rm el}} -
    \displaystyle\frac{m_{\rm p}}{m_{A}}\, e^{i\, c_{p}\, \vb{k_{\rm ph}} \vb{r}}\, \vb{D}_{A 2,\, {\rm el}}
  \biggr\} \cdot
  \Phi_{\rm p - nucl, i} (\vb{r})\; \vb{dr},
\end{array}
\label{eq.13.1.10}
\end{equation}
\begin{equation}
\begin{array}{lll}
  M_{\Delta M} & = &
  -\, i\, (2\pi)^{3}\,  \mu_{N}\,
  \displaystyle\sum\limits_{\alpha=1,2}
  \displaystyle\int\limits_{}^{}
    \Phi_{\rm p - nucl, f}^{*} (\vb{r})\;
  \biggl\{
    e^{i\, c_{p}\, \vb{k_{\rm ph}} \vb{r}}\; D_{A 1,\, {\rm mag}} (\vb{e}^{(\alpha)}) -
    e^{i\, c_{p}\, \vb{k_{\rm ph}} \vb{r}}\; D_{A 2,\, {\rm mag}} (\vb{e}^{(\alpha)})
  \biggr\} \cdot
  \Phi_{\rm p - nucl, i} (\vb{r})\; \vb{dr}.
\end{array}
\label{eq.13.1.11}
\end{equation}
%

\subsection{Matrix elements of bremsstrahlung without form factors of the scattered proton
\label{sec.13}}

The matrix elements of coherent and incoherent bremsstrahlung without contribution form internal structure of the scattered proton are calculated in Appendix~\ref{sec.13.2}.
We obtain for the coherent bremsstrahlung (in the dipole approximation)
\begin{equation}
\begin{array}{lll}
\vspace{1.5mm}
  M_{p}^{(E,\, {\rm dip})} =
  i \hbar\, (2\pi)^{3}
  \displaystyle\frac{2\, \mu_{N}\,  m_{\rm p}}{\mu}\;
  Z_{\rm eff}^{\rm (dip)}\;
  \displaystyle\sum\limits_{\alpha=1,2} \vb{e}^{(\alpha)} \cdot \vb{I}_{1}, \\

  M_{p}^{(M,\, {\rm dip})} =
  \hbar\, (2\pi)^{3}\, \displaystyle\frac{\mu_{N}}{\mu} \cdot \alpha_{M} \cdot
  (\vb{e}_{\rm x} + \vb{e}_{\rm z})\,
  \displaystyle\sum\limits_{\alpha=1,2}
  \Bigl[ \vb{I}_{1} \times \vb{e}^{(\alpha)} \Bigr],
\end{array}
\label{eq.resultingformulas.1}
\end{equation}
and for incoherent bremsstrahlung
\begin{equation}
\begin{array}{lll}
\vspace{1.4mm}
  M_{\Delta E} = 0, \\
\vspace{1.4mm}
  M_{\Delta M} = i\, \hbar\, (2\pi)^{3}\, \mu_{N}\, f_{1} \cdot |\vb{k}_{\rm ph}| \cdot Z_{\rm A} (\vb{k}_{\rm ph}) \cdot I_{2}, \\

  M_{k} =
    -\, i\, \hbar\, (2\pi)^{3}\, \mu_{N} \cdot k_{\rm ph}\, z_{\rm p}\: \mu_{\rm p}^{\rm (an)} \cdot I_{3} -
    \displaystyle\frac{\bar{\mu}_{\rm pn}^{\rm (an)}}{f_{1}} \cdot M_{\Delta M}.
\end{array}
\label{eq.resultingformulas.2}
\end{equation}
Integrals are
\begin{equation}
\begin{array}{lllll}
\vspace{0.5mm}
  \vb{I}_{1} = \biggl\langle\: \Phi_{\rm p - nucl, f} (\vb{r})\; \biggl|\, e^{-i\, \vb{k}_{\rm ph} \vb{r}}\; \vb{\displaystyle\frac{d}{dr}} \biggr|\: \Phi_{\rm p - nucl, i} (\vb{r})\: \biggr\rangle, \\
\vspace{0.5mm}
  I_{2} = \Bigl\langle \Phi_{\rm p - nucl, f} (\vb{r})\; \Bigl|\, e^{i\, c_{p}\, \vb{k_{\rm ph}} \vb{r}}\, \Bigr|\, \Phi_{\rm p - nucl, i} (\vb{r})\: \Bigr\rangle, \\
  I_{3} = \Bigl\langle \Phi_{\rm p - nucl, f} (\vb{r})\; \Bigl|\, e^{-i\, c_{A}\, \vb{k_{\rm ph}} \vb{r}}\, \Bigr|\, \Phi_{\rm p - nucl, i} (\vb{r})\: \Bigr\rangle.
\end{array}
\label{eq.resultingformulas.3}
\end{equation}
In the dipole approximation, the effective electric charge and magnetic momentum (\ref{eq.13.1.8}) are
\begin{equation}
\begin{array}{lll}
  Z_{\rm eff}^{\rm (dip)} = \displaystyle\frac{m_{A}\, F_{p,\, {\rm el}} - m_{p}\, F_{A,\, {\rm el}} }{m_{p} + m_{A}}, &
  \vb{M}_{\rm eff}^{\rm (dip)} = \displaystyle\frac{m_{A}\, \vb{F}_{p,\, {\rm mag}}  - m_{p}\,\vb{F}_{A,\, {\rm mag}}}{m_{p} + m_{A}},
\end{array}
\label{eq.resultingformulas.4}
\end{equation}
\begin{equation}
\begin{array}{lll}
  \alpha_{M} =
    \Bigl[ Z_{\rm A} (\vb{k}_{\rm ph})\: m_{p}\, \bar{\mu}_{\rm pn}^{\rm (an)} - z_{\rm p}\, m_{A}\, \mu_{\rm p}^{\rm (an)} \Bigr] \cdot \displaystyle\frac{m_{p}}{m_{p} + m_{A}}, &
  f_{1} = \displaystyle\frac{A-1}{2A}\: \bar{\mu}_{\rm pn}^{\rm (an)}.
\end{array}
\label{eq.resultingformulas.5}
\end{equation}
Also we have
\begin{equation}
\begin{array}{lll}
  \displaystyle\frac{\bar{\mu}_{\rm pn}^{\rm (an)}}{f_{1}} =
  \displaystyle\frac{\bar{\mu}_{\rm pn}^{\rm (an)}}{\displaystyle\frac{A-1}{2A}\: \bar{\mu}_{\rm pn}^{\rm (an)}} =
  \displaystyle\frac{2\, A}{A-1}.
\end{array}
\label{eq.resultingformulas.6}
\end{equation}
Here,
$\bar{\mu}_{\rm pn}^{\rm (an)} = (\mu_{\rm p}^{\rm (an)} + \kappa\,\mu_{\rm n}^{\rm (an)})$,
$\kappa = (A-N)/N$, $A$ and $N$ are numbers of nucleons and neutrons in nucleus,
$\mu_{\rm p}^{\rm (an)}$ and $\mu_{\rm n}^{\rm (an)}$ are anomalous magnetic moments of proton and neutron.

Integral $\vb{I}_{1}$ is calculated in Appendix~\ref{sec.app.integrals} [see Eq.~(\ref{eq.app.integrals.4})].
According to Eqs.~(\ref{eq.app.integrals.2.1}), we obtain:
\begin{equation}
\begin{array}{ll}
  \displaystyle\sum\limits_{\alpha=1,2} \vb{e}^{(\alpha)} \cdot \vb{I}_{1} =
  \sqrt{\displaystyle\frac{\pi}{2}}\:
  \displaystyle\sum\limits_{l_{\rm ph}=1}\, (-i)^{l_{\rm ph}}\, \sqrt{2l_{\rm ph}+1}\;
  \displaystyle\sum\limits_{\mu=\pm 1} \mu\,h_{\mu}\, \bigl(p_{l_{\rm ph}, \mu}^{M} + p_{l_{\rm ph}, -\mu}^{E} \bigr), \\

  (\vb{e}_{\rm x} + \vb{e}_{\rm z})\,  \displaystyle\sum\limits_{\alpha=1,2} \Bigl[ \vb{I}_{1} \times \vb{e}^{(\alpha)} \Bigr] =
  \sqrt{\displaystyle\frac{\pi}{2}}\:
  \displaystyle\sum\limits_{l_{\rm ph}=1}\, (-i)^{l_{\rm ph}}\, \sqrt{2l_{\rm ph}+1}\;
  \displaystyle\sum\limits_{\mu=\pm 1} \mu\,h_{\mu}\, \bigl(p_{l_{\rm ph}, \mu}^{M} - p_{l_{\rm ph}, -\mu}^{E} \bigr).
\end{array}
\label{eq.resultingformulas.7}
\end{equation}
and matrix elements are transformed to
\begin{equation}
\begin{array}{lll}
\vspace{1.5mm}
  M_{p}^{(E,\, {\rm dip})} =
  i \hbar\, (2\pi)^{3}
  \displaystyle\frac{2\, \mu_{N}\,  m_{\rm p}}{\mu}\;
  Z_{\rm eff}^{\rm (dip)}\;
  \sqrt{\displaystyle\frac{\pi}{2}}\:
    \displaystyle\sum\limits_{l_{\rm ph}=1}\, (-i)^{l_{\rm ph}}\, \sqrt{2l_{\rm ph}+1}\;
    \displaystyle\sum\limits_{\mu=\pm 1} \mu\,h_{\mu}\, \bigl(p_{l_{\rm ph}, \mu}^{M} + p_{l_{\rm ph}, -\mu}^{E} \bigr), \\

  M_{p}^{(M,\, {\rm dip})} =
  \hbar\, (2\pi)^{3}\, \displaystyle\frac{\mu_{N}}{\mu} \cdot \alpha_{M} \cdot
  \sqrt{\displaystyle\frac{\pi}{2}}\:
    \displaystyle\sum\limits_{l_{\rm ph}=1}\, (-i)^{l_{\rm ph}}\, \sqrt{2l_{\rm ph}+1}\;
    \displaystyle\sum\limits_{\mu=\pm 1} \mu\,h_{\mu}\, \bigl(p_{l_{\rm ph}, \mu}^{M} - p_{l_{\rm ph}, -\mu}^{E} \bigr).
\end{array}
\label{eq.resultingformulas.8}
\end{equation}

\subsection{Matrix element of emission of bremsstrahlung photons with form factors
\label{sec.17}}

Now we define the matrix element of emission, related with form factors.
%
In Sec.~\ref{sec.11} we found operator of emission with form factors.
According to Eqs.~(\ref{eq.11.1.6}), (\ref{eq.11.1.7}), (\ref{eq.11.1.3}), 
we have
\begin{equation}
\begin{array}{llllll}
  \hat{H}_{\gamma 1} =
  \displaystyle\sum\limits_{i=1}^{A+1} \bar{h}_{\gamma 1} (\vb{r}_{i}) =
  \bar{h}_{\gamma 1}^{\rm (proton)} (\vb{r}_{i}) + \bar{h}_{\gamma 1}^{\rm (nucleus)} (\vb{r}_{i}),
\end{array}
\label{eq.17.1.1}
\end{equation}
From Eqs.~(\ref{eq.17.2.5}), (\ref{eq.17.3.2}), (\ref{eq.17.5.4})
we obtain:
\begin{equation}
\begin{array}{lll}
  M_{\rm form factor} = M_{\rm form factor}^{(1)} + M_{\rm form factor}^{(2)},
\end{array}
\label{eq.17.resultingformulas.1}
\end{equation}
\begin{equation}
\begin{array}{lll}
  M_{\rm form factor}^{(1)} & = &
  (2\pi)^{3}\: \displaystyle\frac{\hbar\, e}{\sqrt{2}}\;
  \Bigl\{
    \displaystyle\frac{z_{p}\, F_{2p}}{f_{p}}\: \bigl( p_{\rm q,1}^{\rm (p)} + p_{\rm q,2}^{\rm (p)} \bigr) +
  \displaystyle\sum\limits_{i=1}^{A}
    \displaystyle\frac{z_{i}\, F_{2i}}{f_{i}}\:
    \bigl( p_{\rm q,1}^{\rm (Ai)} + p_{\rm q,2}^{\rm (Ai)} \bigr)
  \Bigr\}, \\
  M_{\rm form factor}^{(2)} & = &
  (2\pi)^{3}\; \sqrt{\displaystyle\frac{\hbar w_{\rm ph}}{2\pi\, c^{2}}}\,
  \displaystyle\frac{\pi\hbar}{w_{\rm ph}}\, 4e^{2}\,
  \Bigl\{
    \displaystyle\frac{z_{p}^{2}\, F_{2p}}{f_{p}\,m_{p}}\:
    p_{q,3}^{\rm (p)} +
    \displaystyle\sum\limits_{j=1}^{A}
    \displaystyle\frac{z_{i}^{2}\, F_{2i}}{f_{i}\,m_{i}}\,
    p_{q,3}^{\rm (Ai)}
  \Bigr\},
\end{array}
\label{eq.17.resultingformulas.2}
\end{equation}
\begin{equation}
\begin{array}{lcl}
\vspace{1.3mm}
  \tilde{p}_{\rm q,1}^{\rm (p)} & = &
    A_{1}^{\rm (p)} (Q, F_{1}, F_{2})\, I_{3} (\vb{k}_{\rm ph}) +
    B_{1}^{\rm (p)} (Q, F_{1}, F_{2})\,  I_{4} (\vb{k}_{\rm ph}), \\

\vspace{1.3mm}
  \tilde{p}_{\rm q,2}^{\rm (p)} & = &
    A_{2}^{\rm (p)} (Q, F_{1}, F_{2})\, I_{3} (\vb{k}_{\rm ph}) +
    B_{2}^{\rm (p)} (Q, F_{1}, F_{2})\,  I_{4} (\vb{k}_{\rm ph}), \\

  \tilde{p}_{\rm q,3}^{\rm (p)} & = &
    A_{3}^{\rm (p)} (Q, F_{1}, F_{2})\,
    I_{3} (2\, \vb{k}_{\rm ph}), \\

\vspace{1.3mm}
  \tilde{p}_{\rm q,1,0}^{\rm (Ai)} & = &
    \Bigl( A_{1}^{\rm (Ai)} (Q, F_{1}, F_{2}) + B_{1}^{\rm (Ai)} (Q, F_{1}, F_{2}) \Bigr)\,
    I_{2} (\vb{k}_{\rm ph}), \\

\vspace{1.3mm}
  \tilde{p}_{\rm q,2,0}^{\rm (Ai)} & = &
    \Bigl( A_{2}^{\rm (Ai)} (Q, F_{1}, F_{2}) + B_{2}^{\rm (Ai)} (Q, F_{1}, F_{2}) \Bigr)\,
      I_{2} (\vb{k}_{\rm ph}), \\

  \tilde{p}_{\rm q,3,0}^{\rm (Ai)} & = &
    A_{3}^{\rm (Ai)} (Q, F_{1}, F_{2})\,
    I_{2} (2\,\vb{k}_{\rm ph}),
\end{array}
\label{eq.17.resultingformulas.3}
\end{equation}
\begin{equation}
\begin{array}{lcl}
\vspace{1.5mm}
  A_{1}^{\rm (p)} (Q, F_{1p}, F_{2p}) & = &
    \displaystyle\frac{4Q}{\pi}
    \Bigl\{ -i\, \Bigl[ F_{1p}^{2} (1 - F_{1p}) - F_{1p} F_{2p}^{2}\, Q^{2} \Bigr] +
       \displaystyle\frac{F_{2p}\, Q^{2}}{m_{\rm p}c}\, \bigl( 2F_{1p}^{2} - F_{2p}^{2}\, Q^{2} \bigr) \Bigr\}, \\
\vspace{1.5mm}
  B_{1}^{\rm (p)} (Q, F_{1p}, F_{2p}) & = & i\,8 Q\, \displaystyle\frac{F_{1p}^{3}}{\pi m_{\rm p}c^{2}}, \\

\vspace{1.5mm}
  A_{2}^{\rm (p)} (Q, F_{1p}, F_{2p}) & = &
    2\, \bigl( F_{1p}^{2} - F_{2p}^{2}\, Q^{2} \bigr)\,
    \sqrt{2}\, \varepsilon_{mjl}\, q^{j}\, \sigma_{l} \displaystyle\sum\limits_{\alpha=1,2} \vb{e}_{m}^{(\alpha)}, \\

\vspace{0.9mm}
  B_{2}^{\rm (p)} (Q, F_{1p}, F_{2p}) & = &
    -\, 2\, \bigl( F_{1p}^{2} - F_{2p}^{2} Q^{2} \bigr)\,
    \displaystyle\frac{\sqrt{2}\, F_{1p}}{m_{\rm p}c^{2}}\,
    \varepsilon_{mjl}\, q^{j}\, \sigma_{l} \displaystyle\sum\limits_{\alpha=1,2} \vb{e}_{m}^{(\alpha)}, \\

  A_{3}^{\rm (p)} (Q, F_{1p}, F_{2p}) & = &
    \displaystyle\frac{F_{2p}Q^{2}}{2}\, \bigl( 2F_{1p}^{2} - F_{2p}^{2}\, Q^{2} \bigr),
\end{array}
\label{eq.17.resultingformulas.4}
\end{equation}
and solutions for
$A_{1}^{\rm (Ai)}$, $B_{1}^{\rm (Ai)}$, $A_{2}^{\rm (Ai)}$, $B_{2}^{\rm (Ai)}$, $A_{3}^{\rm (Ai)}$
are obtained from Eqs.~(\ref{eq.17.resultingformulas.4}) by changing bottom indexes $p \to i$.
We have introduced a new integral
\begin{equation}
\begin{array}{lllll}
  I_{4} (\vb{k}_{\rm ph}) =
    \Bigl\langle \Phi_{\rm p - nucl, f} (\vb{r})\; \Bigl|\,
    e^{- i\, c_{A}\, \vb{k_{\rm ph}} \vb{r}}\, V(\vb{r})\, \Bigr|\, \Phi_{\rm p - nucl, i} (\vb{r})\: \Bigr\rangle.
\end{array}
\label{eq.17.resultingformulas.6}
\end{equation}

In order to perform the first calculations of the spectra including incoherent bremsstrahlung and form factors of the scattered proton,
we introduce the following approximation:
We will neglect virtual photons between nucleons of nucleus-target, but we will study virtual photons between the scattered proton and nucleons of nucleus-target.

Such an approximation is explained by the following.
In QED, full potential of electromagnetic field $A_{\mu}(x)$ can be represented as summation of potential of external classic electromagnetic field $A_{\mu}^{\rm (ext)}$ and quantum electromagnetic field $A_{\mu}^{\rm (quant)}$
[see book~\cite{Ahiezer.1981}, p.~140--158]].
Operator $A_{\mu}^{\rm (quant)}$ describes (bremsstrahlung real and virtual) photons via quantization of electromagnetic field.
Potential $A_{\mu}^{\rm (ext)}$ takes into account evolution of nucleon under influence of other nucleons that gives different physical picture (corresponding wave functions, properties of proton-nucleus system in scattering) from picture of one nucleon moving inside vacuum [where quantum properties of nuclear forces are omitted].
In particular, in simple understanding, potential $A_{\mu}^{\rm (ext)}$ describes Coulomb potential between the scattered proton and nucleus-target
(this nucleus is not-point-like, we plan to include nuclear deformations to formalism in the future steps).
So, in framework of this approximation, Eq.~(\ref{eq.17.resultingformulas.2}) is transformed to
\begin{equation}
\begin{array}{lll}
\vspace{1.0mm}
  M_{\rm form factor}^{(1)} & = &
  (2\pi)^{3}\: \displaystyle\frac{\hbar\, e}{\sqrt{2}} \cdot
    \displaystyle\frac{z_{p}\, F_{2p}}{f_{p}}\: \bigl( p_{\rm q,1}^{\rm (p)} + p_{\rm q,2}^{\rm (p)} \bigr), \\

  M_{\rm form factor}^{(2)} & = &
  (2\pi)^{3}\; \sqrt{\displaystyle\frac{\hbar w_{\rm ph}}{2\pi\, c^{2}}}\,
  \displaystyle\frac{\pi\hbar}{w_{\rm ph}}\, 4e^{2} \cdot
    \displaystyle\frac{z_{p}^{2}\, F_{2p}}{f_{p}\,m_{p}}\: p_{q,3}^{\rm (p)}.
\end{array}
\label{eq.17.resultingformulas.7}
\end{equation}
These formulas can be used for starting calculations of bremsstrahlung cross-sections with incoherent terms
(that was not included in the previous research~\cite{Maydanyuk_Zhang_Zou.2019.PRC.microscopy}).

\section{Matrix elements at $l_{i}=0$, $l_{f}=1$, $l_{\rm ph}=1$
\label{sec.18}}

To obtain the spectra, we need to calculate the following integrals
[see Eqs.~(\ref{eq.resultingformulas.3}), (\ref{eq.17.resultingformulas.6})]
\begin{equation}
\begin{array}{lllll}
\vspace{0.5mm}
  \vb{I}_{1} = \biggl\langle\: \Phi_{\rm p - nucl, f} (\vb{r})\; \biggl|\, e^{-i\, \vb{k}_{\rm ph} \vb{r}}\; \vb{\displaystyle\frac{d}{dr}} \biggr|\: \Phi_{\rm p - nucl, i} (\vb{r})\: \biggr\rangle_\mathbf{r}, \\
\vspace{0.5mm}
  I_{2} = \Bigl\langle \Phi_{\rm p - nucl, f} (\vb{r})\; \Bigl|\, e^{i\, c_{p}\, \vb{k_{\rm ph}} \vb{r}}\, \Bigr|\, \Phi_{\rm p - nucl, i} (\vb{r})\: \Bigr\rangle_\mathbf{r}, \\
\vspace{0.5mm}
  I_{3} = \Bigl\langle \Phi_{\rm p - nucl, f} (\vb{r})\; \Bigl|\, e^{-i\, c_{A}\, \vb{k_{\rm ph}} \vb{r}}\, \Bigr|\, \Phi_{\rm p - nucl, i} (\vb{r})\: \Bigr\rangle_\mathbf{r}, \\
  I_{4} = \Bigl\langle \Phi_{\rm p - nucl, f} (\vb{r})\; \Bigl|\, e^{- i\, c_{A}\, \vb{k_{\rm ph}} \vb{r}}\, V(\vb{r})\, \Bigr|\, \Phi_{\rm p - nucl, i} (\vb{r})\: \Bigr\rangle_\mathbf{r}.
\end{array}
\label{eq.18.1.1}
\end{equation}
We define wave function of relative motion $\Phi_{\rm p - nucl} (\vb{r})$ as
\begin{equation}
  \Phi_{\rm p - nucl,\, l} (\vb{r}) = R_{l}\,(r)\: \displaystyle\sum\limits_{m=-l}^{l} Y_{lm}(\mathbf{n}_{\rm r}),
\label{eq.18.1.2}
\end{equation}
where $R\,(r)$ is radial scalar function (not dependent on $m$ at the same $l$), $\vb{n}_{\rm r} = \vb{r} / r$ is unit vector directed along $\mathbf{r}$,
$Y_{lm}(\vb{n}_{\rm r})$ are spherical functions (we use definition (28,7)--(28,8), p.~119 in~\cite{Landau.v3.1989}).
Here, $\bigl< \Phi_{\rm p - nucl, f} \bigl|\, \ldots \bigr| \, \Phi_{\rm p - nucl, i} \bigr>_\mathbf{r}$ is one-component matrix element
\begin{equation}
\begin{array}{lcl}
  \Bigl< \Phi_{\rm p - nucl, f} (\vb{r}) \Bigl|\, \hat{f}  \Bigr| \,\Phi_{\rm p - nucl, i} (\vb{r}) \Bigr>_\mathbf{r} & \equiv &
  \displaystyle\sum\limits_{m_{f}=-l_{f}}^{l_{f}}
  \displaystyle\sum\limits_{m_{i}=-l_{i}}^{l_{i}}
  \displaystyle\int
    R_{f}^{*}\,(r)\: Y_{l_{f}m_{f}}(\mathbf{n}_{\rm r})^{*}\;
    \hat{f} \:
    R_{i}\,(r)\: Y_{l_{i}m_{i}}(\mathbf{n}_{\rm r})\; \mathbf{dr},
\end{array}
\label{eq.18.1.3}
\end{equation}
where integration should be performed over space coordinates only.
In order to calculate integrals, we apply multipole expansion of wave function of photons.
Calculations of these integrals for a general case are given in Appendix~\ref{sec.app.integrals} [see Eqs.~(\ref{eq.app.integrals.4})--(\ref{eq.app.integrals.9})].
Calculation of radial integrals [see Eqs.~(\ref{eq.app.integrals.9}), (\ref{eq.app.integrals.6})]
\begin{equation}
\begin{array}{lcl}
  J_{1}(l_{i},l_{f},n) & = & \displaystyle\int\limits^{+\infty}_{0} \displaystyle\frac{dR_{i}(r, l_{i})}{dr}\: R^{*}_{f}(l_{f},r)\, j_{n}(kr)\; r^{2} dr, \\
  \tilde{J}\,(c, l_{i},l_{f},n) & = & \displaystyle\int\limits^{+\infty}_{0} R_{i}(r)\, R^{*}_{f}(l,r)\, j_{n}(c\, kr)\; r^{2} dr, \\
  \breve{J}\,(c_{A}, l_{i}, l_{f},n) & = & \displaystyle\int\limits^{+\infty}_{0} R_{i}(r)\, R^{*}_{l,f}(r)\, V(\mathbf{r})\, j_{n}(c_{A}\,kr)\; r^{2} dr,
\end{array}
\label{eq.18.1.6}
\end{equation}
is the most difficult numeric part in this research.
But, they do not depend on $\mu$, and also $m_{i}$, $m_{f}$.
The angular integrals are calculated in Appendix~B in Ref.~\cite{Maydanyuk.2012.PRC} [see Eqs.~(B1)--(B10) in the paper].

We define the probability of the emitted bremsstrahlung photons on the basis of the full matrix element $p_{fi}$ 
in frameworks of formalism given in \cite{Maydanyuk_Zhang_Zou.2016.PRC,Maydanyuk.2012.PRC,Maydanyuk_Zhang.2015.PRC} and we do not repeat it in this paper.
Finally, we obtain the bremsstrahlung probability as%
%
%
\begin{equation}
\begin{array}{ccl}
  \displaystyle\frac{d\,P }{dw_{\rm ph}} & = &
  \displaystyle\frac{e^{2}}{2\pi\,c^{5}}\: \displaystyle\frac{w_{\rm ph}\,E_{i}}{m_{\rm p}^{2}\,k_{i}}\: \bigl| p_{fi} \bigr|^{2}.
\end{array}
\label{eq.model.bremprobability.1}
\end{equation}
%
%
In analysis we calculate the different contributions of the emitted photons to the full bremsstrahlung spectrum.
For estimation of the interesting contribution, we use the corresponding matrix element of emission.

The matrix elements with orbital quantum numbers $l_{i}=0$, $l_{f}=1$ for wave functions and $l_{\rm ph}=1$ are
[see Appendix~\ref{sec.app.model.simplestcase}, Eqs.~(\ref{eq.app.simplestcase.10}), (\ref{eq.app.simplestcase.13}), (\ref{eq.app.simplestcase.formfactors.1});
here, $l_{i}$ and $l_{f}$ are orbital quantum numbers of wave function $\Phi_{\rm p - nucl} (\vb{r})$ defined in Eq.~(\ref{eq.18.1.2}) for states before emission of photon and after this emission,
$l_{\rm ph}$ is orbital quantum number of photon in the multipole approach defined in Eq.~(\ref{eq.app.integrals.3})]
\begin{equation}
\begin{array}{lll}
\vspace{1.5mm}
  M_{p}^{(E,\, {\rm dip})} & \simeq &
  -\, \hbar\, (2\pi)^{3}\, \displaystyle\frac{2\, \mu_{N}\,  m_{\rm p}}{\mu}\; Z_{\rm eff}^{\rm (dip)} \cdot \displaystyle\frac{\sqrt{3}}{6} \cdot J_{1}(0,1,0), \\

\vspace{1.5mm}
  M_{p}^{(M,\, {\rm dip})} & \simeq &
  -\, i\, \hbar\, (2\pi)^{3}\, \displaystyle\frac{\mu_{N}\, \alpha_{M}}{\mu} \cdot \displaystyle\frac{\sqrt{3}}{6} \cdot J_{1}(0,1,0), \\

\vspace{1.5mm}
  M_{\Delta M} & \simeq  &
   \hbar\, (2\pi)^{3}\, \mu_{N}\, f_{1}\, k_{\rm ph}\, Z_{\rm A} (\vb{k}_{\rm ph}) \cdot \displaystyle\frac{\sqrt{3}}{2} \cdot \tilde{J}\, (-c_{p}, 0,1,1), \\

  M_{k} & \simeq &
    -\, \hbar\, (2\pi)^{3}\, \mu_{N} \cdot k_{\rm ph}\, z_{\rm p}\: \mu_{\rm p}^{\rm (an)} \cdot \displaystyle\frac{\sqrt{3}}{2} \cdot \tilde{J}\, (c_{A}, 0,1,1) -
    \displaystyle\frac{\bar{\mu}_{\rm pn}^{\rm (an)}}{f_{1}} \cdot M_{\Delta M},
\end{array}
\label{eq.simplestcase.14}
\end{equation}
\begin{equation}
\begin{array}{llllllll}
\vspace{1.0mm}
  M_{\rm form factor}^{(1)} & = &
  (2\pi)^{3}\: \displaystyle\frac{\hbar\, e}{\sqrt{2}} \cdot
  \displaystyle\frac{z_{p}\, F_{2p}}{f_{p}}\:
  \Bigl\{
    \Bigl[ A_{1}^{\rm (p)} (Q, F_{1}, F_{2}) + A_{2}^{\rm (p)} (Q, F_{1}, F_{2}) \Bigr]\, I_{3} (\vb{k}_{\rm ph})\; + \\
  & + &
    \Bigl[ B_{1}^{\rm (p)} (Q, F_{1}, F_{2}) + B_{1}^{\rm (p)} (Q, F_{1}, F_{2}) \Bigr]\, I_{4} (\vb{k}_{\rm ph})
  \Bigr\}, \\

  M_{\rm form factor}^{(2)} & = &
  (2\pi)^{3}\; \sqrt{\displaystyle\frac{\hbar w_{\rm ph}}{2\pi\, c^{2}}}\,
  \displaystyle\frac{\pi\hbar}{w_{\rm ph}}\, 4e^{2} \cdot
    \displaystyle\frac{z_{p}^{2}\, F_{2p}}{f_{p}\,m_{p}} \cdot
    A_{3}^{\rm (p)} (Q, F_{1}, F_{2})\,
    I_{3} (2\, \vb{k}_{\rm ph}).
\end{array}
\label{eq.simplestcase.formfactors.1}
\end{equation}
The effective electric charge and other parameters are
\begin{equation}
\begin{array}{llllll}
  Z_{\rm eff}^{\rm (dip)} = \displaystyle\frac{m_{A}\, F_{p,\, {\rm el}} - m_{p}\, F_{A,\, {\rm el}} }{m_{p} + m_{A}}, &
  \alpha_{M} = \Bigl[ Z_{\rm A} (\vb{k}_{\rm ph})\: m_{p}\, \bar{\mu}_{\rm pn}^{\rm (an)} - z_{\rm p}\, m_{A}\, \mu_{\rm p}^{\rm (an)} \Bigr] \cdot \displaystyle\frac{m_{p}}{m_{p} + m_{A}}, &
  f_{1} = \displaystyle\frac{A-1}{2A}\: \bar{\mu}_{\rm pn}^{\rm (an)},
\end{array}
\label{eq.simplestcase.15}
\end{equation}
and
\begin{equation}
\begin{array}{lll}
  M_{p}^{(E,\, {\rm dip})} + M_{p}^{(M,\, {\rm dip})} =
  M_{p}^{(E,\, {\rm dip})} \cdot \Bigl\{ 1 + i \displaystyle\frac{\alpha_{M}}{2\, m_{\rm p}\, Z_{\rm eff}^{\rm (dip)}} \Bigr\}.
\end{array}
\label{eq.simplestcase.16}
\end{equation}
Here, factor $i \displaystyle\frac{\alpha_{M}}{2\, m_{\rm p}\, Z_{\rm eff}^{\rm (dip)}}$ describes suppression or reinforcement of electric emission of the bremsstrahlung photons
via additional inclusion of magnetic emission [that factor is different for different nuclei].
%
Coefficients $A_{1}^{\rm (p)}$, $B_{1}^{\rm (p)}$, $A_{2}^{\rm (p)}$, $B_{2}^{\rm (p)}$, $A_{3}^{\rm (p)}$ are given in Eqs.~(\ref{eq.17.resultingformulas.4}),
function $f_{\rm p}$ is given in Eq.~(\ref{eq.7.1.4}),
integrals are calculated in Eqs.~(\ref{eq.app.simplestcase.formfactors.2}).

The Dirac $F_{1}$ and Pauli $F_{2}$ form factors are related with the Sachs form factors $G^{E}$ and $G^{M}$ as [for example, see Eqs.~(5) in Ref.~\cite{A1Collaboration.2014.PRCv90p015206}]:
\begin{equation}
\begin{array}{lllll}
  F_{1} = \displaystyle\frac{\tau\,G^{M} + G^{E}}{1 + \tau}, &
  F_{2} = \displaystyle\frac{G^{M} - G^{E}}{(1 + \tau)},
\end{array}
\label{eq.simplestcase.formfactors.4}
\end{equation}
where $\tau = Q^{2}/(4m_{\rm p}^{2})$.
In the static limit $Q^{2} = 0$, these form factors normalize to
the charge and magnetic momentum of the proton in units of the
electron charge and the nuclear magneton $\mu_{K}$, $G^{E(0)} = 1$ and $G^{M(0)} = \mu_{p}$.
Authors of Ref.~\cite{A1Collaboration.2014.PRCv90p015206} collected different parametrizations of Sachs form factors.
We take some of those parametrizations for our analysis as
\begin{equation}
\begin{array}{lll}
\vspace{1.0mm}
  G_{\rm standard\, dipole}^{E} (Q^{2}) & = &
  \displaystyle\frac{G_{\rm standard\, dipole}^{Mp} (Q^{2})}{\mu_{\rm p}} =
  \displaystyle\frac{G_{\rm standard\, dipole}^{Mn} (Q^{2})}{\mu_{\rm n}} =
  \Bigl(1 + \displaystyle\frac{Q^{2}}{0.71\, GeV^{2}} \Bigr)^{-2}, \\

\vspace{1.0mm}
  G_{\rm dipole}^{E,M} (Q^{2}) & = &
  \Bigl(1 + \displaystyle\frac{Q^{2}}{a^{E,M}} \Bigr)^{-2}, \\

\vspace{1.0mm}
  G_{\rm double\, dipole}^{E,M} (Q^{2}) & = &
  a_{0}^{E,M}\, \Bigl(1 + \displaystyle\frac{Q^{2}}{a_{1}^{E,M}} \Bigr)^{-2} +
  (1 - a_{0}^{E,M})\, \Bigl(1 + \displaystyle\frac{Q^{2}}{a_{2}^{E,M}} \Bigr)^{-2}, \\

\end{array}
\label{eq.simplestcase.formfactors.5}
\end{equation}
Here, $\mu_{\rm p}$ and $\mu_{\rm n}$ are magnetic momentums of proton and neutron.
Parameters can be found from Ref.~\cite{Friedrich.2003.EPJA}
(see  Table~2 in p.~611).
Our approach allows to obtain such parameters independently (see next section).

\section{Analysis
\label{sec.results}}

\subsection{Role of incoherent emission in bremsstrahlung
\label{sec.results.1}}

We calculate the bremsstrahlung probability of emission of photons accompanying the scattering of protons off nuclei.
For analysis we choose scattering of $p + ^{197}{\rm Au}$ at proton beam energy of 190~MeV,
where experimental bremsstrahlung data \cite{Goethem.2002.PRL} were obtained with good accuracy.
Wave function of relative motion between proton and center-of-mass of nucleus is calculated numerically
concerning to the proton-nucleus potential in form of $V (r) = v_{c}(r) + v_{N}(r) + v_{\rm so}(r) + v_{l} (r)$,
where $v_{c}(r)$, $v_{N}(r)$, $v_{\rm so}(r)$, and $v_{l} (r)$ are Coulomb, nuclear, spin-orbital, and centrifugal components, respectivelly.
Parameters of that potential are defined in Eqs.~(46)--(47) in Ref.~\cite{Maydanyuk_Zhang.2015.PRC}.
At first, we reconstruct the spectrum without contribution of internal structure of the scattered proton, but with inclusion of incoherent bremsstrahlung emission.
We calculate the bremsstrahlung cross-section by Eq.~(\ref{eq.model.bremprobability.1}),
where we include
matrix elements of coherent emission $M_{p}^{(E,\, {\rm dip})}$, $M_{p}^{(M,\, {\rm dip})}$ by Eqs.~(\ref{eq.resultingformulas.1}),
and matrix elements of incoherent emission $ M_{\Delta M}$, $M_{k}$ by Eqs.~(\ref{eq.resultingformulas.2}).
We do not include internal structure of the scattered proton $M_{\rm form factor}^{(1)}$, $M_{\rm form factor}^{(2)}$ by Eqs.~(\ref{eq.17.resultingformulas.7}).

%
\begin{figure}[htbp]
\centerline{\includegraphics[width=90mm]{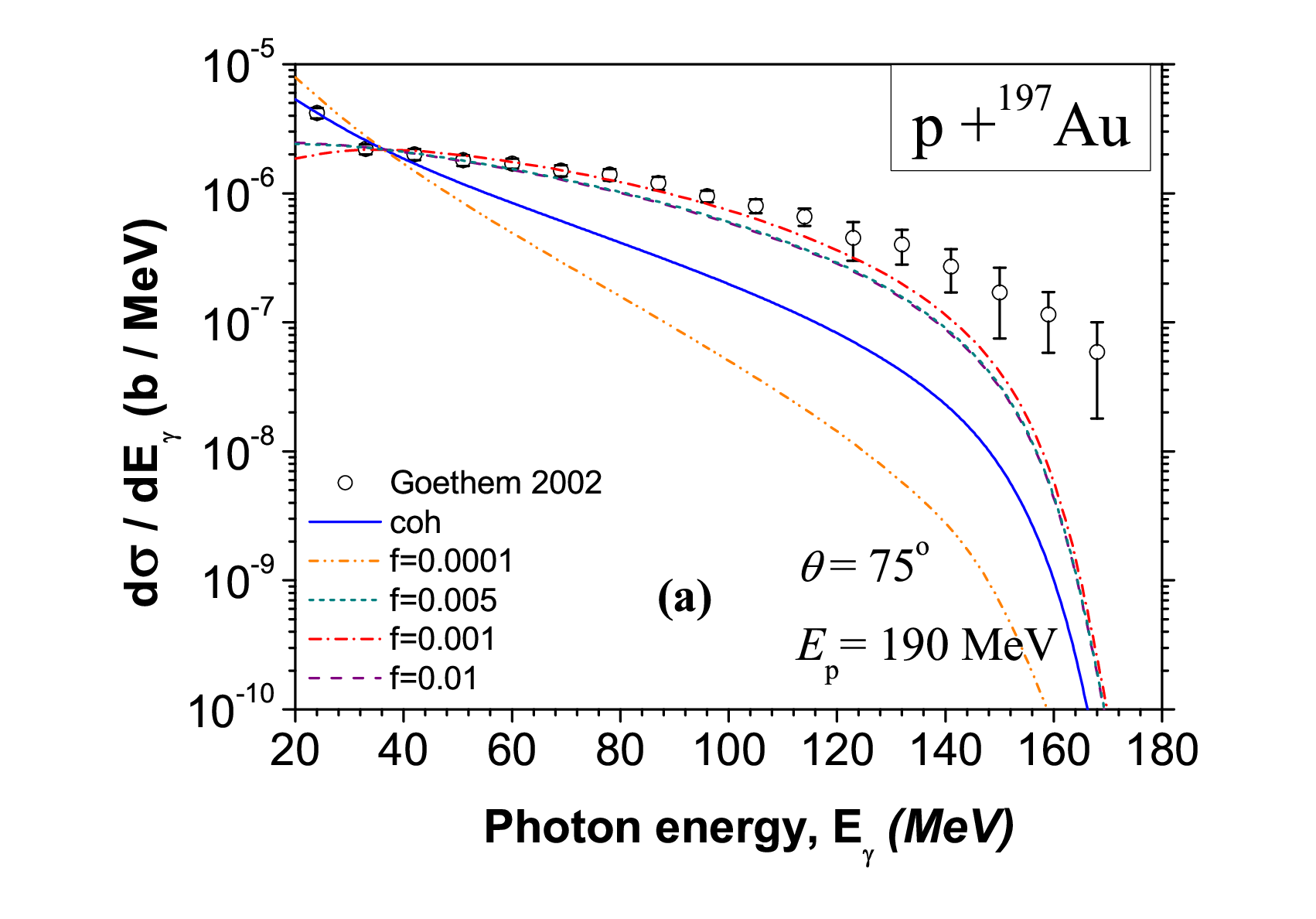}
\hspace{-1mm}\includegraphics[width=86mm]{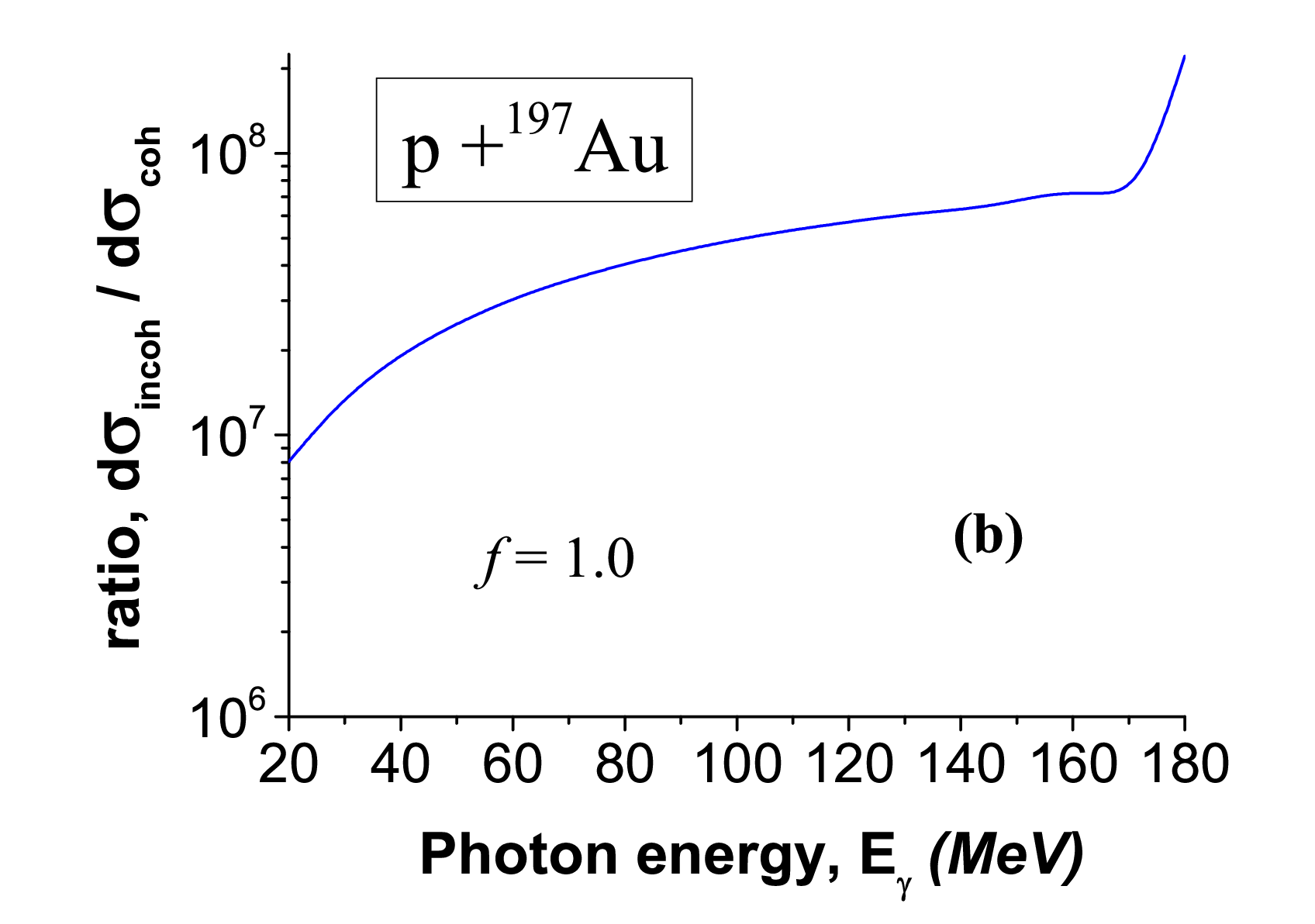}}
\vspace{-4mm}
\caption{\small (Color online)
The calculated bremsstrahlung spectra with coherent and incoherent terms in the scattering of protons off the $^{197}{\rm Au}$ nuclei at energy of proton beam of $E_{\rm p}=190$~MeV
in comparison with experimental data~\cite{Goethem.2002.PRL}
[matrix elements are defined in Eqs.~(\ref{eq.results.1}),
$Z_{A} (k_{\rm ph})\simeq Z_{A}$ is electric charge of nucleus,
we normalize each calculation of the second point of experimental data].
Panel (a):
The full spectra in dependence on factor $f_{\rm incoh}$ which suppresses the intensity of incoherent processes.
One can see clear changes of the full spectrum in dependence on factor $f_{\rm incoh}$,
red dash-dotted solid line (at $f=0.001$) corresponds to the best agreement with experimental data.
This result confirms the important (and not small) role of incoherent emission in bremsstrahlung.
Panel (b):
Ration between incoherent and coherent contributions in dependence on energy of emitted bremsstrahlung photon
[we define ratio as $\varepsilon = \sigma_{\rm incoh}/\sigma_{\rm coh}$,
in calculations we use factor of incoherence of $f_{\rm incoh}=1.0$]
One can see that
(a) incoherent emission is essentially more intensive than the coherent emission,
(b) role of incoherent processes is increased at increasing of energy of photon,
(c) better agreement with experimental data at $f_{\rm incoh}=0.001$ (than at $f_{\rm incoh}=1$) indicates on presence of some unknown effect,
which highly suppresses the incoherent processes.
%
\label{fig.result.1}}
\end{figure}

An important role of incoherent bremsstrahlung emission in the proton-nucleus scattering can be seen from simple calculations.
One can use only the main contributions in both coherent and incoherent terms at $l_{i}=0$, $l_{f}=1$, $l_{\rm ph}=1$,
i.e., we define the coherent term $M_{\rm coh}$ on the basis of term $M_{p}^{(E,\, {\rm dip})}$ and
incoherent term $M_{\rm incoh}$ on the basis of term $M_{\Delta M}$ as
%
\begin{equation}
\begin{array}{llllll}
  M_{\rm full} = M_{\rm coh} + M_{\rm incoh}, &

  M_{\rm coh} = M_{p}^{(E,\, {\rm dip})}, &

  M_{\rm incoh} = f_{incoh} \cdot M_{\Delta M},
\end{array}
\label{eq.results.1}
\end{equation}
where
$M_{p}^{(E,\, {\rm dip})}$ and
$M_{\Delta M}$ are defined in Eqs.~(\ref{eq.simplestcase.14}),
$f_{\rm incoh}$ is a new factor.
Such calculations in comparison with experimental data are presented in Fig.~\ref{fig.result.1}~(a).%
\footnote{In previous study of incoherent emission in the proton-nucleus scattering in Ref.~\cite{Maydanyuk_Zhang.2015.PRC} we have calculated angular integrals numerically with approximations
(over the angle between direction of emission of bremsstrahlung photon and direction of the scattered proton).
Also we have calculated the bremsstrahlung cross-sections with fixed value of this angle, according to experiments (i.e., $\theta=75^{\circ}$ for the studied process $p+\isotope[197]{Au}$).
In current analysis we find analytical solutions for angular integrals at first values of quantum numbers $l_{i}$, $l_{f}$
[however, we do not analyze dependence of cross-sections on this angle, see Eq.~(\ref{eq.model.bremprobability.1})].
Analytic integration shows that
$I_{M}\, (0, 1, 1, 1, \mu) = 0$,
$\tilde{I}\, (0, 1, 1, 0, \mu) = 0$,
$\tilde{I}\, (0, 1, 1, 2, \mu) = 0$ [see Eqs.~(\ref{eq.app.simplestcase.3})],
so we take into account next correction in calculations of matrix elements, that explains difference between spectra in Ref.~\cite{Maydanyuk_Zhang.2015.PRC} and in Fig.~1 in current paper.
Nevertheless, the dependence of the full bremsstrahlung spectrum on the incoherent contribution exits in any case.
}
One can see that full spectra are visibly changed if to suppress incoherent emission via changing factor $f_{\rm incoh}$.
The full spectrum with included both coherent and incoherent contributions at $f_{\rm incoh}=1.0$ is very close to the purple dashed line at $f_{\rm incoh}=0.01$ in this picture, while blue solid line describes only the coherent emission without incoherent contribution.

One can see essential difference between these two spectra, that confirms large role of incoherent emission in the full bremsstrahlung.
Comparing calculations with experimental data, we extract coefficient of ratio of $f=0.001$ for the best agreement.
In Fig.~\ref{fig.result.1}~(b) we show ratio between incoherent and coherent emissions at $f=1.0$.
A main conclusion from this analysis is that incoherent processes are essentially more intensive than coherent processes, this difference is increased at increasing of energy of photon emitted.

\subsection{Internal structure of the scattered proton extracted from bremsstrahlung analysis
\label{sec.results.2}}

\begin{figure}[htbp]
\centerline{\includegraphics[width=90mm]{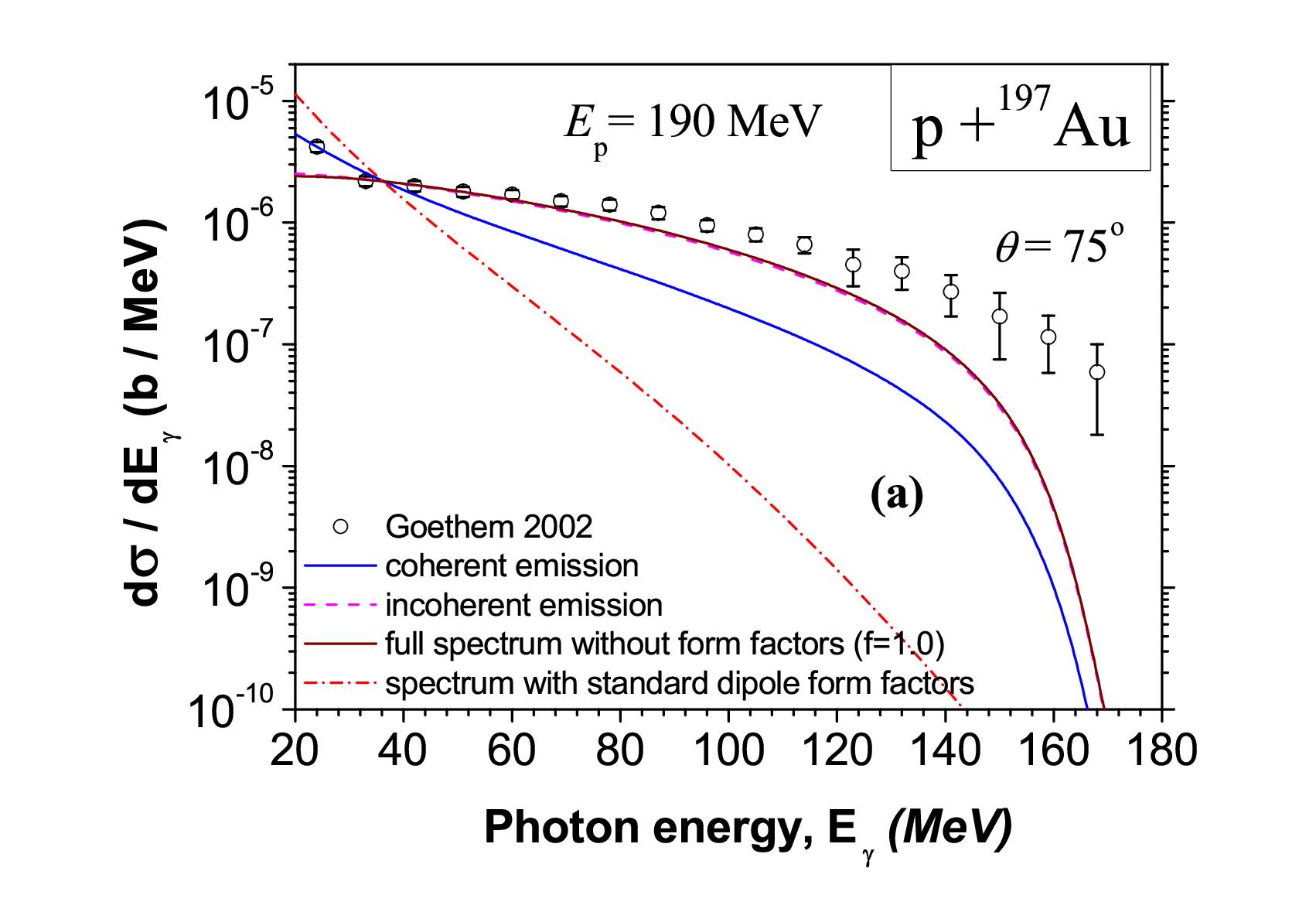}
}
\vspace{-4mm}
\caption{\small (Color online)
The calculated bremsstrahlung spectra with form factors in standard dipole parametrization (see red dash-dotted line) in the scattering of protons off the \isotope[197]{Au} nuclei at energy of proton beam of $E_{\rm p}=190$~MeV in comparison with experimental data~\cite{Goethem.2002.PRL}.
Here,
red dash-double dotted line is the new spectrum with form factors $G_{\rm standard\, dipole}^{E,M} (Q^{2})$ in standard dipole parametrization defined in Eqs.~(\ref{eq.simplestcase.formfactors.5}) (without free amplitudes),
brown solid line is the spectrum (with coherent and incoherent contributions, at $f_{\rm incoh}=1$) without form factors of proton,
blue solid line represents coherent contribution,
purple dashed line characterizes incoherent contribution
[all spectra are normalized on 1 point of experimental data].
\label{fig.result.3}}
\end{figure}

Now we calculate the spectra with inclusion of matrix element with form factors of the scattered proton.
We check, if the full spectrum is changed after inclusion of those form factors.
The simplest analysis can be done on the basis of the standard dipole parametrization of form factors [here there is no any additional free parameters].
In analysis of the internal structure of the scattered proton we restrict ourselves by contribution $M_{\rm form factor}^{(1)}$ with terms at $A_{1}$ in Eqs.~(\ref{eq.simplestcase.formfactors.1}).
In Fig.~\ref{fig.result.3} we show results of such calculations where we add form factors $G_{\rm standard\, dipole}^{E,M} (Q^{2})$ in the standard dipole parametrization defined in Eqs.~(\ref{eq.simplestcase.formfactors.5}).
From this figure one can see that change of the bremsstrahlung spectrum is essential with inclusion of such form factors
[see red dash-dotted line for the spectrum with form factors in standard dipole parametrization in comparison with solid brown line for the spectrum without form factors (and with coherent and incoherent terms) in this figure].
This confirms that contribution of matrix element with form factors of the scattered proton to the full bremsstrahlung spectrum is large.
Note that this result is in agreement with our previous estimations made in Ref.~\cite{Maydanyuk_Zhang_Zou.2019.PRC.microscopy} without analysis of incoherent bremsstrahlung emission in this reaction.
Next conclusion from such a result is that our model is able to estimate form factors of proton in reactions from analysis of bremsstrahlung (this approach is sensitive).
%

\begin{figure}[htbp]
\centerline{\includegraphics[width=90mm]{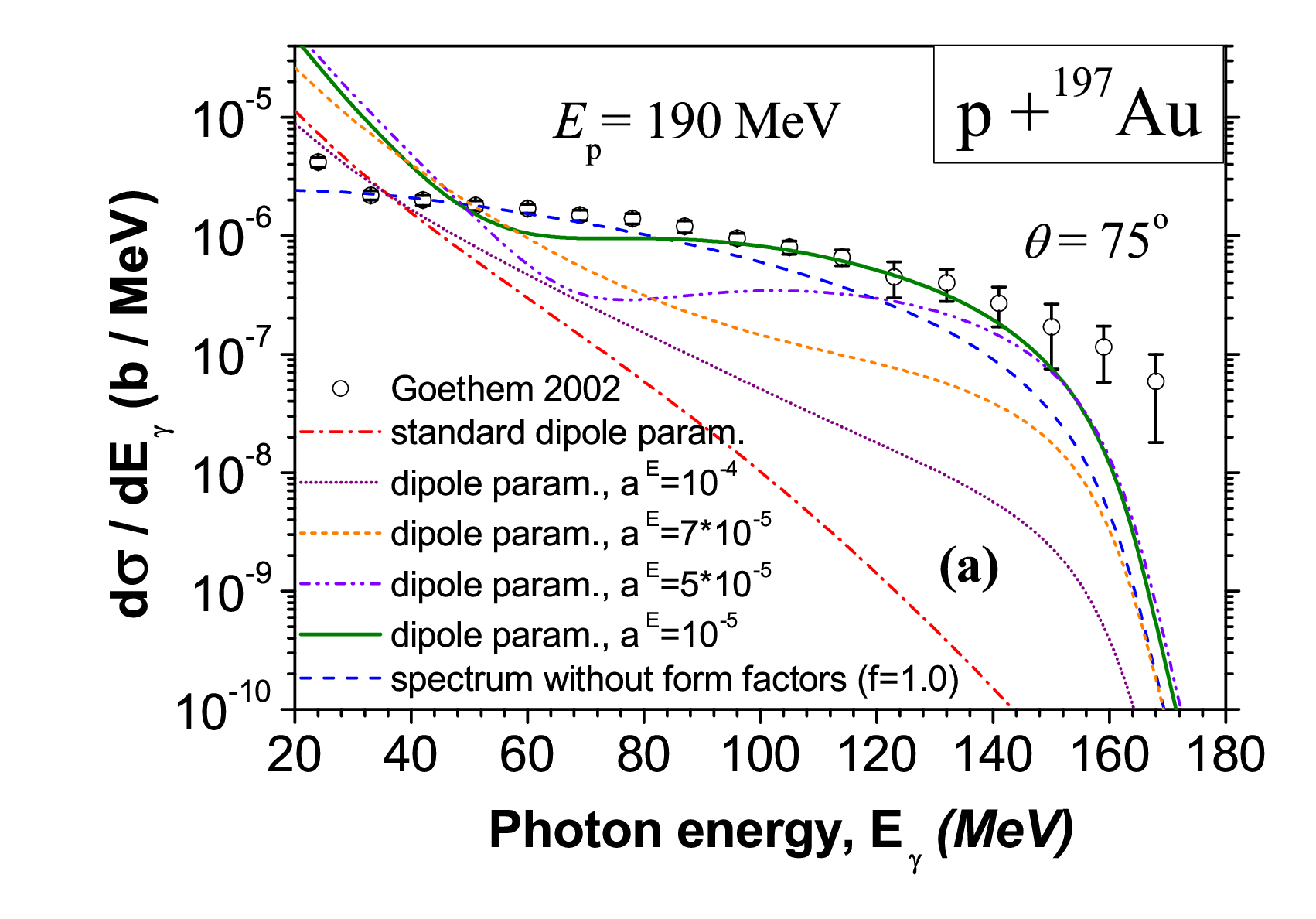}}
\vspace{-4mm}
\caption{\small (Color online)
The calculated bremsstrahlung spectra with form factors in dipole parametrization with one free amplitude $a^{E}$ in the scattering of protons off the $^{197}{\rm Au}$ nuclei at energy of proton beam of $E_{\rm p}=190$~MeV
in comparison with experimental data~\cite{Goethem.2002.PRL}
[form factors in standard dipole parametrization $G_{\rm standard\, dipole}^{E,M} (Q^{2})$ and
form factor in dipole parametrization $G_{\rm dipole}^{E} (Q^{2})$ are defines in Eqs.~(\ref{eq.simplestcase.formfactors.5})
where $G_{\rm dipole}^{M} (Q^{2}) = G_{\rm dipole}^{E} (Q^{2}) / \mu_{\rm p}$].
\label{fig.result.4}}
\end{figure}

In Fig.~\ref{fig.result.4} we show the spectra with form factor in dipole parametrization $G_{\rm dipole}^{E} (Q^{2})$ defined in Eqs.~(\ref{eq.simplestcase.formfactors.5}) with one free parameter $a^{E}$ [here we fix $G_{\rm dipole}^{M} (Q^{2}) = G_{\rm dipole}^{E} (Q^{2})/\mu_{\rm p}$ and $a^{M}=a^{E}$].
One can see that such a dipole parametrization allows to improve essentially the agreement between calculations and experimental data
in comparison with previous calculations with form factors in standard dipole parametrization
(see green solid line for dipole parametrization at $a^{E}=10^{-5}$ in comparison with red dash-dotted line for standard dipole parametrization in this figure).
From this figure one can see that sensitivity in determination of parameters for form factors by such an approach is really good.

\begin{figure}[htbp]
\centerline{\includegraphics[width=90mm]{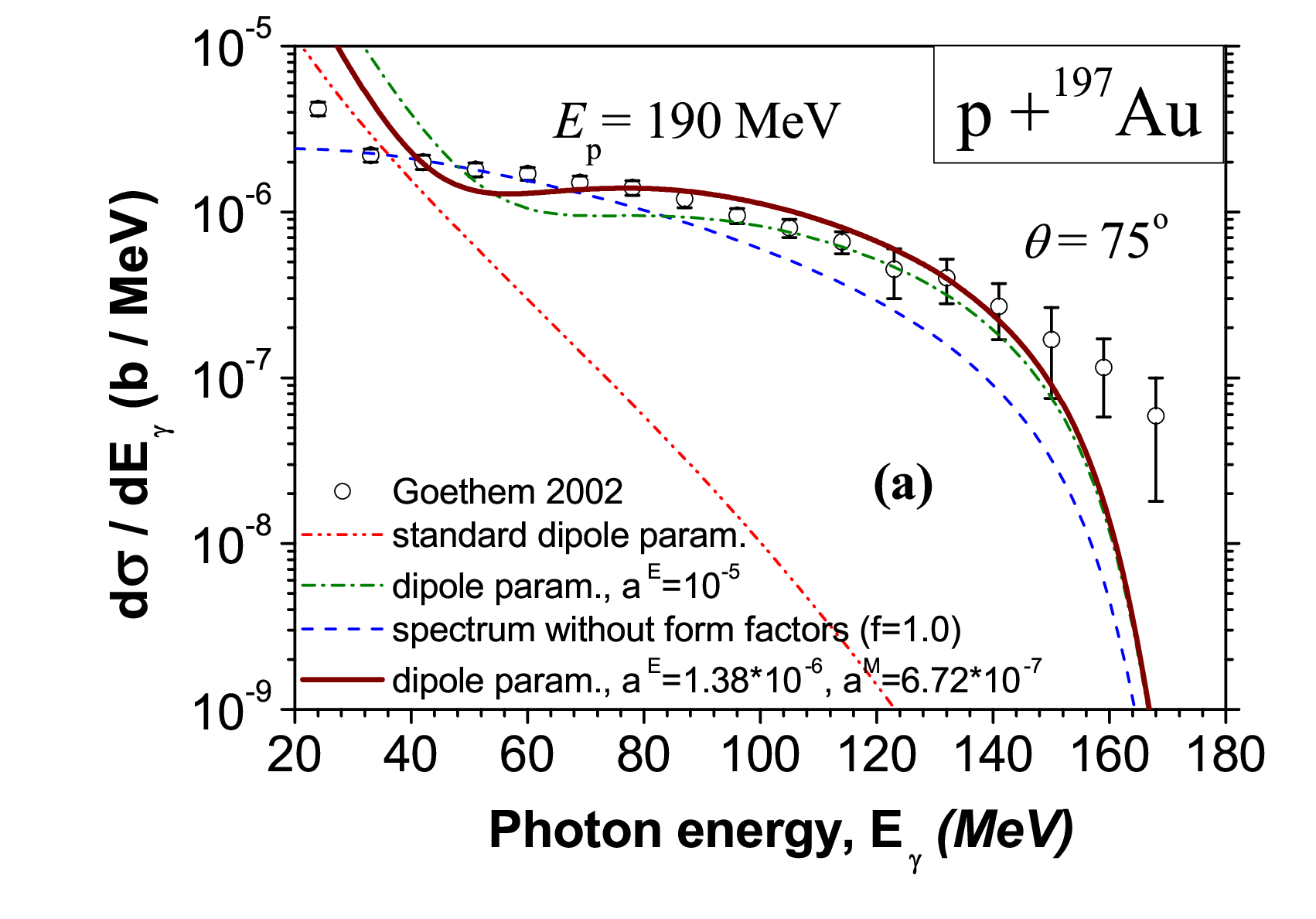}
\hspace{-1mm}\includegraphics[width=86mm]{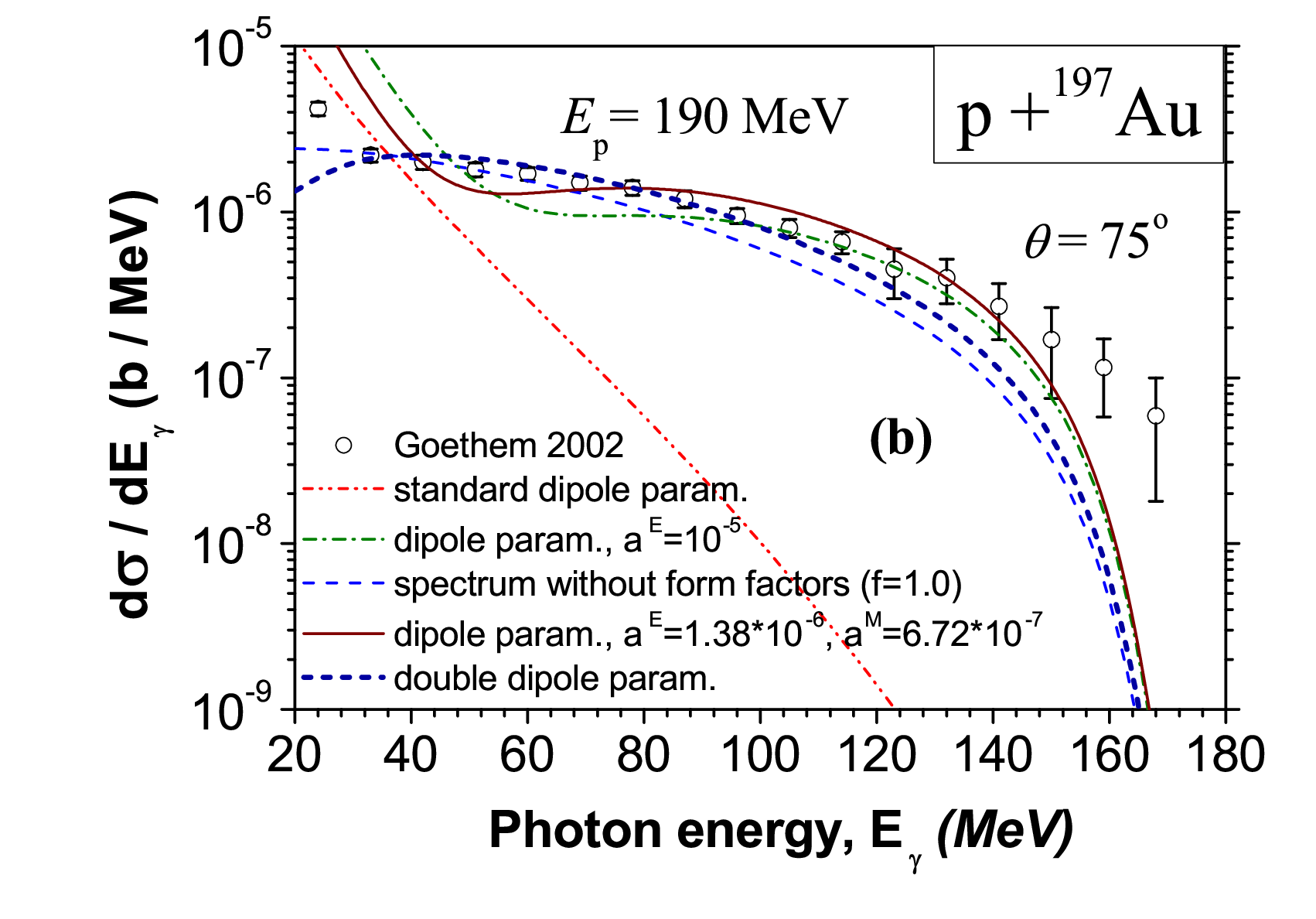}}
\vspace{-4mm}
\caption{\small (Color online)
The calculated bremsstrahlung spectra with form factors in the dipole parametrization with two free amplitudes (see brown solid line) in the scattering of protons off the \isotope[197]{Au} nuclei
at energy of proton beam of $E_{\rm p}=190$~MeV in comparison with experimental data~\cite{Goethem.2002.PRL}.
Panel (a):
Calculations of spectrum with form factors in the dipole parametrization with two free amplitudes.
Here,
blue dashed line is the spectrum (with coherent and incoherent contributions) without form factors of proton (at $f_{\rm incoh}=1$),
green dash-dotted line is the spectrum with form factors in dipole parametrization at one free amplitude $a^{E}$
(here $G_{\rm dipole}^{E} (Q^{2})$ is defined in Eqs.~(\ref{eq.simplestcase.formfactors.5}) and $G_{\rm dipole}^{M} (Q^{2}) = G_{\rm dipole}^{E} (Q^{2}) /\mu_{\rm p}$),
red dash-double dotted line is the spectrum with form factors $G_{\rm standard\, dipole}^{E,M} (Q^{2})$ in standard dipole parametrization defined in Eqs.~(\ref{eq.simplestcase.formfactors.5})
(i.e., without free parameters),
brown solid line is the spectrum with form factors in dipole parametrization defined in Eqs.~(\ref{eq.simplestcase.formfactors.5}) with two free amplitudes $a^{E}$ and $a^{M}$
[we obtain
$a^{E} = 1.38(18)\, 10^{-6}$,
$a^{M} = 6.(72)\, 10^{-7}$,
the corresponding accuracy functions are $\varepsilon_{1} = 0.0310890$,
$\varepsilon_{2} = 0.2414291$,
$\varepsilon_{3} = 0.0264169$
].
Panel (b):
Calculations of spectrum with form factors in double dipole parametrization (see new dark blue short dashed line)
[we obtain
$a_{0}^{E} = 0.1$,
$a_{1}^{E} = 1.53795\, 10^{-5}$,
$a_{2}^{E} = 10^{-7}$,
$a_{0}^{M} = 0.1$,
$a_{1}^{M} = 6.588235\, 10^{-7}$,
$a_{2}^{M} = 10^{-4}$,
and accuracy functions are $\varepsilon_{1} = 0.027315$,
$\varepsilon_{2} = 0.280900$,
$\varepsilon_{3} = 0.037561$.
We calculate above the amplitudes, analyzing minimums of functions of errors $\varepsilon_{1}$, $\varepsilon_{2}$, $\varepsilon_{3}$, following the definitions~(20) in Ref.~\cite{Maydanyuk.2015.NPA}].
\label{fig.result.5}}
\end{figure}

In Fig.~\ref{fig.result.5}~(a) we show the spectrum if two independent amplitudes in dipole parametrization defined in Eqs.~(\ref{eq.simplestcase.formfactors.5}) are used.
Analysis shows that form factors in the double dipole parametrization defined in Eqs.~(\ref{eq.simplestcase.formfactors.5})
allow to improve agreement between calculations and experimental data (in comparison with all calculated spectra above), but not much
[see solid line in Fig.~\ref{fig.result.5}~(b)].
In this paper we would not like to analyze in details different parametrizations of form factors on the basis of fitting calculations,
as a main aim of this research is to focus on aspects of formalism presented above by including incoherent bremsstrahlung emission.

\section{Conclusions and perspective
\label{sec.conclusions}}

In this paper we have improved the bremsstrahlung model~\cite{Maydanyuk_Zhang_Zou.2019.PRC.microscopy} for the proton-nucleus scattering with electromagnetic form factors of the scattered proton, where a new formalism of incoherent bremsstrahlung emission is included.
For analysis, experimental data for $p + \isotope[197]{Au}$ scattering at energy of proton beam of 190~MeV obtained by TAPS collaboration is chosen~\cite{Goethem.2002.PRL}.
Conclusions from analysis on the basis of this model are the following.

\begin{enumerate}
\item
Inclusion of incoherent emission to the model and calculations improves essentially agreement with experimental data~\cite{Goethem.2002.PRL}, in comparison with the coherent bremsstrahlung.
Incoherent bremsstrahlung emission in the analyzed reaction is larger than coherent one (see Fig.~\ref{fig.result.1}).
Such a conclusion looks to be unexpected, as in previous bremsstrahlung calculations in nuclear reactions by different authors the largest contribution
was supposed comes from coherent bremsstrahlung,
usually the coherent bremsstrahlung was studied only in many tasks.
From this result it follows that an united system of evolving nucleons in scattering provides more accurate bremsstrahlung,
than relative motion of two nuclear fragments as two bodies (within nucleon distribution inside each fragment).
It turns out that the difference between these two pictures in calculations of bremsstrahlung is essential.

\item
Inclusion of internal structure of proton with form factors in nuclear scattering allows to improve agreement with experimental data~\cite{Goethem.2002.PRL}, in comparison with calculations with coherent and incoherent bremsstrahlung emission without form factors of the scattered proton (see solid green line for spectrum with form factors in dipole parametrization in comparison with blue dashed line for spectrum without form factors in Fig.~\ref{fig.result.4}, Fig.~\ref{fig.result.5}).
This confirms our idea that one can develop a model to study internal structure of protons via bremsstrahlung analysis.

\item
A sensitivity of the model in study of form factors of the scattered proton is really high
(see the difference between brown solid line for spectrum without for factors and brown dash-dotted line for spectrum with form factors in standard dipole parametrization in Fig.~\ref{fig.result.3}).
This demonstrates a good opportunity to study the internal structure of protons under influence of nuclear forces in nuclear scattering in wide energy region.
This allows to reanalyze other experimental data in proton-nucleus scattering obtained in the past.
In this article, however, our aim is to find new information about internal structure of protons under strong influence of nuclei in scattering via bremsstrahlung analysis.

\item
Note that the agreement between the last calculations with inclusion of form factors of the scattered proton [see the solid brown line for dipole parametrization and blue dashed line for spectrum with form factors in double dipole parametrization in Fig.~\ref{fig.result.5}~(b)] and experimental data~\cite{Goethem.2002.PRL} is the most accurate,
in comparison with other models applied for that reaction.
This indicates on a good perspective on further investigations on this research line.

\item
Estimation on the form factors of proton in our model are very preliminary.
More reliable estimation for these form factors can be realized via more accurate calculations with including appropriate model approximations.
\end{enumerate}

\section*{Acknowledgements
\label{sec.acknowledgements}}

Authors are highly appreciated to
Prof.~M.~Ya.~Amusia for his interesting insight on general aspects of bremsstrahlung topic,
Prof.~V.~S.~Vasilevsky for interesting discussions concerning to peculiarities of Dirac equation,
Profs.~V.~A.~Plujko, S.~N.~Fedotkin, A.~G.~Magner, F.~A.~Ivanyuk, A.~P.~Ilyin, A.~Ya.~Dzyublik for fruitful and useful discussions.
S.~P.~M. thanks the
Sun Yat-Sen University for warm hospitality and support.


\appendix

\section{Matrix element of emission of bremsstrahlung photons without form factors
\label{sec.app.1}}

\subsection{Form factors
\label{sec.app.form_factors}}

In this Appendix we present explicite form of form factors
%
%
[details of calculation can be found in Ref.~\cite{Maydanyuk_Zhang_Zou.2019.brem_alpha_nucleus.arxiv}, see Eqs.~(C2), (C7), (C8), (C10) in that paper]:
\begin{equation}
\begin{array}{llllll}
\vspace{1mm}
  F_{p,\, {\rm el}} (\vb{k}_{\rm ph}) & = &
  D_{p, P\, {\rm el}} =
    z_{n}, \\

\vspace{1mm}
  F_{A,\, {\rm el}} (\vb{k}_{\rm ph}) & = &
  D_{A,P\, {\rm el}} =
    \displaystyle\sum\limits_{m=1}^{A}
    \Bigl\langle \psi_{\rm nucl, f} (\beta_{A}) \Bigl|\, z_{m}\, e^{-i \vb{k}_{\rm ph} \rhobf_{A m} } \Bigr|\, \psi_{\rm nucl, i} (\beta_{A}) \Bigr\rangle , \\

\vspace{1mm}
  \vb{F}_{p,\, {\rm mag}} (\vb{k}_{\rm ph}) & = &
  \vb{D}_{p, P\, {\rm mag}} =
    \mu_{\rm p}^{\rm (an)}\, m_{\rm p}\; \bigl\langle \psi_{p, f}\, \bigl|\, \sigmabf \bigr| \psi_{p, i} \bigr\rangle, \\

  \vb{F}_{A,\, {\rm mag}} (\vb{k}_{\rm ph}) & = &
  \vb{D}_{A,P\, {\rm mag}} =
    \displaystyle\sum_{j=1}^{A}
    \Bigl\langle \psi_{\rm nucl, f} (\beta_{A})\, \Bigl|\, \mu_{j}^{\rm (an)}\, m_{Aj}\; e^{-i\, \vb{k_{\rm ph}} \rhobf_{Aj}}\, \sigmabf \Bigr| \psi_{\rm nucl, i} (\beta_{A}) \Bigr\rangle,
\end{array}
\label{eq.app.13.1.9.a}
\end{equation}
%
%
\begin{equation}
\begin{array}{lll}
\vspace{1mm}
  \vb{D}_{p 1,\, {\rm el}} = &
      0, \\

\vspace{1mm}
  \vb{D}_{A 1,\, {\rm el}} = &
    \displaystyle\sum\limits_{i=1}^{A-1}
      \displaystyle\frac{z_{j} m_{\rm p}}{m_{Aj}}\,
      \Bigl\langle \psi_{A, f} (\beta_{A})\, \Bigl|\,
        e^{-i \vb{k}_{\rm ph} \rhobf_{Aj}} \vb{\tilde{p}}_{Aj}
      \Bigr|\,  \psi_{Aj} (\beta_{A}) \Bigr\rangle , \\

\vspace{1mm}
  \vb{D}_{p 2,\, {\rm el}} = &
      0, \\

  \vb{D}_{A 2,\, {\rm el}} = &
    \displaystyle\sum\limits_{i=1}^{A}
      z_{j}\,
      \Bigl\langle \psi_{A, f} (\beta_{A})\, \Bigl|\,
        e^{-i \vb{k}_{\rm ph} \rhobf_{Aj}}
        \Bigl( \displaystyle\sum_{k=1}^{A-1} \vb{\tilde{p}}_{Ak} \Bigr)
      \Bigr|\,  \psi_{Aj} (\beta_{A}) \Bigr\rangle,
\end{array}
\label{eq.app.13.1.9.b}
\end{equation}
\begin{equation}
\begin{array}{lll}
\vspace{1mm}
  D_{p 1,\, {\rm mag}} (\vb{e}^{(\alpha)}) = &
      0, \\

\vspace{1mm}
  D_{A 1,\, {\rm mag}} (\vb{e}^{(\alpha)}) = &
    \displaystyle\sum\limits_{j=1}^{A-1}
    \mu_{j}^{\rm (an)}\,
    \Bigl\langle \psi_{A, f} (\beta_{A})\, \Bigl|\,
      e^{-i\, \vb{k_{\rm ph}} \rhobf_{Aj}}\; \sigmabf \cdot
      \bigl[ \vb{\tilde{p}}_{Aj} \times \vb{e}^{(\alpha)} \bigr]
    \Bigr|\,  \psi_{Aj} (\beta_{A}) \Bigr\rangle, \\

\vspace{1mm}
  D_{p 2,\, {\rm mag}} (\vb{e}^{(\alpha)}) = &
      0, \\

  D_{A 2,\, {\rm mag}} (\vb{e}^{(\alpha)}) = &
    \displaystyle\sum\limits_{j=1}^{A}
      \mu_{j}^{\rm (an)}\,
      \displaystyle\frac{m_{Aj}}{m_{A}}\,
      \Bigl\langle \psi_{A, f} (\beta_{A})\, \Bigl|\,
      e^{-i\, \vb{k_{\rm ph}} \rhobf_{Aj}}\; \sigmabf \cdot
      \Bigl[ \Bigl( \displaystyle\sum_{k=1}^{A-1} \vb{\tilde{p}}_{Ak} \Bigr) \times \vb{e}^{(\alpha)} \Bigr]
      \Bigr|\,  \psi_{A, i} (\beta_{A}) \Bigr\rangle,
\end{array}
\label{eq.app.13.1.9.c}
\end{equation}
\begin{equation}
\begin{array}{lll}
\vspace{1mm}
  \vb{D}_{p,\, {\rm k}} = &
  \mu_{\rm p}^{\rm (an)}\,
    \bigl\langle \psi_{p, f}\, \bigl|\, \sigmabf \bigr|\, \psi_{p, i} \bigr\rangle, \\

  \vb{D}_{A,\, {\rm k}} = &
  \displaystyle\sum\limits_{j=1}^{A}
    \mu_{j}^{\rm (an)}\, \Bigl\langle \psi_{A, f} (\beta_{A})\, \Bigl|\, e^{-i\, \vb{k_{\rm ph}} \rhobf_{Aj}}\, \sigmabf \Bigr|\, \psi_{A, j} (\beta_{A}) \Bigr\rangle.
\end{array}
\label{eq.app.13.1.9.d}
\end{equation}

\subsection{Matrix element of the coherent bremsstrahlung
\label{sec.13.2}}

We will find matrix elements, which describes the coherent emission of bremsstrahlung photons and should be the largest.
These are terms~$M_{p}^{(E)}$ and $M_{p}^{(M)}$ determined in Eqs.~(\ref{eq.13.1.3}) which we rewrite as
\begin{equation}
\begin{array}{lll}
\vspace{-0.2mm}
  M_{p}^{(E)} & = &
  i \hbar\, (2\pi)^{3} \displaystyle\frac{2\, \mu_{N}\,  m_{\rm p}}{\mu}\;
  \displaystyle\sum\limits_{\alpha=1,2}
    \vb{e}^{(\alpha)}
    \biggl\langle\: \Phi_{\rm p - nucl, f} (\vb{r})\; \biggl|\,
    Z_{\rm eff} (\vb{k}_{\rm ph}, \vb{r}) \,
    e^{-i\, \vb{k}_{\rm ph} \vb{r}}\;
    \vb{\displaystyle\frac{d}{dr}}
    \biggr|\: \Phi_{\rm p - nucl, i} (\vb{r})\: \biggr\rangle, \\

  M_{p}^{(M)} & = &
  -\, \hbar\, (2\pi)^{3} \displaystyle\frac{ \mu_{N}}{\mu}\;
  \displaystyle\sum\limits_{\alpha=1,2}
  \biggl\langle\: \Phi_{\rm p - nucl, f} (\vb{r})\; \biggl|\,
  \vb{M}_{\rm eff} (\vb{k}_{\rm ph}, \vb{r}) \cdot
  e^{-i\, \vb{k}_{\rm ph} \vb{r}} \cdot
  \Bigl[ \vb{\displaystyle\frac{d}{dr}} \times \vb{e}^{(\alpha)} \Bigr]
  \biggr|\: \Phi_{\rm p - nucl, i} (\vb{r})\: \biggr\rangle.
\end{array}
\label{eq.13.2.1}
\end{equation}
In the dipole approximation, the effective electric charge and magnetic momentum (\ref{eq.13.1.8}) are transformed to
\begin{equation}
\begin{array}{lll}
  Z_{\rm eff}^{\rm (dip)} = \displaystyle\frac{m_{A}\, F_{p,\, {\rm el}} - m_{p}\, F_{A,\, {\rm el}} }{m_{p} + m_{A}}, &
  \vb{M}_{\rm eff}^{\rm (dip)} = \displaystyle\frac{m_{A}\, \vb{F}_{p,\, {\rm mag}}  - m_{p}\,\vb{F}_{A,\, {\rm mag}}}{m_{p} + m_{A}},
\end{array}
\label{eq.13.2.2}
\end{equation}
and matrix elements (\ref{eq.13.2.1}) are transformed to
[see Eqs.~(\ref{eq.app.matr_el.coh_mag.8}), (\ref{eq.app.matr_el.coh_mag.9}) in Appendix~\ref{sec.app.matr_el.coh_mag}, for calculations]
\begin{equation}
\begin{array}{lll}
  M_{p}^{(E,\, {\rm dip})} =
  i \hbar\, (2\pi)^{3}
  \displaystyle\frac{2\, \mu_{N}\,  m_{\rm p}}{\mu}\;
  Z_{\rm eff}^{\rm (dip)}\;
  \displaystyle\sum\limits_{\alpha=1,2} \vb{e}^{(\alpha)} \cdot \vb{I}_{1},
\end{array}
\label{eq.13.2.3}
\end{equation}
\begin{equation}
\begin{array}{lll}
  M_{p}^{(M,\, {\rm dip})} =
  \hbar\, (2\pi)^{3}\, \displaystyle\frac{\mu_{N}}{\mu} \cdot \alpha_{M} \cdot
  (\vb{e}_{\rm x} + \vb{e}_{\rm z})\,
  \displaystyle\sum\limits_{\alpha=1,2}
  \Bigl[ \vb{I}_{1} \times \vb{e}^{(\alpha)} \Bigr], &

  \alpha_{M} =
  \Bigl[
    Z_{\rm A} (\vb{k}_{\rm ph})\: m_{p}\, \bar{\mu}_{\rm pn}^{\rm (an)} -
    z_{\rm p}\, m_{A}\, \mu_{\rm p}^{\rm (an)}
  \Bigr] \cdot
  \displaystyle\frac{m_{p}}{m_{p} + m_{A}},
\end{array}
\label{eq.13.2.4}
\end{equation}
where
\begin{equation}
\begin{array}{lll}
  \vb{I}_{1} = \biggl\langle\: \Phi_{\rm p - nucl, f} (\vb{r})\;
  \biggl|\, e^{-i\, \vb{k}_{\rm ph} \vb{r}}\;
  \vb{\displaystyle\frac{d}{dr}} \biggr|\:
  \Phi_{\rm p - nucl, i} (\vb{r})\: \biggr\rangle
\end{array}
\label{eq.13.2.5}
\end{equation}
and
$\bar{\mu}_{\rm pn}^{\rm (an)} = \mu_{\rm p}^{\rm (an)} + \kappa\,\mu_{\rm n}^{\rm (an)}$,
$\kappa = (A-N)/N$, $A$ and $N$ are numbers of nucleons and neutrons in nucleus,
$\mu_{\rm p}^{\rm (an)}$ and $\mu_{\rm n}^{\rm (an)}$ are anomalous magnetic moments of proton and neutron
[measured in units of nuclear magneton $\mu_{N}$].

\subsection{Matrix element of incoherent emission of electric and magnetic types 
\label{sec.14}}

We calculate the matrix elements $M_{\Delta E}$ in Eq.~(\ref{eq.13.1.10}).
As functions $\vb{D}_{A 1,\, {\rm el}}$, $\vb{D}_{A 2,\, {\rm el}}$ do not depend on variable $\vb{r}$, we rewrite formula above as multiplication of two independent integrals as
\begin{equation}
\begin{array}{lll}
  M_{\Delta E} & = &
  -\, (2\pi)^{3}\; 2\, \mu_{N}
  \displaystyle\sum\limits_{\alpha=1,2} \vb{e}^{(\alpha)}\;
  \Bigl\{
  \Bigl\langle
    \Phi_{\rm p - nucl, f} (\vb{r})\; \Bigl|\, e^{i\, c_{p}\, \vb{k_{\rm ph}} \vb{r}}\, \Bigr|\, \Phi_{\rm p - nucl, i} (\vb{r})\:
  \Bigr\rangle\,
  \Bigl( \vb{D}_{A 1,\, {\rm el}} - \displaystyle\frac{m_{\rm p}}{m_{A}}\, \vb{D}_{A 2,\, {\rm el}} \Bigr)
  \Bigr\}.
\end{array}
\label{eq.14.1.1}
\end{equation}
%
Calculation of $\vb{D}_{A 1,\, {\rm el}}$ and $\vb{D}_{A 2,\, {\rm el}}$ are given in Appendix~\ref{sec.app.4}
[see Eqs.~(\ref{eq.app.3.1.3.6})].
For such solutions and taking into account $\vb{e}^{(\alpha)} \cdot \vb{k}_{\rm ph} = 0$ (as ortogonality of vectors $\vb{e}^{(\alpha)}$ and $\vb{k}_{\rm ph}$),
we obtain [see Eqs.~(\ref{eq.app.3.1.3.8})]:
\begin{equation}
\begin{array}{llllllll}
  \vb{e}^{(\alpha)} \cdot \vb{D}_{A 1,\, {\rm el}} = 0, &
  \vb{e}^{(\alpha)} \cdot \vb{D}_{A 2,\, {\rm el}} = 0, &
  M_{\Delta E} = 0.
\end{array}
\label{eq.14.1.7}
\end{equation}


We will consider the matrix element $M_{\Delta M}$ in Eq.~(\ref{eq.13.1.11}).
As functions $D_{A 1,\, {\rm mag}}$, $D_{A 2,\, {\rm mag}}$ do not depend on variable $\vb{r}$, we rewrite formula above as multiplication of two independent integrals as
\begin{equation}
\begin{array}{lll}
\vspace{1.0mm}
  M_{\Delta M} & = &
  -\, i\, (2\pi)^{3}\: \mu_{N}
  \displaystyle\sum\limits_{\alpha=1,2}\;
  \Bigl\{
  \Bigl\langle
    \Phi_{\rm p - nucl, f} (\vb{r})\; \Bigl|\, e^{i\, c_{p}\, \vb{k_{\rm ph}} \vb{r}}\, \Bigr|\, \Phi_{\rm p - nucl, i} (\vb{r})\:
  \Bigr\rangle\,
  \Bigl( D_{A 1,\, {\rm mag}} (\vb{e}^{(\alpha)}) - D_{A 2,\, {\rm mag}} (\vb{e}^{(\alpha)}) \Bigr)
  \Bigr\}.
\end{array}
\label{eq.15.1.1}
\end{equation}
%
After summation over spin states, for even number of spin states we obtain:
\begin{equation}
\begin{array}{lll}
  D_{A 2,\, {\rm mag}} (\vb{e}^{(\alpha)}) = 0
\end{array}
\label{eq.15.1.2}
\end{equation}
and the matrix element (\ref{eq.15.1.1}) is simplified as
\begin{equation}
\begin{array}{lll}
\vspace{1.0mm}
  M_{\Delta M} & = &
  -\, i\, (2\pi)^{3}\, \mu_{N}
  \displaystyle\sum\limits_{\alpha=1,2}\;
  \Bigl\{
  \Bigl\langle
    \Phi_{\rm p - nucl, f} (\vb{r})\; \Bigl|\, e^{i\, c_{p}\, \vb{k_{\rm ph}} \vb{r}}\, \Bigr|\, \Phi_{\rm p - nucl, i} (\vb{r})\:
  \Bigr\rangle \cdot
  D_{A 1,\, {\rm mag}} (\vb{e}^{(\alpha)})
  \Bigr\}.
\end{array}
\label{eq.15.1.3}
\end{equation}

We calculate summation of form factors for even-even nuclei [see Eq.~(\ref{eq.app.matr_el.mag.1.4.7})]:
\begin{equation}
\begin{array}{lll}
  \displaystyle\sum\limits_{\alpha=1,2} D_{A 1,\, {\rm mag}} (\vb{e}^{(\alpha)}) & = &
  -\, \displaystyle\frac{\hbar\, (A-1)}{2\,A}\; \bar{\mu}_{\rm pn}^{\rm (an)}\, k_{\rm ph} \cdot Z_{\rm A} (\vb{k}_{\rm ph}),
\end{array}
\label{eq.15.1.4}
\end{equation}
where $\bar{\mu}_{\rm pn}^{\rm (an)} = \mu_{\rm n}^{\rm (an)} + \kappa\,\mu_{\rm n}^{\rm (an)}$,
$\kappa = (A-N)/N$,
$A$ and $N$ are numbers of nucleons and neutrons in nucleus.
Taking solution
\begin{equation}
\begin{array}{lll}
  \vb{D}_{A 1,\, {\rm el}} = &
    \displaystyle\sum\limits_{j=1}^{A-1}
      \displaystyle\frac{z_{j} m_{\rm p}}{m_{Aj}}\,
      \Bigl\langle \psi_{A, f} (\beta_{A})\, \Bigl|\,
        e^{-i \vb{k}_{\rm ph} \rhobf_{Aj}} \vb{\tilde{p}}_{Aj}
      \Bigr|\,  \psi_{A, j} (\beta_{A}) \Bigr\rangle =
  \displaystyle\frac{\hbar}{2}\; \displaystyle\frac{A-1}{A}\; \vb{k}_{\rm ph}\; Z_{\rm A} (\vb{k}_{\rm ph})
\end{array}
\label{eq.15.1.5}
\end{equation}
%
into account [see Eqs.~(\ref{eq.app.3.1.3.6})], we find final solution for the matrix element (\ref{eq.15.1.3})
[assuming $\bar{s}_{k} = s_{k}$]:
\begin{equation}
\begin{array}{lll}
  M_{\Delta M} & = &
  i\, \hbar\, (2\pi)^{3}\, \mu_{N}\, f_{1} \cdot |\vb{k}_{\rm ph}| \cdot Z_{\rm A} (\vb{k}_{\rm ph}) \cdot I_{2},
\end{array}
\label{eq.15.1.6}
\end{equation}
where
\begin{equation}
\begin{array}{lll}
  f_{1} = \displaystyle\frac{A-1}{2A}\: \bar{\mu}_{\rm pn}^{\rm (an)}, &
  I_{2} =
  \Bigl\langle \Phi_{\rm p - nucl, f} (\vb{r})\; \Bigl|\, e^{i\, c_{p}\, \vb{k_{\rm ph}} \vb{r}}\, \Bigr|\, \Phi_{\rm p - nucl, i} (\vb{r})\: \Bigr\rangle.
\end{array}
\label{eq.15.1.7}
\end{equation}

\subsection{Matrix element of incoherent emission of magnetic type $M_{k}$
\label{sec.16}}

We will consider the matrix element $M_{k}$ in Eq.~(\ref{eq.13.1.5}),
where form factors are calculated in Appendix~\ref{sec.app.matr_el.k.1} [see Eqs.~(\ref{eq.app.matr_el.k.1.4.2}), (\ref{eq.app.matr_el.k.1.4.6}) and have form:
\begin{equation}
\begin{array}{lcl}
  \vb{D}_{A,\, {\rm k}} = \bar{\mu}_{\rm pn}^{\rm (an)}\; (\vb{e}_{\rm x} + \vb{e}_{\rm z}) \cdot Z_{\rm A} (\vb{k}_{\rm ph}), &
  \vb{D}_{p,\, {\rm k}} = \mu_{\rm p}^{\rm (an)}\; (\vb{e}_{\rm x} + \vb{e}_{\rm z}) \cdot z_{\rm p}.
\end{array}
\label{eq.16.1.1}
\end{equation}
As functions $\vb{D}_{A,\, {\rm k}}$ and $\vb{D}_{p,\, {\rm k}}$ do not depend on variable $\vb{r}$, we rewrite formula~(\ref{eq.13.1.5}) as multiplication of two independent integrals as
\begin{equation}
\begin{array}{lcl}
\vspace{0.5mm}
  M_{k} & = &
  i\, \hbar\, (2\pi)^{3}\, \mu_{N}
  \displaystyle\sum\limits_{\alpha=1,2}
    \bigl[ \vb{k_{\rm ph}} \cp \vb{e}^{(\alpha)} \bigr]\; \times \\

  & \times &
  \Bigl\{
    \vb{D}_{p,\, {\rm k}}\; \Bigl\langle \Phi_{\rm p - nucl, f} (\vb{r})\; \Bigl|\, e^{-i\, c_{A}\, \vb{k_{\rm ph}} \vb{r}}\; \Bigl|\, \Phi_{\rm p - nucl, i} (\vb{r})\; \Bigr\rangle +
    \vb{D}_{A,\, {\rm k}}\; \Bigl\langle \Phi_{\rm p - nucl, f} (\vb{r})\; \Bigl|\, e^{i\, c_{p}\, \vb{k_{\rm ph}} \vb{r}}\; \Bigl|\, \Phi_{\rm p - nucl, i} (\vb{r})\; \Bigr\rangle
  \Bigr\}.
\end{array}
\label{eq.16.1.2}
\end{equation}
%
Summation of these form factors over polarizations is calculated in Appendix~\ref{sec.app.matr_el.k.1} [see Eq.~(\ref{eq.app.matr_el.k.1.4.5})]
\begin{equation}
\begin{array}{lll}
  \displaystyle\sum\limits_{\alpha=1,2} \bigl[ \vb{k_{\rm ph}} \cp \vb{e}^{(\alpha)} \bigr] \cdot \vb{D}_{p,\, {\rm k}} =
  - k_{\rm ph}\; \mu_{\rm p}^{\rm (an)}\; z_{\rm p}, &

  \displaystyle\sum\limits_{\alpha=1,2} \bigl[ \vb{k_{\rm ph}} \cp \vb{e}^{(\alpha)} \bigr] \cdot \vb{D}_{A,\, {\rm k}} =
  - k_{\rm ph}\; \bar{\mu}_{\rm pn}^{\rm (an)}\; Z_{\rm A} (\vb{k}_{\rm ph}).
\end{array}
\label{eq.16.1.3}
\end{equation}
We substitute these solutions to Eq.~(\ref{eq.16.1.2}) and obtain
\begin{equation}
\begin{array}{lcl}
\vspace{1.5mm}
  M_{k} & = &
  -\, i\, \hbar\, (2\pi)^{3}\, \mu_{N} \cdot k_{\rm ph}\; \times \\

  & \times &
  \Bigl\{
    z_{\rm p}\; \mu_{\rm p}^{\rm (an)}\:
      \Bigl\langle \Phi_{\rm p - nucl, f} (\vb{r})\; \Bigl|\, e^{-i\, c_{A}\, \vb{k_{\rm ph}} \vb{r}}\; \Bigl|\, \Phi_{\rm p - nucl, i} (\vb{r})\; \Bigr\rangle +
    Z_{\rm A} (\vb{k}_{\rm ph})\; \bar{\mu}_{\rm pn}^{\rm (an)}\:
      \Bigl\langle \Phi_{\rm p - nucl, f} (\vb{r})\; \Bigl|\, e^{i\, c_{p}\, \vb{k_{\rm ph}} \vb{r}}\; \Bigl|\, \Phi_{\rm p - nucl, i} (\vb{r})\; \Bigr\rangle
  \Bigr\}.
\end{array}
\label{eq.16.1.4}
\end{equation}
Taking solution (\ref{eq.15.1.6})--(\ref{eq.15.1.7}) for the incoherent magnetic emission, we obtain
\begin{equation}
  M_{k} =
  -\, i\, \hbar\, (2\pi)^{3}\, \mu_{N} \cdot k_{\rm ph}\, z_{\rm p}\: \mu_{\rm p}^{\rm (an)} \cdot I_{3} -
  \displaystyle\frac{\bar{\mu}_{\rm pn}^{\rm (an)}}{f_{1}} \cdot M_{\Delta M},
\label{eq.16.1.6}
\end{equation}
where
\begin{equation}
\begin{array}{lllll}
  I_{3} = \Bigl\langle \Phi_{\rm p - nucl, f} (\vb{r})\; \Bigl|\, e^{-i\, c_{A}\, \vb{k_{\rm ph}} \vb{r}}\, \Bigr|\, \Phi_{\rm p - nucl, i} (\vb{r})\: \Bigr\rangle.
\end{array}
\label{eq.16.1.7}
\end{equation}
Also we have
\begin{equation}
\begin{array}{lll}
  \displaystyle\frac{\bar{\mu}_{\rm pn}^{\rm (an)}}{f_{1}} =
  \displaystyle\frac{\bar{\mu}_{\rm pn}^{\rm (an)}}{\displaystyle\frac{A-1}{2A}\: \bar{\mu}_{\rm pn}^{\rm (an)}} =
  \displaystyle\frac{2\, A}{A-1}.
\end{array}
\label{eq.16.1.8}
\end{equation}


\section{Matrix element of emission of bremsstrahlung photons with form factors
\label{sec.app.2}}

Now we define the matrix element of emission, related with form factors.
%
In Sec.~\ref{sec.11} we found operator of emission with form factors.
According to Eqs.~(\ref{eq.11.1.6}), (\ref{eq.11.1.7}), (\ref{eq.11.1.3}) (see p.~\pageref{eq.11.1.6}), we have
\begin{equation}
\begin{array}{llllll}
  \hat{H}_{\gamma 1} =
  \displaystyle\sum\limits_{i=1}^{A+1} \bar{h}_{\gamma 1} (\vb{r}_{i}) =
  \bar{h}_{\gamma 1}^{\rm (proton)} (\vb{r}_{i}) + \bar{h}_{\gamma 1}^{\rm (nucleus)} (\vb{r}_{i}),
\end{array}
\label{eq.17.1.1}
\end{equation}
where
\vspace{0.5mm}
\begin{equation}
\begin{array}{llllll}
\vspace{1.5mm}
  \bar{h}_{\gamma 1}^{\rm (proton)} (\vb{r}_{i}) =
  e^{-i\, \vb{k}_{\rm ph}\, \vb{R}}\, e^{-i\, \vb{k}_{\rm ph}\, c_{A}\, \vb{r}}\, S_{1p} +
  e^{-2i\, \vb{k}_{\rm ph}\, \vb{R}}\, e^{-2i\, \vb{k}_{\rm ph}\, c_{A}\, \vb{r}}\, S_{2p}, \\

  \bar{h}_{\gamma 1}^{\rm (nucleus)} (\vb{r}_{i}) =
  e^{-i\, \vb{k}_{\rm ph}\, \vb{R}}
    e^{i\, \vb{k}_{\rm ph} c_{\rm p}\, \vb{r}} \displaystyle\sum\limits_{j=1}^{A} S_{1j}\, e^{-i\, \vb{k}_{\rm ph} \rhobf_{A j}} +
  e^{-2i\, \vb{k}_{\rm ph}\, \vb{R}}
    e^{2i\, \vb{k}_{\rm ph} c_{\rm p}\, \vb{r}} \displaystyle\sum\limits_{j=1}^{A} S_{2j}\, e^{-2i\, \vb{k}_{\rm ph} \rhobf_{A j}},
\end{array}
\label{eq.17.1.2}
\end{equation}
%
%
\vspace{0.5mm}
\begin{equation}
\begin{array}{llllll}
\vspace{1.0mm}
  S_{1i} =
    \displaystyle\frac{z_{i}e\, F_{2i}}{f_{i}} \cdot
  \sqrt{\displaystyle\frac{\pi\hbar c^{2}}{w_{\rm ph}}}\;
  \Bigl\{
    2 Q\, \sin \varphi_{ph,i}\, \Bigl[ -\, i\, b_{1i} + \displaystyle\frac{F_{2i}\, Q^{2}}{m_{i}c}\, \bigl( 2F_{1i}^{2} - F_{2i}^{2}\, Q^{2} \bigr) \Bigr] -
    i\, \sqrt{2}\, b_{2i}\, \varepsilon_{mjl}\, q^{j}\, \sigma_{l} \sum\limits_{\alpha=1,2} \mathbf{e}_{m}^{(\alpha),\,*}
  \Bigr\}, \\

  S_{2i} =
    \displaystyle\frac{4z_{i}^{2}e^{2}\, F_{2i}}{f_{i}\,m_{i}} \cdot
    \displaystyle\frac{\pi\hbar}{w_{\rm ph}} \cdot
    F_{2i}Q^{2}\, \bigl( 2F_{1i}^{2} - F_{2i}^{2}\, Q^{2} \bigr)\;
    \sin^{2} \varphi_{ph,i}.
\end{array}
\label{eq.17.1.3}
\end{equation}
Here, functions $S_{1i}$ and $S_{2i}$ do not depend on internal coordinates (they depend on relative distance $\vb{r}$).

We defined the wave function of the full nuclear system, which after ignoring correction $\Delta \Psi$ obtains form:
\begin{equation}
\begin{array}{lcl}
  \Psi = \Phi (\vb{R}) \cdot F (\vb{r}, \beta_{A}, \beta_{\rm p}), &
  F (\vb{r}, \beta_{A}, \beta_{\rm p}) =
  \Phi_{\rm p - nucl} (\vb{r}) \cdot
  \psi_{\rm nucl} (\beta_{A}) \cdot
  \psi_{\rm p} (\beta_{\rm p}),
\end{array}
\label{eq.17.1.4}
\end{equation}
Here, $\Phi (\vb{R})$ is wave function describing evolution of center of mass of the full nuclear system.
We shall assume approximated form for the function $\Phi_{\bar{s}}$ before and after emission of photons as
\begin{equation}
  \Phi_{\bar{s}} (\vb{R}) =  e^{-i\,\vb{K}_{\bar{s}}\cdot\vb{R}},
\label{eq.17.1.5}
\end{equation}
where $\bar{s} = i$ or $f$ (indexes $i$ and $f$ denote the initial state,
i.e. the state before emission of photon,
and the final state, i.e. the state after emission of photon),
$\vb{K}_{s}$ is momentum of the total system. 
%

Let us calculate the matrix element $M_{\rm form factor}$ on the basis of operator (\ref{eq.17.1.1}).
Taking into account Eqs.~(\ref{eq.13.1.1})--(\ref{eq.13.1.2}) [\pageref{eq.13.1.1}], we obtain it as
\begin{equation}
\begin{array}{lll}
\vspace{0.5mm}
  M_{\rm form factor} & = &
  (2\pi)^{3}\; \sqrt{\displaystyle\frac{\hbar w_{\rm ph}}{2\pi\, c^{2}}}\, 

  \Bigl[
  \delta (\vb{K}_{f} - \vb{K}_{i} - \vb{k}_{\rm ph})\;
  \biggl\langle F_{f}\, \biggl|\,
    e^{-i\, \vb{k}_{\rm ph}\, c_{A}\, \vb{r}}\, S_{1p} +
    e^{i\, \vb{k}_{\rm ph} c_{\rm p}\, \vb{r}} \displaystyle\sum\limits_{j=1}^{A} S_{1j}\, e^{-i\, \vb{k}_{\rm ph} \rhobf_{A j}}
  \biggr|\, F_{i}\, \biggr\rangle\; + \\

  & + &
  \delta (\vb{K}_{f} - \vb{K}_{i} - 2\vb{k}_{\rm ph})\;
  \biggl\langle F_{f}\, \biggl|\,
    e^{-2i\, \vb{k}_{\rm ph}\, c_{A}\, \vb{r}}\, S_{2p} +
    e^{2i\, \vb{k}_{\rm ph} c_{\rm p}\, \vb{r}} \displaystyle\sum\limits_{j=1}^{A} S_{2j}\, e^{-2i\, \vb{k}_{\rm ph} \rhobf_{A j}}\;
  \biggr|\, F_{i}\, \biggr\rangle
  \Bigr].
\end{array}
\label{eq.17.1.9}
\end{equation}
In this formula we have integration over space variables $\vb{r}$, $\rhobf_{Aj}$ ($i = 1 \ldots n-1$, $j = 1 \ldots A$).

\subsection{Integration over momentum $\vb{K}$
\label{sec.17.2}}

We will calculate cross-sections of emission of photons, not dependent on vector $\vb{K}_{f}$ (i.e., momentum of the full nuclear system after emission of photon in the laboratory frame). Therefore, we have to average all matrix elements over all degrees of freedom related with $\mathbf{K}_{f}$, i.e. we integrate these matrix elements over $\vb{K}_{f}$.
We integrate matrix element (\ref{eq.17.1.9}) over momentum $\vb{K}_{f}$ as
\begin{equation}
\begin{array}{lll}
\vspace{0.5mm}
  M_{\rm form factor} & = &
  (2\pi)^{3}\; \sqrt{\displaystyle\frac{\hbar w_{\rm ph}}{2\pi\, c^{2}}}\, 

  \Bigl[
  \biggl\langle F_{f}\, \biggl|\,
    e^{-i\, \vb{k}_{\rm ph}\, c_{A}\, \vb{r}}\, S_{1p} +
    e^{i\, \vb{k}_{\rm ph} c_{\rm p}\, \vb{r}} \displaystyle\sum\limits_{j=1}^{A} S_{1j}\, e^{-i\, \vb{k}_{\rm ph} \rhobf_{A j}}
  \biggr|\, F_{i}\, \biggr\rangle_{\vb{K}_{i} = \vb{K}_{f} + \vb{k}_{\rm ph}}\; + \\

  & + &
  \biggl\langle F_{f}\, \biggl|\,
    e^{-2i\, \vb{k}_{\rm ph}\, c_{A}\, \vb{r}}\, S_{2p} +
    e^{2i\, \vb{k}_{\rm ph} c_{\rm p}\, \vb{r}} \displaystyle\sum\limits_{j=1}^{A} S_{2j}\, e^{-2i\, \vb{k}_{\rm ph} \rhobf_{A j}}\;
  \biggr|\, F_{i}\, \biggr\rangle_{\vb{K}_{i} = \vb{K}_{f} + 2\vb{k}_{\rm ph}}
  \Bigr].
\end{array}
\label{eq.17.2.1}
\end{equation}
We group terms with internal integration over internal variables as
\begin{equation}
\begin{array}{lll}
\vspace{0.5mm}
  M_{\rm form factor} & = &  M_{\rm form factor}^{(1)} + M_{\rm form factor}^{(2)}, \\

\vspace{0.5mm}
  M_{\rm form factor}^{(1)} & = &
  (2\pi)^{3}\; \sqrt{\displaystyle\frac{\hbar w_{\rm ph}}{2\pi\, c^{2}}}\,

  \displaystyle\int\limits_{}^{}
    \Phi_{\rm p - nucl, f}^{*} (\vb{r})\;
    \Bigl[
      e^{-i\, \vb{k}_{\rm ph}\, c_{A}\, \vb{r}}\, S_{1p} +
      e^{i\, \vb{k}_{\rm ph} c_{\rm p}\, \vb{r}} \displaystyle\sum\limits_{j=1}^{A} S_{1j}\, e^{-i\, \vb{k}_{\rm ph} \rhobf_{A j}}
    \Bigr]\,
    \Phi_{\rm p - nucl, i} (\vb{r})\; \vb{dr}\, \biggl|_{\vb{K}_{i} = \vb{K}_{f} + \vb{k}_{\rm ph}}, \\

  M_{\rm form factor}^{(2)} & = &
  (2\pi)^{3}\; \sqrt{\displaystyle\frac{\hbar w_{\rm ph}}{2\pi\, c^{2}}}\,
  \displaystyle\int\limits_{}^{}
    \Phi_{\rm p - nucl, f}^{*} (\vb{r})\;
    \Bigl[
      e^{-i\, 2\vb{k}_{\rm ph}\, c_{A}\, \vb{r}}\, S_{2p} +
      e^{i\, 2\vb{k}_{\rm ph} c_{\rm p}\, \vb{r}} \displaystyle\sum\limits_{j=1}^{A} S_{2j}\, e^{-i\, 2\vb{k}_{\rm ph} \rhobf_{A j}}
    \Bigr]\,
    \Phi_{\rm p - nucl, i} (\vb{r})\; \vb{dr}\, \biggl|_{\vb{K}_{i} = \vb{K}_{f} + 2\vb{k}_{\rm ph}}.
\end{array}
\label{eq.17.2.2}
\end{equation}

At first, we shall calculate $ M_{\rm form factor}^{(1)}$.
We use explicit form (\ref{eq.17.1.3}) of $S_{1i}$ as it dependent on $\vb{r}$ (inside functions $b_{1i}$, $b_{2i}$),
substitute it to Eq.~(\ref{eq.17.2.2}) and obtain:
\begin{equation}
\begin{array}{lll}
  M_{\rm form factor}^{(1)} =
  (2\pi)^{3}\: \displaystyle\frac{\hbar\, e}{\sqrt{2}}\;
  \Bigl\{
    \displaystyle\frac{z_{p}\, F_{2p}}{f_{p}}\: \bigl( p_{\rm q,1}^{\rm (p)} + p_{\rm q,2}^{\rm (p)} \bigr) +

  \displaystyle\sum\limits_{i=1}^{A}
    \displaystyle\frac{z_{i}\, F_{2i}}{f_{i}}\:
    \bigl( p_{\rm q,1}^{\rm (Ai)} + p_{\rm q,2}^{\rm (Ai)} \bigr)
  \Bigr\}\Bigl|_{\vb{K}_{i} = \vb{K}_{f} + \vb{k}_{\rm ph}},
\end{array}
\label{eq.17.2.5}
\end{equation}
where
\begin{equation}
\begin{array}{lll}
\vspace{0.5mm}
  p_{\rm q,1}^{\rm (p)} & = &
    2 Q\, \sin \varphi_{ph,p}\,
    \Bigl\langle F_{f}\, \Bigl|\,
      e^{-i\, \vb{k}_{\rm ph}\, c_{A}\, \vb{r}}\;
      \Bigl[ -\, i\, b_{1p} + \displaystyle\frac{F_{2p}\, Q^{2}}{m_{p}c}\, \bigl( 2F_{1p}^{2} - F_{2p}^{2}\, Q^{2} \bigr) \Bigr]
    \Bigr|\, F_{i}\, \Bigr\rangle, \\

\vspace{0.5mm}
  p_{\rm q,2}^{\rm (p)}  & = & -
    i\, \sqrt{2}\, \varepsilon_{mjl}\, q^{j}\, \sigma_{l} \sum\limits_{\alpha=1,2} \mathbf{e}_{m}^{(\alpha)}
    \Bigl\langle F_{f}\, \Bigl|\,
      e^{-i\, \vb{k}_{\rm ph}\, c_{A}\, \vb{r}}\; b_{2p}\,
    \Bigr|\, F_{i}\, \Bigr\rangle
  \Bigr], \\

\vspace{0.5mm}
  p_{\rm q,1}^{\rm (Ai)} & = &
    2 Q\, \sin \varphi_{ph,i}\,
    \Bigl\langle F_{f}\, \Bigl|\,
      e^{i\, \vb{k}_{\rm ph}\, c_{\rm p}\, \vb{r}}\;
      \Bigl[ -\, i\, b_{1i} + \displaystyle\frac{F_{2i}\, Q^{2}}{m_{i}c}\, \bigl( 2F_{1i}^{2} - F_{2i}^{2}\, Q^{2} \bigr) \Bigr]
      e^{-i\, \vb{k}_{\rm ph} \rhobf_{A i}}
    \Bigr|\, F_{i}\, \Bigr\rangle, \\

  p_{\rm q,2}^{\rm (Ai)}  & = & -
    i\, \sqrt{2}\, \varepsilon_{mjl}\, q^{j}\, \sigma_{l} \sum\limits_{\alpha=1,2} \mathbf{e}_{m}^{(\alpha)}
    \Bigl\langle F_{f}\, \Bigl|\,
      e^{i\, \vb{k}_{\rm ph}\, c_{\rm p}\, \vb{r}}\; b_{2i}\,
      e^{-i\, \vb{k}_{\rm ph} \rhobf_{A i}}
    \Bigr|\, F_{i}\, \Bigr\rangle
  \Bigr].
\end{array}
\label{eq.17.2.6}
\end{equation}


Now we calculate the second term $M_{\rm form factor}^{(2)}$.
We substitute $S_{2i}$ in form~(\ref{eq.17.1.3}) to Eq.~(\ref{eq.17.2.2})
and obtain ($\vb{K}_{i} = \vb{K}_{f} + 2\vb{k}_{\rm ph}$):
\begin{equation}
\begin{array}{lll}
\vspace{0.5mm}
  M_{\rm form factor}^{(2)} & = &
  (2\pi)^{3}\; \sqrt{\displaystyle\frac{\hbar w_{\rm ph}}{2\pi\, c^{2}}}\,
  \displaystyle\frac{\pi\hbar}{w_{\rm ph}}\, 4e^{2}\,
  \Bigl\{
    \displaystyle\frac{z_{p}^{2}\, F_{2p}}{f_{p}\,m_{p}}\:
    p_{q,3}^{\rm (p)} +
    \displaystyle\sum\limits_{j=1}^{A}
    \displaystyle\frac{z_{i}^{2}\, F_{2i}}{f_{i}\,m_{i}}\,
    p_{q,3}^{\rm (Ai)}
  \Bigr\},
\end{array}
\label{eq.17.3.2}
\end{equation}
where
\begin{equation}
\begin{array}{lll}
\vspace{0.5mm}
  p_{q,3}^{\rm (p)} & = &
     F_{2p}Q^{2}\, \bigl( 2F_{1p}^{2} - F_{2p}^{2}\, Q^{2} \bigr)\; \sin^{2} \varphi_{ph,p}
    \Bigl\langle F_{f}\, \Bigl|\,
      e^{-i\, 2\vb{k}_{\rm ph}\, c_{A}\, \vb{r}}\,
    \Bigr|\, F_{i}\, \Bigr\rangle, \\

  p_{q,3}^{\rm (Ai)} & = &
    F_{2i}Q^{2}\, \bigl( 2F_{1i}^{2} - F_{2i}^{2}\, Q^{2} \bigr)\; \sin^{2} \varphi_{ph,i}\,
    \Bigl\langle F_{f}\, \Bigl|\,
      e^{i\, 2\vb{k}_{\rm ph} c_{\rm p}\, \vb{r}}
      e^{-i\, 2\vb{k}_{\rm ph} \rhobf_{A j}}
    \Bigr|\, F_{i}\, \Bigr\rangle.
\end{array}
\label{eq.17.3.3}
\end{equation}

\subsection{Calculations of $p_{\rm q,1}$, $p_{\rm q,2}$ at elastic scattering of virtual photons between nucleons of nucleus and scattered proton
\label{sec.17.4}}

Coefficients $b_{1}$, $b_{2}$ and $b_{3}$ are defined in Eqs.~(A27) in Ref.~\cite{Maydanyuk_Zhang_Zou.2019.PRC.microscopy}
[here, we use them for each nucleon with number $i$].
At $A_{0} = 0$ and elastic scattering of virtual photon
[i.e., at $q_{4} = 0$, $\vb{q}^{2} = - q^{2} = Q^{2}$, see Eqs.~(\ref{eq.6.1.2})]:
\begin{equation}
\begin{array}{lll}
\vspace{0.9mm}
  b_{1} & = &
  F_{1}^{2} (1 - F_{1}) -
  F_{1} F_{2}^{2}\, Q^{2} -
  \displaystyle\frac{2\,F_{1}^{3}}{mc^{2}}\, V(\mathbf{r}), \\

  b_{2} & = &
    i\, \Bigl[
      2\, F_{1}^{2} -
      2 F_{2}^{2}\, Q^{2} -
      \displaystyle\frac{2\,F_{1}}{mc^{2}} \bigl( F_{1}^{2} - F_{2}^{2} Q^{2} \bigr)\, V(\mathbf{r})
    \Bigr].
\end{array}
\label{eq.17.4.2}
\end{equation}
Now we calculate the matrix elements as
\begin{equation}
\begin{array}{lll}
\vspace{0.3mm}
  p_{\rm q,1}^{\rm (p)} & = &
    2 Q\, \sin \varphi_{ph,p}\,
    \Bigl\{
      -\, i\, \Bigl[F_{1p}^{2} (1 - F_{1p}) - F_{1p} F_{p2}^{2}\, Q^{2} \Bigr] +
      \displaystyle\frac{F_{2p}\, Q^{2}}{m_{\rm p}c}\, \bigl( 2F_{1p}^{2} - F_{2p}^{2}\, Q^{2} \bigr)
    \Bigr\}
    \Bigl\langle F_{f}\, \Bigl|\, e^{-i\, \vb{k}_{\rm ph}\, c_{A}\, \vb{r}}\: \Bigr|\, F_{i}\, \Bigr\rangle\; + \\
\vspace{1.5mm}
  & + &
    2 Q\, \sin \varphi_{ph,p}\,
    i\, \displaystyle\frac{2\,F_{1p}^{3}}{m_{\rm p}c^{2}}\,
    \Bigl\langle F_{f}\, \Bigl|\, e^{-i\, \vb{k}_{\rm ph}\, c_{A}\, \vb{r}}\: V(\vb{r}_{\rm p}) \Bigr|\, F_{i}\, \Bigr\rangle, \\

\vspace{0.3mm}
  p_{\rm q,2}^{\rm (p)} & = &
    \sqrt{2}\, \varepsilon_{mjl}\, q^{j}\, \sigma_{l}\:
    \Bigl[ 2\, F_{1}^{2} - 2 F_{2}^{2}\, Q^{2} \Bigr]
    \displaystyle\sum\limits_{\alpha=1,2} \mathbf{e}_{m}^{(\alpha)}
    \Bigl\langle F_{f}\, \Bigl|\,
      e^{-i\, \vb{k}_{\rm ph}\, c_{A}\, \vb{r}}\:
    \Bigr|\, F_{i}\, \Bigr\rangle\; - \\
\vspace{1.5mm}
  & - &
    \sqrt{2}\, \varepsilon_{mjl}\, q^{j}\, \sigma_{l}
    \displaystyle\frac{2\,F_{1}}{m_{\rm p}c^{2}} \bigl( F_{1}^{2} - F_{2}^{2} Q^{2} \bigr)\,
    \displaystyle\sum\limits_{\alpha=1,2} \mathbf{e}_{m}^{(\alpha)}
    \Bigl\langle F_{f}\, \Bigl|\, e^{-i\, \vb{k}_{\rm ph}\, c_{A}\, \vb{r}}\: V(\vb{r}_{\rm p}) \Bigr|\, F_{i}\, \Bigr\rangle, \\

\vspace{0.3mm}
  p_{\rm q,1}^{\rm (Ai)} & = &
    2 Q\, \sin \varphi_{ph,i}\,
    \Bigl\{
      -\, i\, \Bigl[F_{1i}^{2} (1 - F_{1i}) - F_{1i} F_{2i}^{2}\, Q^{2} \Bigr] +
      \displaystyle\frac{F_{2i}\, Q^{2}}{m_{i}c}\, \bigl( 2F_{1i}^{2} - F_{2i}^{2}\, Q^{2} \bigr)
    \Bigr\}
    \Bigl\langle F_{f}\, \Bigl|\,
      e^{i\, \vb{k}_{\rm ph}\, c_{\rm p}\, \vb{r}}\:
      e^{-i\, \vb{k}_{\rm ph} \rhobf_{A i}}
    \Bigr|\, F_{i}\, \Bigr\rangle\; + \\
\vspace{1.5mm}
  & + &
    2 Q\, \sin \varphi_{ph,i}\,
    i\, \displaystyle\frac{2\,F_{1i}^{3}}{m_{i}c^{2}}\,
    \Bigl\langle F_{f}\, \Bigl|\,
      e^{i\, \vb{k}_{\rm ph}\, c_{\rm p}\, \vb{r}}\:
      e^{-i\, \vb{k}_{\rm ph} \rhobf_{A i}}
      V(\vb{r}_{i})
    \Bigr|\, F_{i}\, \Bigr\rangle, \\

\vspace{0.1mm}
  p_{\rm q,2}^{\rm (Ai)} & = &
    \sqrt{2}\, \varepsilon_{mjl}\, q^{j}\, \sigma_{l}\:
    \Bigl[ 2\, F_{1i}^{2} - 2 F_{2i}^{2}\, Q^{2} \Bigr]
    \displaystyle\sum\limits_{\alpha=1,2} \mathbf{e}_{m}^{(\alpha)}
    \Bigl\langle F_{f}\, \Bigl|\,
      e^{i\, \vb{k}_{\rm ph} c_{\rm p}\, \vb{r}}\:
      e^{-i\, \vb{k}_{\rm ph} \rhobf_{A i}}
    \Bigr|\, F_{i}\, \Bigr\rangle\; - \\
  & - &
    \sqrt{2}\, \varepsilon_{mjl}\, q^{j}\, \sigma_{l}
    \displaystyle\frac{2\,F_{1i}}{m_{i}c^{2}} \bigl( F_{1i}^{2} - F_{2i}^{2} Q^{2} \bigr)
    \displaystyle\sum\limits_{\alpha=1,2} \mathbf{e}_{m}^{(\alpha)}
    \Bigl\langle F_{f}\, \Bigl|\,
      e^{i\, \vb{k}_{\rm ph} c_{\rm p}\, \vb{r}}\:
      e^{-i\, \vb{k}_{\rm ph} \rhobf_{A i}}
      V(\vb{r}_{i})
    \Bigr|\, F_{i}\, \Bigr\rangle.
\end{array}
\label{eq.17.4.4}
\end{equation}

\subsection{Averaging over polarization of virtual photons
\label{sec.17.5}}

We assume that there is no way to fix direction of polarization of the virtual photons (concerning to vectors of polarization of the bremsstrahlung photons) experimentally.
So, we have to integrate the matrix elements $p_{\rm q,i}$ over all such a possible directions (i.e. we integrate over angle $\varphi_{\rm ph}$, $i = 1,2,3$):
\begin{equation}
\begin{array}{lcl}
  \tilde{p}_{\rm q,i} = N \cdot \displaystyle\int\limits_{0}^{\pi} p_{\rm q,i}\; d \varphi_{\rm ph}, &
  N = \displaystyle\frac{1}{\pi}.
\end{array}
\label{eq.17.5.1}
\end{equation}
Taking into account that
\begin{equation}
\begin{array}{lcl}
  \displaystyle\int\limits_{0}^{\pi} \sin\varphi_{\rm ph}\; d \varphi_{\rm ph} =
  2, &

  \displaystyle\int\limits_{0}^{\pi} d \varphi_{\rm ph} =
  \pi, &

  \displaystyle\int\limits_{0}^{\pi} \sin^{2}\varphi_{\rm ph}\; d \varphi_{\rm ph} =
  \displaystyle\frac{\pi}{2},
\end{array}
\label{eq.17.5.2}
\end{equation}
from Eqs.~(\ref{eq.17.4.4}) and (\ref{eq.17.3.3}) we obtain:
\begin{equation}
\begin{array}{lcl}
\vspace{1.3mm}
  \tilde{p}_{\rm q,1}^{\rm (p)} & = &
    A_{1}^{\rm (p)} (Q, F_{1}, F_{2})\, \Bigl< k_{f} \Bigl|\, e^{-i\, \vb{k_{\rm ph}} c_{A} \vb{r}}\; \Bigr| \,k_{i} \Bigr> +
    B_{1}^{\rm (p)} (Q, F_{1}, F_{2})\, \Bigl< k_{f} \Bigl|\, e^{-i\, \vb{k_{\rm ph}} c_{A} \vb{r}}\; V(\vb{r}_{\rm p}) \Bigr| \,k_{i} \Bigr>, \\

\vspace{1.3mm}
  \tilde{p}_{\rm q,2}^{\rm (p)} & = &
    A_{2}^{\rm (p)} (Q, F_{1}, F_{2})\, \Bigl< k_{f} \Bigl|\, e^{-i\, \vb{k_{\rm ph}} c_{A} \vb{r}}\; \Bigr| \,k_{i} \Bigr> +
    B_{2}^{\rm (p)} (Q, F_{1}, F_{2})\, \Bigl< k_{f} \Bigl|\, e^{-i\, \vb{k_{\rm ph}} c_{A} \vb{r}}\; V(\vb{r}_{\rm p}) \Bigr| \,k_{i} \Bigr>, \\

  \tilde{p}_{\rm q,3}^{\rm (p)} & = &
    A_{3}^{\rm (p)} (Q, F_{1}, F_{2})\, \Bigl< k_{f} \Bigl|\, e^{-i\, 2\vb{k_{\rm ph}} c_{A} \vb{r}}\; \Bigr| \,k_{i} \Bigr>, \\

\vspace{1.3mm}
  \tilde{p}_{\rm q,1}^{\rm (Ai)} & = &
    A_{1}^{\rm (Ai)} (Q, F_{1}, F_{2})\,
      \Bigl< k_{f} \Bigl|\, e^{i\, \vb{k_{\rm ph}} c_{\rm p} \vb{r}}\; e^{-i\, \vb{k}_{\rm ph} \rhobf_{A j}} \Bigr| \,k_{i} \Bigr> +
    B_{1}^{\rm (Ai)} (Q, F_{1}, F_{2})\,
      \Bigl< k_{f} \Bigl|\, e^{i\, \vb{k_{\rm ph}} c_{\rm p} \vb{r}}\; e^{-i\, \vb{k}_{\rm ph} \rhobf_{A j}} V(\vb{r}_{i}) \Bigr| \,k_{i} \Bigr>, \\

\vspace{1.3mm}
  \tilde{p}_{\rm q,2}^{\rm (Ai)} & = &
    A_{2}^{\rm (Ai)} (Q, F_{1}, F_{2})\,
      \Bigl< k_{f} \Bigl|\, e^{i\, \vb{k_{\rm ph}} c_{\rm p} \vb{r}}\; e^{-i\, \vb{k}_{\rm ph} \rhobf_{A j}} \Bigr| \,k_{i} \Bigr> +
    B_{2}^{\rm (Ai)} (Q, F_{1}, F_{2})\,
      \Bigl< k_{f} \Bigl|\, e^{i\, \vb{k_{\rm ph}} c_{\rm p} \vb{r}}\; e^{-i\, \vb{k}_{\rm ph} \rhobf_{A j}} V(\vb{r}_{i}) \Bigr| \,k_{i} \Bigr>, \\

  \tilde{p}_{\rm q,3}^{\rm (Ai)} & = &
    A_{3}^{\rm (Ai)} (Q, F_{1}, F_{2})\,
    \Bigl< k_{f} \Bigl|\, e^{i\, 2\vb{k_{\rm ph}} c_{\rm p} \vb{r}}\; e^{-i\, 2\vb{k}_{\rm ph} \rhobf_{A j}} \Bigr| \,k_{i} \Bigr>,
\end{array}
\label{eq.17.5.3}
\end{equation}
where
\begin{equation}
\begin{array}{lcl}
\vspace{1.5mm}
  A_{1}^{\rm (p)} (Q, F_{1p}, F_{2p}) & = &
    \displaystyle\frac{4Q}{\pi}
    \Bigl\{ -i\, \Bigl[ F_{1p}^{2} (1 - F_{1p}) - F_{1p} F_{2p}^{2}\, Q^{2} \Bigr] +
       \displaystyle\frac{F_{2p}\, Q^{2}}{m_{\rm p}c}\, \bigl( 2F_{1p}^{2} - F_{2p}^{2}\, Q^{2} \bigr) \Bigr\}, \\
\vspace{1.5mm}
  B_{1}^{\rm (p)} (Q, F_{1p}, F_{2p}) & = & i\,8 Q\, \displaystyle\frac{F_{1p}^{3}}{\pi m_{\rm p}c^{2}}, \\

\vspace{1.5mm}
  A_{2}^{\rm (p)} (Q, F_{1p}, F_{2p}) & = &
    2\, \bigl( F_{1p}^{2} - F_{2p}^{2}\, Q^{2} \bigr)\,
    \sqrt{2}\, \varepsilon_{mjl}\, q^{j}\, \sigma_{l} \displaystyle\sum\limits_{\alpha=1,2} \vb{e}_{m}^{(\alpha)}, \\

\vspace{0.9mm}
  B_{2}^{\rm (p)} (Q, F_{1p}, F_{2p}) & = &
    -\, 2\, \bigl( F_{1p}^{2} - F_{2p}^{2} Q^{2} \bigr)\,
    \displaystyle\frac{\sqrt{2}\, F_{1p}}{m_{\rm p}c^{2}}\,
    \varepsilon_{mjl}\, q^{j}\, \sigma_{l} \displaystyle\sum\limits_{\alpha=1,2} \vb{e}_{m}^{(\alpha)}, \\

  A_{3}^{\rm (p)} (Q, F_{1p}, F_{2p}) & = &
    \displaystyle\frac{F_{2p}Q^{2}}{2}\, \bigl( 2F_{1p}^{2} - F_{2p}^{2}\, Q^{2} \bigr),
\end{array}
\label{eq.17.5.4}
\end{equation}
and solutions for
$A_{1}^{\rm (Ai)}$, $B_{1}^{\rm (Ai)}$, $A_{2}^{\rm (Ai)}$, $B_{2}^{\rm (Ai)}$, $A_{3}^{\rm (Ai)}$
are obtained from Eqs.~(\ref{eq.17.5.4}) by changing bottom indexes $p \to i$.

\subsection{Integration over internal variables $\rhobf_{Aj}$
\label{sec.17.6}}

The matrix elements in Eqs.~(\ref{eq.17.5.3}) should be integrated over variables $\rhobf_{Aj}$.
But one can see that these matrix elements have additional functions $e^{-i\, \vb{k}_{\rm ph} \rhobf_{A j}}$ and $e^{-i\, 2\vb{k}_{\rm ph} \rhobf_{A j}}$.
The simplest way to go to first calculations of the bremsstrahlung spectra is to apply approximation.
The simples approximation is
\emph{Approximation at zero energy of emitted photon, $\vb{k}_{\rm ph} \to 0$ ($\bar{s}_{k}=1$)}.
In such a case we have
%
\begin{equation}
\begin{array}{lll}
  N_{A} (\vb{k_{\rm ph}}) =
    \Bigl\langle \psi_{A, f} (\beta_{A})\, \Bigl|\, e^{-i\, \vb{k_{\rm ph}} \rhobf_{Ai}}\; \Bigr|\,  \psi_{A,i} (\beta_{A}) \Bigr\rangle \Bigl|_{\vb{k}_{\rm ph} \to 0} = 1, &
  N_{A} (2\vb{k_{\rm ph}}) =
    \Bigl\langle \psi_{A, f} (\beta_{A})\, \Bigl|\, e^{-i\, 2\vb{k_{\rm ph}} \rhobf_{Ai}}\; \Bigr|\,  \psi_{A,i} (\beta_{A}) \Bigr\rangle \Bigl|_{\vb{k}_{\rm ph} \to 0} = 1
\end{array}
\label{eq.17.6.1}
\end{equation}
and from Eqs.~(\ref{eq.17.5.3}) we obtain
\begin{equation}
\begin{array}{lcl}
\vspace{1.3mm}
  \tilde{p}_{\rm q,1}^{\rm (p)} & = &
    A_{1}^{\rm (p)} (Q, F_{1}, F_{2})\, \Bigl< k_{f} \Bigl|\, e^{-i\, \vb{k_{\rm ph}} c_{A} \vb{r}}\; \Bigr| \,k_{i} \Bigr> +
    B_{1}^{\rm (p)} (Q, F_{1}, F_{2})\, \Bigl< k_{f} \Bigl|\, e^{-i\, \vb{k_{\rm ph}} c_{A} \vb{r}}\; V(\vb{r}_{\rm p}) \Bigr| \,k_{i} \Bigr>, \\

\vspace{1.3mm}
  \tilde{p}_{\rm q,2}^{\rm (p)} & = &
    A_{2}^{\rm (p)} (Q, F_{1}, F_{2})\, \Bigl< k_{f} \Bigl|\, e^{-i\, \vb{k_{\rm ph}} c_{A} \vb{r}}\; \Bigr| \,k_{i} \Bigr> +
    B_{2}^{\rm (p)} (Q, F_{1}, F_{2})\, \Bigl< k_{f} \Bigl|\, e^{-i\, \vb{k_{\rm ph}} c_{A} \vb{r}}\; V(\vb{r}_{\rm p}) \Bigr| \,k_{i} \Bigr>, \\

  \tilde{p}_{\rm q,3}^{\rm (p)} & = &
    A_{3}^{\rm (p)} (Q, F_{1}, F_{2})\, \Bigl< k_{f} \Bigl|\, e^{-i\, 2\vb{k_{\rm ph}} c_{A} \vb{r}}\; \Bigr| \,k_{i} \Bigr>, \\

\vspace{1.3mm}
  \tilde{p}_{\rm q,1,0}^{\rm (Ai)} & = &
    A_{1}^{\rm (Ai)} (Q, F_{1}, F_{2})\,
      \Bigl< k_{f} \Bigl|\, e^{i\, \vb{k_{\rm ph}} c_{\rm p} \vb{r}}\; 
        \Bigr| \,k_{i} \Bigr> +
    B_{1}^{\rm (Ai)} (Q, F_{1}, F_{2})\,
      \Bigl< k_{f} \Bigl|\, e^{i\, \vb{k_{\rm ph}} c_{\rm p} \vb{r}}\; 
        V(\vb{r}_{i}) \Bigr| \,k_{i} \Bigr>, \\

\vspace{1.3mm}
  \tilde{p}_{\rm q,2,0}^{\rm (Ai)} & = &
    A_{2}^{\rm (Ai)} (Q, F_{1}, F_{2})\,
      \Bigl< k_{f} \Bigl|\, e^{i\, \vb{k_{\rm ph}} c_{\rm p} \vb{r}}\; 
      \Bigr| \,k_{i} \Bigr> +
    B_{2}^{\rm (Ai)} (Q, F_{1}, F_{2})\,
      \Bigl< k_{f} \Bigl|\, e^{i\, \vb{k_{\rm ph}} c_{\rm p} \vb{r}}\; 
      V(\vb{r}_{i}) \Bigr| \,k_{i} \Bigr>, \\

  \tilde{p}_{\rm q,3,0}^{\rm (Ai)} & = &
    A_{3}^{\rm (Ai)} (Q, F_{1}, F_{2})\,
    \Bigl< k_{f} \Bigl|\, e^{i\, 2\vb{k_{\rm ph}} c_{\rm p} \vb{r}}\; 
    \Bigr| \,k_{i} \Bigr>.
\end{array}
\label{eq.17.6.2}
\end{equation}
We added bottom new index ``0'' indicating such approximation.
A general solution is
%
\begin{equation}
\begin{array}{lll}
  N_{\rm A} (\vb{k}_{\rm ph}) & = &
  \displaystyle\frac{2\kappa}{A}\, e^{-\, (a^{2} k_{x}^{2} + b^{2} k_{y}^{2} + c^{2} k_{z}^{2})\,/4}\; \cdot
  f_{1}\, (\vb{k}_{\rm ph}, n_{1} \ldots n_{\rm A}), \\

  f_{1}\, (\vb{k}_{\rm ph}, n_{1} \ldots n_{\rm A}) & = &
  \displaystyle\sum\limits_{n_{x}, n_{y},n_{z} = 0}^{n_{x} + n_{y} + n_{z} \le [N/2]}
    L_{n_{x}} \Bigl[a^{2} k_{x}^{2}/2\Bigr]\:
    L_{n_{y}} \Bigl[b^{2} k_{y}^{2}/2\Bigr]\:
    L_{n_{z}} \Bigl[c^{2} k_{z}^{2}/2\Bigr],
\end{array}
\label{eq.17.6.3}
\end{equation}
where $\kappa = 2 P/N + (N-P)/N$,
$P$ and $N$ are numbers of protons and neutrons of nucleus-target.
Upper limit in summation $[N/2]$ describes number of space states of nucleus with number of neutrons $N$
(for example, for nuclei with even number of nucleons $[N/2] = N/2$).
We will not analyze this case in current paper.

\subsection{Role of interactions for the scattered proton and nucleons of nucleus-target in matrix elements
\label{sec.17.7}}

For further calculations of matrix elements (\ref{eq.17.6.2})
we should take into account additional role of potentials $V(\vb{r}_{i})$ and $V(\vb{r}_{\rm p})$ for nucleons of nucleus and the scattered proton.
For that, we use perturbative approach.
Only in calculations of such matrix elements we assume the following.
\begin{enumerate}
\item
For scattering of proton two regions are important: internal region of nucleus-target and external region of scattering off this nucleus (as this scattered proton is in unbound state).
So, in calculations of matrix element,
the radius-vector $\vb{r}_{\rm p}$ of the scattered proton in lab. frame is close to relative distance $\vb{r}$ between this proton and center-of-mass of nucleus ($\vb{r}_{\rm p} \simeq \vb{r}$).
\item
For nucleons of the nucleus-target, external region of scattering is less important than internal region inside nucleus (as these nucleons are in bound states).
So, in calculations of matrix element we will omit potential of interactions between scattered proton and nucleus.
\footnote{We suppose to take into account corrections of this approximation in further study.}
\end{enumerate}
On such a basis, one can write:
\begin{equation}
\begin{array}{lcl}
\vspace{1.3mm}
  \Bigl< k_{f} \Bigl|\, e^{i\, \vb{k_{\rm ph}} c_{\rm p} \vb{r}}\; V(\vb{r}_{\rm p}) \Bigr| \,k_{i} \Bigr> \simeq
  \Bigl< k_{f} \Bigl|\, e^{i\, \vb{k_{\rm ph}} c_{\rm p} \vb{r}}\; V(\vb{r}) \Bigr| \,k_{i} \Bigr>, \\

\vspace{1.3mm}
  \Bigl< k_{f} \Bigl|\, e^{i\, \vb{k_{\rm ph}} c_{\rm p} \vb{r}}\; V(\vb{r}_{i}) \Bigr| \,k_{i} \Bigr> =
  \Bigl< k_{f} \Bigl|\, e^{i\, \vb{k_{\rm ph}} c_{\rm p} \vb{r}}\; \Bigr| \,k_{i} \Bigr> +
  \Bigl< k_{f} \Bigl|\, e^{i\, \vb{k_{\rm ph}} c_{\rm p} \vb{r}}\; v^{(1)}(\vb{r}) \Bigr| \,k_{i} \Bigr> \simeq
  \Bigl< k_{f} \Bigl|\, e^{i\, \vb{k_{\rm ph}} c_{\rm p} \vb{r}}\; \Bigr| \,k_{i} \Bigr>, \\

  v^{(1)}(\vb{r}) = 1 - V(\vb{r}_{i}).
\end{array}
\label{eq.17.7.1}
\end{equation}
Applying such an approximation, from Eqs.~(\ref{eq.17.6.2}) we obtain:
\begin{equation}
\begin{array}{lcl}
\vspace{1.3mm}
  \tilde{p}_{\rm q,1}^{\rm (p)} & = &
    A_{1}^{\rm (p)} (Q, F_{1}, F_{2})\, I_{3} (\vb{k}_{\rm ph}) +
    B_{1}^{\rm (p)} (Q, F_{1}, F_{2})\,  I_{4} (\vb{k}_{\rm ph}), \\

\vspace{1.3mm}
  \tilde{p}_{\rm q,2}^{\rm (p)} & = &
    A_{2}^{\rm (p)} (Q, F_{1}, F_{2})\, I_{3} (\vb{k}_{\rm ph}) +
    B_{2}^{\rm (p)} (Q, F_{1}, F_{2})\,  I_{4} (\vb{k}_{\rm ph}), \\

  \tilde{p}_{\rm q,3}^{\rm (p)} & = &
    A_{3}^{\rm (p)} (Q, F_{1}, F_{2})\,
    I_{3} (2\, \vb{k}_{\rm ph}), \\

\vspace{1.3mm}
  \tilde{p}_{\rm q,1,0}^{\rm (Ai)} & = &
    \Bigl( A_{1}^{\rm (Ai)} (Q, F_{1}, F_{2}) + B_{1}^{\rm (Ai)} (Q, F_{1}, F_{2}) \Bigr)\,
    I_{2} (\vb{k}_{\rm ph}), \\

\vspace{1.3mm}
  \tilde{p}_{\rm q,2,0}^{\rm (Ai)} & = &
    \Bigl( A_{2}^{\rm (Ai)} (Q, F_{1}, F_{2}) + B_{2}^{\rm (Ai)} (Q, F_{1}, F_{2}) \Bigr)\,
      I_{2} (\vb{k}_{\rm ph}), \\

  \tilde{p}_{\rm q,3,0}^{\rm (Ai)} & = &
    A_{3}^{\rm (Ai)} (Q, F_{1}, F_{2})\,
    I_{2} (2\,\vb{k}_{\rm ph}),
\end{array}
\label{eq.17.7.3}
\end{equation}
where integrals are defined in Eqs.~(\ref{eq.resultingformulas.3})and we introduce new one
\begin{equation}
  I_{4} (\vb{k}_{\rm ph}) = \Bigl\langle \Phi_{\rm p - nucl, f} (\vb{r})\; \Bigl|\, e^{- i\, c_{A}\, \vb{k_{\rm ph}} \vb{r}}\, V(\vb{r})\, \Bigr|\, \Phi_{\rm p - nucl, i} (\vb{r})\: \Bigr\rangle.
\label{eq.17.7.4}
\end{equation}

\section{Calculations of integrals
\label{sec.app.integrals}}

\subsection{A general case
\label{sec.app.integrals.1}}

In this Appendix we calculate integrals~(\ref{eq.18.1.1}).
%
%
We apply multipole expansion of wave function of photons, following to formalism in Sect.~D in Ref.~\cite{Maydanyuk.2012.PRC}
[see Eqs.~(29)--(31) and (24)--(28) in that paper].
Here, we obtain the following formulas for matrix elements:
%
\begin{equation}
\begin{array}{ll}
  \vspace{1mm}
  \Bigl< k_{f} \Bigl| \,  e^{-i\mathbf{k_{\rm ph}r}} \, \Bigr| \,k_{i} \Bigr>_\mathbf{r} =
  \sqrt{\displaystyle\frac{\pi}{2}}\:
  \displaystyle\sum\limits_{l_{\rm ph}=1}\,
    (-i)^{l_{\rm ph}}\, \sqrt{2l_{\rm ph}+1}\;
  \displaystyle\sum\limits_{\mu = \pm 1}
    \Bigl[ \mu\,\tilde{p}_{l_{\rm ph}\mu}^{M} - i\, \tilde{p}_{l_{\rm ph}\mu}^{E} \Bigr], \\

  \biggl< k_{f} \biggl| \,  e^{-i\mathbf{k_{\rm ph}r}} \displaystyle\frac{\partial}{\partial \mathbf{r}}\,
  \biggr| \,k_{i} \biggr>_\mathbf{r} =
  \sqrt{\displaystyle\frac{\pi}{2}}\:
  \displaystyle\sum\limits_{l_{\rm ph}=1}\,
    (-i)^{l_{\rm ph}}\, \sqrt{2l_{\rm ph}+1}\;
  \displaystyle\sum\limits_{\mu = \pm 1}
    \xibf_{\mu}\, \mu\, \times
    \Bigl[ p_{l_{\rm ph}\mu}^{M} - i\mu\: p_{l_{\rm ph}\mu}^{E} \Bigr].
\end{array}
\label{eq.app.integrals.3}
\end{equation}
On the basis of these formulas, we write solutions for integrals (for simplicity, we write solutions at $l_{i}=0$).

\vspace{1.5mm}
According to second formula in Eqs.~(\ref{eq.app.integrals.3}), the first integral is
\begin{equation}
\begin{array}{ll}
  \vb{I}_{1} =
  \biggl< \Phi_{f} \biggl| \,  e^{-i\mathbf{k_{\rm ph}r}} \displaystyle\frac{\partial}{\partial \vb{r}}\, \biggr| \,\Phi_{i} \biggr>_\mathbf{r} =
  \sqrt{\displaystyle\frac{\pi}{2}}\:
  \displaystyle\sum\limits_{l_{\rm ph}=1}\,
    (-i)^{l_{\rm ph}}\, \sqrt{2l_{\rm ph}+1}\;
  \displaystyle\sum\limits_{\mu = \pm 1}
    \xibf_{\mu}\, \mu\, \times
    \Bigl[ p_{l_{\rm ph}\mu}^{M} - i\mu\: p_{l_{\rm ph}\mu}^{E} \Bigr],
\end{array}
\label{eq.app.integrals.4}
\end{equation}
where
%
\begin{equation}
\begin{array}{lcl}
\vspace{3mm}
  p_{l_{\rm ph}\mu}^{M 0 m_{f}} 
   & = &
    - I_{M}(0, l_{f}, l_{\rm ph}, 1, \mu) \cdot J_{1}(0, l_{f},l_{\rm ph}), \\

\vspace{1mm}
  p_{l_{\rm ph}\mu}^{E 0 m_{f}} & = &
    \sqrt{\displaystyle\frac{l_{\rm ph}+1}{2l_{\rm ph}+1}} \cdot I_{E} (0,l_{f},l_{\rm ph}, 1, l_{\rm ph}-1, \mu) \cdot J_{1}(0,l_{f},l_{\rm ph}-1)\; - \\
  & - &
    \sqrt{\displaystyle\frac{l_{\rm ph}}{2l_{\rm ph}+1}} \cdot I_{E} (0,l_{f}, l_{\rm ph}, 1, l_{\rm ph}+1, \mu) \cdot J_{1}(0,l_{f},l_{\rm ph}+1)
\end{array}
\label{eq.app.integrals.5}
\end{equation}
and
\begin{equation}
\begin{array}{ccl}
  J_{1}(l_{i},l_{f},n) & = & \displaystyle\int\limits^{+\infty}_{0} \displaystyle\frac{dR_{i}(r, l_{i})}{dr}\: R^{*}_{f}(l_{f},r)\, j_{n}(kr)\; r^{2} dr.
\end{array}
\label{eq.app.integrals.6}
\end{equation}
%
Radial integrals are different by factors at exponents and additional factor $V$.
So, from the first formula in Eqs.~(\ref{eq.app.integrals.3}), one can write solution as
\begin{equation}
\begin{array}{lllll}
\vspace{0.5mm}
  I_{2} =
  \sqrt{\displaystyle\frac{\pi}{2}}\:
  \displaystyle\sum\limits_{l_{\rm ph}=1}\,
    (-i)^{l_{\rm ph}}\, \sqrt{2l_{\rm ph}+1}\;
  \displaystyle\sum\limits_{\mu = \pm 1}
    \Bigl[ \mu\,\tilde{p}_{l_{\rm ph}\mu}^{M} (-c_{p}) - i\, \tilde{p}_{l_{\rm ph}\mu}^{E} (-c_{p}) \Bigr], \\

\vspace{0.5mm}
  I_{3} =
  \sqrt{\displaystyle\frac{\pi}{2}}\:
  \displaystyle\sum\limits_{l_{\rm ph}=1}\,
    (-i)^{l_{\rm ph}}\, \sqrt{2l_{\rm ph}+1}\;
  \displaystyle\sum\limits_{\mu = \pm 1}
    \Bigl[ \mu\,\tilde{p}_{l_{\rm ph}\mu}^{M} (c_{A}) - i\, \tilde{p}_{l_{\rm ph}\mu}^{E} (c_{A}) \Bigr], \\

  I_{4} =
  \sqrt{\displaystyle\frac{\pi}{2}}\:
  \displaystyle\sum\limits_{l_{\rm ph}=1}\,
    (-i)^{l_{\rm ph}}\, \sqrt{2l_{\rm ph}+1}\;
  \displaystyle\sum\limits_{\mu = \pm 1}
    \Bigl[ \mu\, \breve{p}_{l_{\rm ph}\mu}^{M} (c_{A}) - i\, \breve{p}_{l_{\rm ph}\mu}^{E} (c_{A}) \Bigr],
\end{array}
\label{eq.app.integrals.7}
\end{equation}
where
%
\begin{equation}
\begin{array}{llll}
\vspace{3mm}
  \tilde{p}_{l_{\rm ph}\mu}^{M0 m_{f}} (c) =
    \tilde{I}\,(0,l_{f},l_{\rm ph}, l_{\rm ph}, \mu) \cdot \tilde{J}\, (c, 0,l_{f},l_{\rm ph}), \\

\vspace{3mm}
  \tilde{p}_{l_{\rm ph}\mu}^{E0 m_{f}} (c) =
    \sqrt{\displaystyle\frac{l_{\rm ph}+1}{2l_{\rm ph}+1}} \tilde{I}\,(0,l_{f},l_{\rm ph},l_{\rm ph}-1,\mu) \cdot \tilde{J}\,(c, 0,l_{f},l_{\rm ph}-1)\; - 
    \sqrt{\displaystyle\frac{l_{\rm ph}}{2l_{\rm ph}+1}} \tilde{I}\,(0,l_{f},l_{\rm ph},l_{\rm ph}+1,\mu) \cdot \tilde{J}\,(c, 0,l_{f},l_{\rm ph}+1), \\

\vspace{3mm}
  \breve{p}_{l_{\rm ph}\mu}^{M0 m_{f}} (c_{A}) =
    \tilde{I}\,(0,l_{f},l_{\rm ph}, l_{\rm ph}, \mu) \cdot \breve{J}\, (c_{A}, 0,l_{f},l_{\rm ph}), \\

  \breve{p}_{l_{\rm ph}\mu}^{E0 m_{f}} (c_{A}) =
    \sqrt{\displaystyle\frac{l_{\rm ph}+1}{2l_{\rm ph}+1}} \tilde{I}\,(0,l_{f},l_{\rm ph},l_{\rm ph}-1,\mu) \cdot \breve{J}\,(c_{A}, 0,l_{f},l_{\rm ph}-1)\; - 
    \sqrt{\displaystyle\frac{l_{\rm ph}}{2l_{\rm ph}+1}} \tilde{I}\,(0,l_{f},l_{\rm ph},l_{\rm ph}+1,\mu) \cdot \breve{J}\,(c_{A}, 0,l_{f},l_{\rm ph}+1)
\end{array}
\label{eq.app.integrals.8}
\end{equation}
and
\begin{equation}
\begin{array}{ccl}
  \tilde{J}\,(c, l_{i},l_{f},n) & = & \displaystyle\int\limits^{+\infty}_{0} R_{i}(r)\, R^{*}_{f}(l,r)\, j_{n}(c\, kr)\; r^{2} dr, \\
  \breve{J}\,(c_{A}, l_{i}, l_{f},n) & = & \displaystyle\int\limits^{+\infty}_{0} R_{i}(r)\, R^{*}_{l,f}(r)\, V(\mathbf{r})\, j_{n}(c_{A}\,kr)\; r^{2} dr.
\end{array}
\label{eq.app.integrals.9}
\end{equation}
%
Also we calculate properties
\begin{equation}
\begin{array}{llll}
  \displaystyle\sum\limits_{\alpha=1,2} \vb{e}^{(\alpha)} \cdot \vb{I}_{1} =
  \sqrt{\displaystyle\frac{\pi}{2}}\:
  \displaystyle\sum\limits_{l_{\rm ph}=1}\, (-i)^{l_{\rm ph}}\, \sqrt{2l_{\rm ph}+1}\;
  \displaystyle\sum\limits_{\mu=\pm 1} \mu\,h_{\mu}\, \bigl(p_{l_{\rm ph}, \mu}^{M} + p_{l_{\rm ph}, -\mu}^{E} \bigr), \\

  (\vb{e}_{\rm x} + \vb{e}_{\rm z})\,  \displaystyle\sum\limits_{\alpha=1,2} \Bigl[ \vb{I}_{1} \times \vb{e}^{(\alpha)} \Bigr] =
  \sqrt{\displaystyle\frac{\pi}{2}}\:
  \displaystyle\sum\limits_{l_{\rm ph}=1}\, (-i)^{l_{\rm ph}}\, \sqrt{2l_{\rm ph}+1}\;
  \displaystyle\sum\limits_{\mu=\pm 1} \mu\,h_{\mu}\, \bigl(p_{l_{\rm ph}, \mu}^{M} - p_{l_{\rm ph}, -\mu}^{E} \bigr).
\end{array}
\label{eq.app.integrals.2.1}
\end{equation}

\subsection{Case of $l_{i}=0$, $l_{f}=1$, $l_{\rm ph}=1$
\label{sec.app.model.simplestcase}}

In a case of $l_{i}=0$, $l_{f}=1$, $l_{\rm ph}=1$ integrals (\ref{eq.app.integrals.4}), (\ref{eq.app.integrals.7}) are simplified to
\begin{equation}
\begin{array}{llllll}
\vspace{0.5mm}
  \vb{I}_{1} =
  -i\, \sqrt{\displaystyle\frac{3\pi}{2}}\:
  \displaystyle\sum\limits_{\mu = \pm 1}
    \xibf_{\mu}\, \mu\, \times
    \Bigl[ p_{l_{\rm ph}\mu}^{M} - i\mu\: p_{l_{\rm ph}\mu}^{E} \Bigr], &

  I_{3} =
  -i\, \sqrt{\displaystyle\frac{3\pi}{2}}\:
  \displaystyle\sum\limits_{\mu = \pm 1}
    \Bigl[ \mu\,\tilde{p}_{l_{\rm ph}\mu}^{M} (c_{A}) - i\, \tilde{p}_{l_{\rm ph}\mu}^{E} (c_{A}) \Bigr], \\

  I_{2} =
  -i\, \sqrt{\displaystyle\frac{3\pi}{2}}\:
  \displaystyle\sum\limits_{\mu = \pm 1}
    \Bigl[ \mu\,\tilde{p}_{l_{\rm ph}\mu}^{M} (-c_{p}) - i\, \tilde{p}_{l_{\rm ph}\mu}^{E} (-c_{p}) \Bigr], &

  I_{4} =
  -i\, \sqrt{\displaystyle\frac{3\pi}{2}}\:
  \displaystyle\sum\limits_{\mu = \pm 1}
    \Bigl[ \mu\, \breve{p}_{l_{\rm ph}\mu}^{M} (c_{A}) - i\, \breve{p}_{l_{\rm ph}\mu}^{E} (c_{A}) \Bigr],
\end{array}
\label{eq.app.simplestcase.1}
\end{equation}
where [see Eqs.~(\ref{eq.app.integrals.8})]
\begin{equation}
\begin{array}{lll}
\vspace{3mm}
  p_{l_{\rm ph}\mu}^{M 0 m_{f}} = - I_{M}(0, 1, 1, 1, \mu) \cdot J_{1}(0, 1, 1), \\

\vspace{3mm}
  p_{l_{\rm ph}\mu}^{E 0 m_{f}} =
    \sqrt{\displaystyle\frac{2}{3}}\, I_{E} (0,1,1,1,0, \mu) \cdot J_{1}(0,1,0) -
    \sqrt{\displaystyle\frac{1}{3}}\, I_{E} (0,1,1,1,2, \mu) \cdot J_{1}(0,1,2), \\

\vspace{3mm}
  \tilde{p}_{l_{\rm ph}\mu}^{M0 m_{f}} (c) = \tilde{I}\,(0,1,1,1, \mu) \cdot \tilde{J}\, (c, 0,1,1), \\

\vspace{3mm}
  \tilde{p}_{l_{\rm ph}\mu}^{E0 m_{f}} (c) =
    \sqrt{\displaystyle\frac{2}{3}} \tilde{I}\,(0,1,1,0,\mu) \cdot \tilde{J}\,(c, 0,1,0) -
    \sqrt{\displaystyle\frac{1}{3}} \tilde{I}\,(0,1,1,2,\mu) \cdot \tilde{J}\,(c, 0,1,2), \\

\vspace{3mm}
  \breve{p}_{l_{\rm ph}\mu}^{M0 m_{f}} (c_{A}) = \tilde{I}\,(0,1,1,1, \mu) \cdot \breve{J}\, (c_{A}, 0,1,1), \\

  \breve{p}_{l_{\rm ph}\mu}^{E0 m_{f}} (c_{A}) =
    \sqrt{\displaystyle\frac{2}{3}} \tilde{I}\,(0,1,1,0,\mu) \cdot \breve{J}\,(c_{A}, 0,1,0) -
    \sqrt{\displaystyle\frac{1}{3}} \tilde{I}\,(0,1,1,2,\mu) \cdot \breve{J}\,(c_{A}, 0,1,2).
\end{array}
\label{eq.app.simplestcase.2}
\end{equation}

Results of calculation of angular integrals are
[we omit details of calculations in this paper]:
%
%
\begin{equation}
\begin{array}{llllll}
\vspace{0.8mm}
  I_{E}\, (0, 1, 1, 1, 0, \mu) = \sqrt{\displaystyle\frac{1}{24\pi}}, &
  I_{M}\, (0, 1, 1, 1, \mu) = 0, &
  I_{E}\, (0, 1, 1, 1, 2, \mu) = \displaystyle\frac{47}{240} \sqrt{\displaystyle\frac{3}{2\pi}}, \\

  \tilde{I}\, (0, 1, 1, 0, \mu) = 0, &
  \tilde{I}\, (0, 1, 1, 1, \mu) = \displaystyle\frac{\mu}{2\sqrt{2\pi}}, &
  \tilde{I}\, (0, 1, 1, 2, \mu) = 0,
\end{array}
\label{eq.app.simplestcase.3}
\end{equation}
and matrix elements (\ref{eq.app.simplestcase.2}) are simplified to
%
\begin{equation}
\begin{array}{lllllllll}
\vspace{2mm}
  p_{l_{\rm ph}\mu}^{M 0 m_{f}} = 0, &
  p_{l_{\rm ph}\mu}^{E 0 m_{f}} =
    \displaystyle\frac{1}{6} \sqrt{\displaystyle\frac{1}{\pi}} \cdot J_{1}(0,1,0) -
    \displaystyle\frac{47}{240} \sqrt{\displaystyle\frac{1}{2\pi}} \cdot J_{1}(0,1,2), \\

\vspace{2mm}
  \tilde{p}_{1 \mu}^{M0 m_{f}} (c) = \displaystyle\frac{\mu}{2\sqrt{2\pi}} \cdot \tilde{J}\, (c, 0,1,1), &
  \tilde{p}_{1 \mu}^{E0 m_{f}} (c) = 0, \\

  \breve{p}_{1\mu}^{M0 m_{f}} (c_{A}) = \displaystyle\frac{\mu}{2\sqrt{2\pi}} \cdot \breve{J}\, (c_{A}, 0,1,1), &
  \breve{p}_{1\mu}^{E0 m_{f}} (c_{A}) = 0.
\end{array}
\label{eq.app.simplestcase.4}
\end{equation}

For the coherent matrix elements from Eqs.~(\ref{eq.resultingformulas.8}) we obtain:
\begin{equation}
\begin{array}{lll}
\vspace{1.5mm}
  M_{p}^{(E,\, {\rm dip})} =
  -\, \hbar\, (2\pi)^{3}\,
  \displaystyle\frac{2\, \mu_{N}\,  m_{\rm p}}{\mu}\;
  Z_{\rm eff}^{\rm (dip)}\: \sqrt{3\pi} \cdot
  p_{l_{\rm ph}=1, -\mu}^{E}, \\

  M_{p}^{(M,\, {\rm dip})} =
  -\, i\, \hbar\, (2\pi)^{3}\, \displaystyle\frac{\mu_{N}\, \alpha_{M}}{\mu}\: \sqrt{3\pi} \cdot
  p_{l_{\rm ph=1}, -\mu}^{E}.
\end{array}
\label{eq.app.simplestcase.8}
\end{equation}

Taking into account Eq.~(\ref{eq.app.simplestcase.4}) and $|J_{1}(0,1,0)| > |J_{1}(0,1,1)|$ and $|J_{1}(0,1,0)| > |J_{1}(0,1,2)|$,
we obtain
\begin{equation}
\begin{array}{lll}
\vspace{1.5mm}
  M_{p}^{(E,\, {\rm dip})} \simeq

  -\, \hbar\, (2\pi)^{3}\,
  \displaystyle\frac{2\, \mu_{N}\,  m_{\rm p}}{\mu}\;
  Z_{\rm eff}^{\rm (dip)} \cdot \displaystyle\frac{\sqrt{3}}{6} \cdot J_{1}(0,1,0), \\

  M_{p}^{(M,\, {\rm dip})} \simeq

  -\, i\, \hbar\, (2\pi)^{3}\, \displaystyle\frac{\mu_{N}\, \alpha_{M}}{\mu} \cdot \displaystyle\frac{\sqrt{3}}{6} \cdot J_{1}(0,1,0)
\end{array}
\label{eq.app.simplestcase.10}
\end{equation}
and we find property
\begin{equation}
\begin{array}{lll}
  M_{p}^{(E,\, {\rm dip})} + M_{p}^{(M,\, {\rm dip})} \simeq
  M_{p}^{(E,\, {\rm dip})} \cdot \Bigl\{ 1 + i \displaystyle\frac{\alpha_{M}}{2\, m_{\rm p}\, Z_{\rm eff}^{\rm (dip)}} \Bigr\}.
\end{array}
\label{eq.app.simplestcase.11}
\end{equation}

For the incoherent matrix elements from Eqs.~(\ref{eq.resultingformulas.2}) we obtain ($k_{\rm ph} = |\vb{k}_{\rm ph}|$):
\begin{equation}
\begin{array}{lll}
\vspace{1.4mm}
  M_{\Delta M} & = &
  \hbar\, (2\pi)^{3}\, \mu_{N}\, f_{1}\, k_{\rm ph}\, Z_{\rm A} (\vb{k}_{\rm ph}) \cdot \displaystyle\frac{\sqrt{3}}{2} \cdot \tilde{J}\, (-c_{p}, 0,1,1), \\

  M_{k} & = &
    -\, \hbar\, (2\pi)^{3}\, \mu_{N} \cdot k_{\rm ph}\, z_{\rm p}\: \mu_{\rm p}^{\rm (an)} \cdot \displaystyle\frac{\sqrt{3}}{2} \cdot \tilde{J}\, (c_{A}, 0,1,1) -
      \displaystyle\frac{\bar{\mu}_{\rm pn}^{\rm (an)}}{f_{1}} \cdot M_{\Delta M}.
\end{array}
\label{eq.app.simplestcase.13}
\end{equation}
In the used approximations, the effective electric charge and other parameters are
\begin{equation}
\begin{array}{llllll}
  Z_{\rm eff}^{\rm (dip)} = \displaystyle\frac{m_{A}\, F_{p,\, {\rm el}} - m_{p}\, F_{A,\, {\rm el}} }{m_{p} + m_{A}}, &
  \alpha_{M} = \Bigl[ Z_{\rm A} (\vb{k}_{\rm ph})\: m_{p}\, \bar{\mu}_{\rm pn}^{\rm (an)} - z_{\rm p}\, m_{A}\, \mu_{\rm p}^{\rm (an)} \Bigr] \cdot \displaystyle\frac{m_{p}}{m_{p} + m_{A}}, &
  f_{1} = \displaystyle\frac{A-1}{2A}\: \bar{\mu}_{\rm pn}^{\rm (an)}.
\end{array}
\label{eq.app.simplestcase.15}
\end{equation}

From Eqs.~(\ref{eq.17.resultingformulas.7}) and (\ref{eq.17.resultingformulas.3}) we obtain the following matrix elements of bremsstrahlung with form factors
\begin{equation}
\begin{array}{llllllll}
\vspace{1.0mm}
  & M_{\rm form factor}^{(1)} =
  (2\pi)^{3}\: \displaystyle\frac{\hbar\, e}{\sqrt{2}} \cdot
  \displaystyle\frac{z_{p}\, F_{2p}}{f_{p}}\:
  \Bigl\{
    \Bigl[ A_{1}^{\rm (p)} (Q, F_{1}, F_{2}) + A_{2}^{\rm (p)} (Q, F_{1}, F_{2}) \Bigr]\, I_{3} (\vb{k}_{\rm ph}) +
    \Bigl[ B_{1}^{\rm (p)} (Q, F_{1}, F_{2}) + B_{1}^{\rm (p)} (Q, F_{1}, F_{2}) \Bigr]\, I_{4} (\vb{k}_{\rm ph})
  \Bigr\}, \\

  & M_{\rm form factor}^{(2)} =
  (2\pi)^{3}\; \sqrt{\displaystyle\frac{\hbar w_{\rm ph}}{2\pi\, c^{2}}}\,
  \displaystyle\frac{\pi\hbar}{w_{\rm ph}}\, 4e^{2} \cdot
    \displaystyle\frac{z_{p}^{2}\, F_{2p}}{f_{p}\,m_{p}} \cdot
    A_{3}^{\rm (p)} (Q, F_{1}, F_{2})\,
    I_{3} (2\, \vb{k}_{\rm ph})
    .
\end{array}
\label{eq.app.simplestcase.formfactors.1}
\end{equation}
Coefficients $A_{1}^{\rm (p)}$, $B_{1}^{\rm (p)}$, $A_{2}^{\rm (p)}$, $B_{2}^{\rm (p)}$, $A_{3}^{\rm (p)}$ are given in Eqs.~(\ref{eq.17.resultingformulas.4});
function $f_{\rm p}$ is given in Eq.~(\ref{eq.7.1.4}).
Integrals are calculated [from Eqs.~(\ref{eq.app.simplestcase.1})]
\begin{equation}
\begin{array}{lllll}
  I_{3} = -\, i\, \displaystyle\frac{\sqrt{3}}{2} \cdot \tilde{J}\, (c_{A}, 0,1,1), &
  I_{4} = -\, i\, \displaystyle\frac{\sqrt{3}}{2} \cdot \breve{J}\, (c_{A}, 0,1,1).
\end{array}
\label{eq.app.simplestcase.formfactors.2}
\end{equation}

\section{Calculations of matrix elements
\label{sec.app.4}}

\subsection{Matrix elements of coherent and incoherent types
\label{sec.app.4.1}}

Logic of calculation of the matrix element $\vb{D}_{A1,\, {\rm el}} (\vb{k}_{\rm ph})$ is given in Appendic~C in Ref.~\cite{Maydanyuk_Zhang.2015.PRC}.
Taking into account upper limit $A-1$ in summation in Eq.~(C2) in that paper,
we obtain:
\begin{equation}
\begin{array}{lll}
  \vb{D}_{A1,\, {\rm el}} (\vb{k}_{\rm ph}) =
  \displaystyle\frac{\hbar}{2}\; \displaystyle\frac{A-1}{A}\; \vb{k}_{\rm ph}\; Z_{\rm A} (\vb{k}_{\rm ph}),
\end{array}
\label{eq.app.3.1.3.6}
\end{equation}
where
$Z_{\rm A} (\vb{k}_{\rm ph})$ is form factor of nucleus defined in Eq.~(A9) in Ref.~\cite{Maydanyuk_Zhang.2015.PRC} and calculated in Appendix A in Ref.~\cite{Maydanyuk_Zhang.2015.PRC}%
\footnote{We correct old formula (A9) in Ref.~\cite{Maydanyuk_Zhang.2015.PRC} by including new factor $s_{k}$.}.
%
%
%
%
%
Taking into account $\vb{e}^{(\alpha)} \cdot \vb{k}_{\rm ph} = 0$ (as ortogonality of vectors $\vb{e}^{(\alpha)}$ and $\vb{k}_{\rm ph}$), we obtain:
\begin{equation}
\begin{array}{llllllll}
  \vb{e}^{(\alpha)} \cdot \vb{D}_{\alpha 1,\, {\rm el}} = 0, &
  \vb{e}^{(\alpha)} \cdot \vb{D}_{A 1,\, {\rm el}} = 0.
\end{array}
\label{eq.app.3.1.3.8}
\end{equation}
Similar calculation for $\vb{D}_{A 2,\, {\rm el}}$ in Eq.~(\ref{eq.app.13.1.9.b}) gives
\begin{equation}
\begin{array}{llllllll}
  \vb{D}_{A 2,\, {\rm el}} \sim \vb{k}_{\rm ph}, &
  \vb{e}^{(\alpha)} \cdot \vb{D}_{A 2,\, {\rm el}} = 0.
\end{array}
\label{eq.app.3.2.5}
\end{equation}


Let's calculate the first matrix element of incoherent type $D_{A 1,\, {\rm mag}} (\vb{e}^{(\alpha)})$ defined in Eqs.~(\ref{eq.app.13.1.9.c}).
We use many-nucleon function of the nucleus $\psi_{\rm nucl} (\beta_{A})$
defined in Eq.~(12) Ref.~\cite{Maydanyuk_Zhang.2015.PRC} on the basis of one-nucleon functions $\psi_{\lambda_{s}}(s)$
[here one-nucleon functions $\psi_{\lambda_{s}}(s)$ represent multiplication of space and spin-isospin
functions as $\psi_{\lambda_{s}} (s) = \varphi_{n_{s}} (\vb{r}_{s})\, \bigl|\, \sigma^{(s)} \tau^{(s)} \bigr\rangle$,
where $\varphi_{n_{s}}$ is the space function of the nucleon with number $s$,
$n_{s}$ is the number of state of the space function of the nucleon with number $s$,
$\bigl|\, \sigma^{(s)} \tau^{(s)} \bigr\rangle$ is the spin-isospin function of the nucleon with number $s$].
%
Using logic of calculation of matrix elements in Eqs.~(A1)--(A4) in Ref.~\cite{Maydanyuk_Zhang.2015.PRC},
we obtain
\begin{equation}
\begin{array}{lll}
  D_{A 1,\, {\rm mag}} (\vb{e}^{(\alpha)}) & = &
    \displaystyle\frac{1}{A}\,
    \displaystyle\sum\limits_{j=1}^{A-1}
    \displaystyle\sum\limits_{k=1}^{A,s}
      \mu_{j}^{\rm (an)}\,
    \Bigl\langle \varphi_{k}(\rhobf_{j}) \cdot \sigma^{(s)} (\rhobf_{j}) \tau^{(s)} (\rhobf_{j}) \bigr\rangle \Bigl|\,
      e^{-i\, \vb{k_{\rm ph}} \rhobf_{Aj}}\; \sigmabf \cdot \bigl[ \vb{\tilde{p}}_{Aj} \times \vb{e}^{(\alpha)} \bigr]
    \Bigl|\, \varphi_{k}(\rhobf_{j}) \cdot \sigma^{(s)} (\rhobf_{j}) \tau^{(s)} (\rhobf_{j}) \Bigr\rangle.
\end{array}
\label{eq.app.matr_el.mag.1.1.2}
\end{equation}

\subsubsection{Summation over isospin and spin states
\label{sec.app.matr_el.mag.1.2}}

We perform summation in Eq.~(\ref{eq.app.matr_el.mag.1.1.2}) over isospin states.
In contrast to calculations of $\vb{D}_{A 1,\, {\rm el}}$, terms with neutron states in Eq.~(\ref{eq.app.matr_el.mag.1.1.2}) are not zero.
So, we obtain
\begin{equation}
\begin{array}{lll}
  D_{A 1,\, {\rm mag}} (\vb{e}^{(\alpha)}) = &
    \displaystyle\frac{\bar{\mu}_{\rm pn}^{\rm (an)}}{A}
    \displaystyle\sum\limits_{j=1}^{A-1}
    \displaystyle\sum\limits_{k=1}^{A}
    \displaystyle\sum\limits_{s_{i}, s_{f} = \pm 1/2}
    \bigl\langle\, \sigma_{s_{f}} (\rhobf_{i}) \bigl|\, \sigmabf\, \bigr|\, \sigma_{s_{i}} (\rhobf_{i}) \bigr\rangle \cdot

    \Bigl\langle\, \varphi_{k, f} (\rhobf_{j})
    \Bigl|\,  e^{-i\, \vb{k_{\rm ph}} \rhobf_{Aj}}\, \bigl[ \vb{\tilde{p}}_{Aj} \times \vb{e}^{(\alpha)} \bigr]
    \Bigr|\, \varphi_{k, i} (\rhobf_{j})
    \Bigr\rangle,
\end{array}
\label{eq.app.matr_el.mag.1.3.1}
\end{equation}
where
$s_{i}, s_{f}$ are indexes of spin states,
$k$ is index of space state,
$\bar{\mu}_{\rm pn}^{\rm (an)} = \mu_{\rm p}^{\rm (an)} + \kappa\,\mu_{\rm n}^{\rm (an)}$
($\mu_{\rm pn}^{\rm (an)} = \mu_{\rm p}^{\rm (an)} +  \mu_{\rm n}^{\rm (an)}$ in previous papers),
$\kappa = (A-N)/N$, $A$ and $N$ are numbers of nucleons and neutrons of nucleus.
%
In contrast to calculation of $\vb{D}_{A 1,\, {\rm el}}$, operator of spin $\hat{\sigmabf}$ in Eq.~(\ref{eq.app.matr_el.mag.1.3.1}) should be taken into account in summation over spin states.
We use property of summation~(19) in Ref.~\cite{Maydanyuk.2012.PRC}:
\begin{equation}
\begin{array}{ccl}
  \displaystyle\sum\limits_{s_{i}, s_{f} = \pm 1/2} v_{\mu_{f}}^{*} (s_{f})\, \hat{\sigmabf}\, v_{\mu_{i}} (s_{i}) & = &
  \vb{e}_{\rm x} + \vb{e}_{\rm y}\, i\, \Bigl\{ \delta_{\mu_{i}, +1/2}\; -\; \delta_{\mu_{i}, -1/2} \Bigr\} + \vb{e}_{\rm z},
\end{array}
\label{eq.app.matr_el.mag.1.3.2}
\end{equation}
where system of unit orthogonal vectors $\vb{e}_{\rm x}$, $\vb{e}_{\rm y}$, $\vb{e}_{\rm z}$ is used (at $\vb{e}_{\rm x} = \vb{e}^{(1)}$, $\vb{e}_{\rm y} = \vb{e}^{(2)}$).
From here we obtain
\begin{equation}
\begin{array}{lll}
  D_{A 1,\, {\rm mag}} (\vb{e}^{(\alpha)}) = &
  \displaystyle\frac{\bar{\mu}_{\rm pn}^{\rm (an)}}{A}
  \displaystyle\sum\limits_{j=1}^{A-1}
  \displaystyle\sum\limits_{k=1}^{A}
    \bar{\sigmabf}_{jk} \cdot
    \Bigl[
      \Bigl\langle \varphi_{k,f}(\rhobf_{j})\, \Bigl|\, e^{-i\, \vb{k_{\rm ph}} \rhobf_{Aj}}\: \vb{\tilde{p}}_{Aj} \Bigr|\, \varphi_{k,i}(\rhobf_{j})\, \Bigr\rangle
  \times \vb{e}^{(\alpha)} \Bigr],
\end{array}
\label{eq.app.matr_el.mag.1.3.3}
\end{equation}
where a new operator is
\begin{equation}
  \bar{\sigmabf}_{jk} = \Bigl[ \vb{e}_{\rm x} + \vb{e}_{\rm y}\, i\, \Bigl\{ \delta_{\mu_{j}, +1/2}\; -\; \delta_{\mu_{j}, -1/2} \Bigr\} + \vb{e}_{\rm z} \Bigr].
\label{eq.app.matr_el.mag.1.3.4}
\end{equation}
In this formula it needs to perform summation over all space and spin states.
Let's analyze even-even nuclei.
In summation over spin states,
terms at $\vb{e}_{\rm y}$ are appeared with opposite signs,
but terms at $\vb{e}_{\rm x}$ and $\vb{e}_{\rm z}$ are the same.
On such a logic, Eq.~(\ref{eq.app.matr_el.mag.1.3.3}) is simplified as
\begin{equation}
\begin{array}{lll}
  D_{A 1,\, {\rm mag}} (\vb{e}^{(\alpha)}) = &
  \displaystyle\frac{2\, \bar{\mu}_{\rm pn}^{\rm (an)}}{A}
  \displaystyle\sum\limits_{j=1}^{A-1}
  \displaystyle\sum\limits_{k=1}^{B}
    \bar{s}_{k}
    (\vb{e}_{\rm x} + \vb{e}_{\rm z}) \cdot
    \Bigl[
      \Bigl\langle \varphi_{k,f}(\rhobf_{j})\, \Bigl|\, e^{-i\, \vb{k_{\rm ph}} \rhobf_{Aj}}\: \vb{\tilde{p}}_{Aj} \Bigr|\, \varphi_{k,i}(\rhobf_{j})\, \Bigr\rangle
  \times \vb{e}^{(\alpha)} \Bigr],
\end{array}
\label{eq.app.matr_el.mag.1.3.6}
\end{equation}
where $\bar{s}_{k}$ is a new factor taking into account different space states for the same spin states (one can take $\bar{s}_{k}=1$ for \isotope[4]{He}, etc.),
$B$ is number of states of the space wave function of protons in nucleus.

\subsubsection{Calculation of one-nucleon matrix element
\label{sec.app.matr_el.mag.1.4}}

We substitute the one-nucleon space wave function defined in Eq.~(A6) of Ref.~\cite{Maydanyuk_Zhang.2015.PRC} (according to Appendix~A.1 in that paper) to the matrix element and obtain:
\begin{equation}
\begin{array}{lll}
  D_{A 1,\, {\rm mag}} (\vb{e}^{(\alpha)}) =

  \displaystyle\frac{2\, \bar{\mu}_{\rm pn}^{\rm (an)}}{A}\;
  \displaystyle\sum\limits_{i=1}^{A-1}
  \displaystyle\sum\limits_{k=1}^{B}
    \bar{s}_{k}\, (\vb{e}_{\rm x} + \vb{e}_{\rm z}) \cdot
    \Bigl[
      \Bigl( \vb{e}_{x} J_{x}(n_{x}) + \vb{e}_{y} J_{y}(n_{y}) + \vb{e}_{z} J_{z}(n_{z}) \Bigr)
  \times \vb{e}^{(\alpha)} \Bigr],
\end{array}
\label{eq.app.matr_el.mag.1.4.1}
\end{equation}
where
\begin{equation}
\begin{array}{lcl}
\vspace{1mm}
  J_{x}(n_{x}) & = &
  - i\, \hbar\; N_{x}^{2}\;
  \displaystyle\int
    \exp[-\,\displaystyle\frac{(x_{i})^{2}}{2a^{2}}] \cdot H_{n_{x}} \Bigl(\displaystyle\frac{x_{i}}{a} \Bigr)\;
    e^{-i k_{x} x_{i}}\;
    \Bigl( \mathbf{e}_{x}\, \displaystyle\frac{d}{d x_{i}} \Bigr)\:
    \Bigl\{ \exp[ -\,\displaystyle\frac{(x_{i})^{2}}{2a^{2}}]\; H_{n_{x}} \Bigl(\displaystyle\frac{x_{i}}{a} \Bigr) \Bigr\}\;
  dx_{i}\; \times \\

& \times &
  N_{y}^{2}\, \displaystyle\int \exp[-\,\displaystyle\frac{(y_{i})^{2}}{b^{2}}]\; H_{n_{y}}^{2} \Bigl(\displaystyle\frac{y_{i}}{b} \Bigr)\; e^{-i k_{y} y_{i}}\; dy_{i} \times
  N_{z}^{2}\, \displaystyle\int \exp[-\,\displaystyle\frac{(z_{i})^{2}}{c^{2}}]\; H_{n_{z}}^{2} \Bigl(\displaystyle\frac{z_{i}}{c} \Bigr)\; e^{-i k_{z} z_{i}}\; dz_{i}
\end{array}
\label{eq.app.matr_el.mag.1.4.2}
\end{equation}
and integrals $J_{y}(n_{y})$, $J_{z}(n_{z})$ are obtained at change of indexes.
Integrating over variable $x$ by parts, we find
\begin{equation}
\begin{array}{lcl}
  J_{x}(n_{x}) & = & \vb{e}_{x}\: \displaystyle\frac{\hbar\, k_{\rm x}}{2} \cdot I_{x} (n_{x}, a)\; I_{y}\; (n_{y}, b)\; I_{z} (n_{z}, c),
\end{array}
\label{eq.app.matr_el.mag.1.4.3}
\end{equation}
where integrals $I_{x}(n_{x}, a)$ and similar ones are calculated in (A10) in Ref.~\cite{Maydanyuk_Zhang.2015.PRC}.
Substituting this solution to
Eq.~(\ref{eq.app.matr_el.mag.1.4.1}), we calculate 
(index $k$ indicates numbers of space states $n_{x}$, $n_{y}$, $n_{z}$)
\begin{equation}
\begin{array}{lll}
  D_{A 1,\, {\rm mag}} (\vb{e}^{(\alpha)}) & = &
  \displaystyle\frac{\hbar\, (A-1)}{A}\;
  \bar{\mu}_{\rm pn}^{\rm (an)}\,
  (\vb{e}_{\rm x} + \vb{e}_{\rm z}) \cdot
  \Bigl[ \vb{k}_{\rm ph} \times \vb{e}^{(\alpha)} \Bigr] \cdot
  \displaystyle\sum\limits_{k=1}^{B}
    \bar{s}_{k}\,
    I_{x} (n_{x}, a)\; I_{y}\; (n_{y}, b)\; I_{z} (n_{z}, c).
\end{array}
\label{eq.app.matr_el.mag.1.4.5}
\end{equation}
Taking properties in Eqs.~(7) in Ref.~\cite{Maydanyuk.2012.PRC} into account for vectors $\vb{k}_{\rm ph}$ and $\vb{e}^{(1,2)}$,
%
%
we obtain
\begin{equation}
\begin{array}{lll}
  \displaystyle\sum\limits_{\alpha=1,2} (\vb{e}_{\rm x} + \vb{e}_{\rm z}) \cdot \Bigl[ \vb{k}_{\rm ph} \times \vb{e}^{(\alpha)} \Bigr] =



  - k_{\rm ph} .
\end{array}
\label{eq.app.matr_el.mag.1.4.6}
\end{equation}
Using such a property, we calculate summation of Eq.~(\ref{eq.app.matr_el.mag.1.4.5}) over states of polarization (at $\bar{s}_{k} \simeq s_{k}$):
\begin{equation}
\begin{array}{lll}
  \displaystyle\sum\limits_{\alpha=1,2} D_{A 1,\, {\rm mag}} (\vb{e}^{(\alpha)}) =
  -\, \displaystyle\frac{\hbar\, (A-1)}{A}\; \bar{\mu}_{\rm pn}^{\rm (an)}\, k_{\rm ph} \cdot
  \displaystyle\sum\limits_{k=1}^{B} \bar{s}_{k}\, I_{x} (n_{x}, a)\; I_{y}\; (n_{y}, b)\; I_{z} (n_{z}, c) =
  -\, \displaystyle\frac{\hbar\, (A-1)}{2\,A}\; \bar{\mu}_{\rm pn}^{\rm (an)}\, k_{\rm ph} \cdot Z_{\rm A} (\vb{k}_{\rm ph}).
\end{array}
\label{eq.app.matr_el.mag.1.4.7}
\end{equation}

\subsection{Matrix elements $\vb{D}_{A,\, {\rm k}}$
\label{sec.app.matr_el.k.1}}

Let's calculate matrix element $\vb{D}_{A,\, {\rm k}}$ defined in Eqs.~(\ref{eq.app.13.1.9.d}).
We perform summation over isospin and spin states, following logic in Eqs.~(\ref{eq.app.matr_el.mag.1.1.2})--(\ref{eq.app.matr_el.mag.1.3.1}) for $D_{A 1,\, {\rm mag}} (\vb{e}^{(\alpha)})$,
and obtain:
\begin{equation}
\begin{array}{lll}
  \vb{D}_{A,\, {\rm k}} = &
  \displaystyle\frac{2\, \bar{\mu}_{\rm pn}^{\rm (an)}}{A}
  \displaystyle\sum\limits_{j=1}^{A}
  \displaystyle\sum\limits_{k=1}^{B}
    \bar{s}_{k}
    (\vb{e}_{\rm x} + \vb{e}_{\rm z}) \cdot
    \Bigl\langle \varphi_{k,f}(\rhobf_{j})\, \Bigl|\, e^{-i\, \vb{k_{\rm ph}} \rhobf_{Aj}}\: \Bigr|\, \varphi_{k,i}(\rhobf_{j})\, \Bigr\rangle.
\end{array}
\label{eq.app.matr_el.k.1.2.1}
\end{equation}
We substitute the one-nucleon space wave function in form~(A6) of Ref.~\cite{Maydanyuk_Zhang.2015.PRC} to the matrix element and calculate
(at $\bar{s}_{k} \simeq s_{k}$)
\begin{equation}
\begin{array}{lll}
  \vb{D}_{A,\, {\rm k}} & = &
  2\; \bar{\mu}_{\rm pn}^{\rm (an)}\,
  (\vb{e}_{\rm x} + \vb{e}_{\rm z}) \cdot
  \displaystyle\sum\limits_{k=1}^{B} \bar{s}_{k}
    I_{x} (n_{x}, a)\; I_{y}\; (n_{y}, b)\; I_{z} (n_{z}, c) =
  \bar{\mu}_{\rm pn}^{\rm (an)}\; (\vb{e}_{\rm x} + \vb{e}_{\rm z}) \cdot Z_{\rm A} (\vb{k}_{\rm ph}),
\end{array}
\label{eq.app.matr_el.k.1.4.2}
\end{equation}
we find
\begin{equation}
\begin{array}{lll}
  \displaystyle\sum\limits_{\alpha=1,2} \bigl[ \vb{k_{\rm ph}} \cp \vb{e}^{(\alpha)} \bigr] \cdot \vb{D}_{A,\, {\rm k}} =
  - k_{\rm ph}\; \bar{\mu}_{\rm pn}^{\rm (an)}\; Z_{\rm A} (\vb{k}_{\rm ph}).
\end{array}
\label{eq.app.matr_el.k.1.4.5}
\end{equation}
Here, we used the found property~(\ref{eq.app.matr_el.mag.1.4.6}).
We determine form factor of the scattered proton as
\begin{equation}
  \vb{D}_{p,\, {\rm k}} = \mu_{\rm p}^{\rm (an)}\; (\vb{e}_{\rm x} + \vb{e}_{\rm z}) \cdot z_{\rm p}.
\label{eq.app.matr_el.k.1.4.6}
\end{equation}

\subsection{Calculation of magnetic form factor of nucleus, effective magnetic moment and matrix element in the dipole approximation
\label{sec.app.matr_el.coh_mag}}

Let's calculate magnetic form factor of nucleus $\vb{F}_{A,\, {\rm mag}}$ given in Eq.~(\ref{eq.app.13.1.9.a}).
Taking $\vb{D}_{A,\, {\rm k}}$ in form~(\ref{eq.app.13.1.9.d}) into account,
we obtain (at $m_{Aj} = m_{\rm p}$):
\begin{equation}
  \vb{F}_{A,\, {\rm mag}} = m_{\rm p}\, \vb{D}_{A,\, {\rm k}}.
\label{eq.app.matr_el.coh_mag.3}
\end{equation}
Solution for $\vb{D}_{A,\, {\rm k}}$ was found [see Eq.~(\ref{eq.app.matr_el.k.1.4.2})].
%
Then from Eq.~(\ref{eq.app.matr_el.coh_mag.3}) we write
\begin{equation}
  \vb{F}_{A,\, {\rm mag}} = m_{\rm p}\, \bar{\mu}_{\rm pn}^{\rm (an)}\; (\vb{e}_{\rm x} + \vb{e}_{\rm z}) \cdot Z_{\rm A} (\vb{k}_{\rm ph}).
\label{eq.app.matr_el.coh_mag.5}
\end{equation}
By the same way, for the scattered proton we obtain [as in Eq.~(\ref{eq.app.13.1.9.a})]:
\begin{equation}
\begin{array}{lll}
\vspace{1.0mm}
  \vb{D}_{p,\, {\rm mag}} & = &  \mu_{\rm p}^{\rm (an)}\; (\vb{e}_{\rm x} + \vb{e}_{\rm z}) \cdot z_{\rm p}, \\

  \vb{F}_{p,\, {\rm mag}} & = &
    \displaystyle\sum_{j=1}^{}
    \mu_{p}^{\rm (an)}\, m_{p}\;
    \Bigl\langle \psi_{\rm p, f} (\beta_{A})\, \Bigl|\,
        e^{-i\, \vb{k_{\rm ph}} \rhobf_{Aj}}\, \sigmabf
    \Bigr| \psi_{\rm p, i} (\beta_{A}) \Bigr\rangle =

    m_{\rm p}\, \mu_{\rm p}^{\rm (an)}\; (\vb{e}_{\rm x} + \vb{e}_{\rm z}) \cdot z_{\rm p}.
\end{array}
\label{eq.app.matr_el.coh_mag.6}
\end{equation}
We calculate effective magnetic moment of system [from Eqs.~(\ref{eq.13.1.8}), (\ref{eq.13.1.10}):
\begin{equation}
\begin{array}{lll}
\vspace{0.7mm}
  \vb{M}_{\rm eff} (\vb{k}_{\rm ph}, \vb{r}) & = &
  e^{i\, \vb{k_{\rm ph}} \vb{r}}\,
  \Bigl[
    e^{-i\, c_{A} \vb{k_{\rm ph}} \vb{r}} \cdot m_{A}\, \mu_{\rm p}^{\rm (an)} \cdot z_{\rm p} -
    e^{i\, c_{p} \vb{k_{\rm ph}} \vb{r}} \cdot m_{p}\, \bar{\mu}_{\rm pn}^{\rm (an)} \cdot Z_{\rm A} (\vb{k}_{\rm ph})
  \Bigr] \cdot
  \displaystyle\frac{m_{p}}{m_{p} + m_{A}}\; (\vb{e}_{\rm x} + \vb{e}_{\rm z}), \\

  \vb{M}_{\rm eff}^{\rm (dip)} (\vb{k}_{\rm ph}) & = &
  \Bigl[
    z_{\rm p}\, m_{A}\, \mu_{\rm p}^{\rm (an)} -
    Z_{\rm A} (\vb{k}_{\rm ph})\: m_{p}\, \bar{\mu}_{\rm pn}^{\rm (an)}
  \Bigr] \cdot
  \displaystyle\frac{m_{p}}{m_{p} + m_{A}}\; (\vb{e}_{\rm x} + \vb{e}_{\rm z}).
\end{array}
\label{eq.app.matr_el.coh_mag.7}
\end{equation}
Now we calculate matrix element (\ref{eq.13.2.1}) as
\begin{equation}
\begin{array}{lll}
\vspace{0.5mm}
  M_{p}^{(M,\, {\rm dip})} & = &


  \hbar\, (2\pi)^{3}\, \displaystyle\frac{\mu_{N}}{\mu} \cdot \alpha_{M} \cdot
  (\vb{e}_{\rm x} + \vb{e}_{\rm z})\,
  \displaystyle\sum\limits_{\alpha=1,2} \Bigl[ \vb{I}_{1} \times \vb{e}^{(\alpha)} \Bigr],
\end{array}
\label{eq.app.matr_el.coh_mag.8}
\end{equation}
where
\begin{equation}
  \alpha_{M} =
  \Bigl[
    Z_{\rm A} (\vb{k}_{\rm ph})\: m_{p}\, \bar{\mu}_{\rm pn}^{\rm (an)} -
    z_{\rm p}\, m_{A}\, \mu_{\rm p}^{\rm (an)}
  \Bigr] \cdot
  \displaystyle\frac{m_{p}}{m_{p} + m_{A}}.
\label{eq.app.matr_el.coh_mag.9}
\end{equation}


\end{document}